\documentclass[12pt]{iopart}

\usepackage{harvard}
\usepackage{graphicx,graphics}

\def\fun#1#2{\lower3.6pt\vbox{\baselineskip0pt\lineskip.9pt
\ialign{$\mathsurround=0pt#1\hfil##\hfil$\crcr#2\crcr\sim\crcr}}}
\def\lap{\mathrel{\mathpalette\fun <}}
\def\gap{\mathrel{\mathpalette\fun >}}
\def\beq{\begin{equation}}
\def\eeq{\end{equation}}
\def\mh{M_{\bullet}}
\def\msun{M_\odot}
\def\ms{m_\star}
\def\m12{M_{12}}
\def\beq{\begin{equation}}
\def\eeq{\end{equation}}

\begin{document}

\title[Dynamics of galaxy cores]{Dynamics of galaxy cores and supermassive black holes}

\author{David Merritt}

\address{Department of Physics, 
Rochester Institute of Technology,
Rochester, NY 14623, USA}
\ead{merritt@astro.rit.edu}
\begin{abstract}
Recent work on the dynamical evolution of galactic nuclei containing
supermassive black holes is reviewed.
Topics include  galaxy structural properties;
collisionless and collisional equilibria;
loss-cone dynamics;
and dynamics of binary and multiple 
supermassive black holes.

\end{abstract}

\maketitle

\section{Introduction}

Galaxy cores are the hosts of supermassive black holes (SBHs), 
the engines of quasars and of active galactic nuclei.  
There is increasing evidence that SBHs play an important
role in the formation and global evolution of galaxies and of the
intergalactic medium (IGM) \cite{growing-05}.  
The energy released by the formation and growth of SBHs
must have had a major impact on how gas cooled to form galaxies and
galaxy clusters \cite{SR-98}.  
While the detailed history of SBH
growth is still being debated, much work has focused on the
possibility that the seeds of SBHs were black holes of much smaller
mass---either remnants of the first generation of stars, so-called
``Population III black holes'' \cite{MR-01}, 
or the (still speculative) ``intermediate-mass
black holes'' (IMBHs), remnants of massive stars that form in dense
star cluster via physical collisions between stars \cite{PZ-04}.

These exciting developments have led to a renewed interest
in the dynamics and evolution of
galactic nuclei.
Early theoretical studies 
\cite{SS-66,SS-67,colgate-67,sanders-70}
emphasized stellar encounters and collisions as the dominant 
physical processes.
In these models, 
the density of a compact ($\rho\gap 10^6\msun {\rm pc}^{-3}$)
stellar system gradually increases as energetic 
stars are scattered into elongated orbits via two-body 
(star-star) gravitational encounters.
The increase in density leads to a higher rate of 
physical collisions between stars; when collision velocities
exceed $\sim 10^3$ km s$^{-1}$, roughly the escape velocity
from a star, collisions
liberate gas that falls to the center of the system and 
condenses into new stars which undergo further collisions.
\citeasnoun{BR-78} argued that the evolution of a dense
nucleus would lead inevitably to the formation of a massive black hole
at the center, either by runaway stellar mergers or by
creation of a massive gas cloud which collapses.
Subsequent studies \cite{duncan-83,QS-87,QS-89,DDC-87b,DDC-87a}
 included ``seed'' BHs which grow via accretion of stars
or gas liberated by stellar collisions or tidal disruptions.

A fundamental time scale in these models is the relaxation 
time determined by the stars, or
\beq
T_r \approx {v_{rms}^3\over G^2m_*\rho\ln\Lambda}
\label{eq:trelax}
\eeq
\cite{spitzer-87},
where $v_{rms}$ is the stellar velocity dispersion,
$m_*$ and $\rho$ are the stellar mass and mass density,
and $\ln\Lambda\approx 10^1$ is the Coulomb logarithm.
In a time $T_r$, gravitational (not physical) encounters 
between stars can exchange orbital energy and angular momentum,
and stars in the high-velocity tail of the 
distribution will be ejected.
The result, after a time $T_{cc} \approx 10^2T_r$,
is ``core collapse'': the core shrinks to zero size
and infinite density.
In order for many of the evolutionary models
cited above to be viable, $T_{cc}$ must be
shorter than the age of the universe, i.e.
$T_r$ must be less than $\sim 10^8$ yr, implying
very high stellar densities.
Physical collisions between stars
also feature prominently in many
of the early models.
Collisions occur on a time scale
that is longer than $T_r$ by a factor 
$(\ln\Lambda)\Theta^2/(1+\Theta)\approx 10^1$ where
$\Theta$, the ``Safronov number'' \cite{safronov-60},
is of order unity for stars in a galactic nucleus.

The necessity of attaining high densities in order for these
evolutionary models to work -- much higher densities than could be confirmed 
via direct observation at the time (or indeed now)
-- was clearly recognized.
For instance, Saslaw (1973) noted that
\begin{quote}
It is an extrapolation from the observations of galaxies
we have discussed to the idea that even more dense stellar
systems exist...Yet this follows naturally enough from
the observations of quasars and the realization that the central
density of massive compact stellar systems increases with age.
\end{quote}
\citeasnoun{spitzer-71} remarked
\begin{quote}
...the rate of dynamical evolution will depend on how compact
is the stellar system resulting from initial gas inflow.
If this rate of evolution is slow, activity will not
begin for a long time.
In fact, in some systems there might be a wait of $10^{12}$ 
years before the fireworks begin.
\end{quote}

\noindent
A number of developments since the 1970s have led
to a qualitatively different picture of the dynamics and
evolution of galactic nuclei.

\begin{itemize}

\item
SBHs are now known to be ubiquitous components of
galactic nuclei, and, at least in the most massive
galaxies, to have been present with roughly their
current masses ($\sim 10^9\msun$) since very early times, 
as soon as $\sim 10^9$ yr after the Big Bang
\cite{fan-01,fan-03}.
Evolution of galactic nuclei during and after
the era of peak quasar activity therefore 
took place with the SBHs already in place, 
and processes like core collapse 
and the buildup of massive stars via collisions
could not have occurred after this time due to the 
inhibiting effect of the SBH's gravitational field.
\item
Observations with the Hubble Space Telescope 
have elucidated the run of stellar density and velocity dispersion
near the centers of nearby galaxies
\cite{crane-93,ferrarese-94,lauer-95}.
In the majority of galaxies massive enough to contain SBHs, 
the implied central relaxation time is much greater than the 
age of the universe,
due both to the (relatively) low stellar densities and also
to the presence of a SBH, which increases $v_{rms}$
\cite{faber-97,acs6}.
Only the smallest galaxies known to harbor SBHs
have nuclear relaxation times shorter than 
$10^{10}$ yr.
The bulge of the Milky Way is in this category, as is the nearby
dwarf elliptical galaxy M32; both have nuclear relaxation times
$T_r\approx 10^{9.5}$ yr,
short  enough that gravitational encounters between stars could
have influenced the stellar distribution near the SBH.
But in the majority of galaxies, the long relaxation times
imply that nuclear structure will still reflect to a large
extent the details of the nuclear formation process.

\item
Since the 1980s, the dominant model for the formation
of elliptical galaxies and bulges -- the stellar
systems that contain SBHs -- has been the merger model
\cite{toomre-77}.
Galaxy mergers are complex phenomena, but an almost certain consequence of
a merger is the infall of the progenitor galaxies' SBHs into the
nucleus of the merged system, resulting in the formation
of a binary SBH \cite{BBR-80}.
Such a massive binary would inject a 
substantial amount of energy into the stellar motions,
enough to determine the structure of the remnant
core.
Indeed this process is probably responsible for
the low densities, and long relaxation times, at the centers
of the brighter elliptical galaxies.

\item
Quasars and active nuclei are now believed to 
be powered by infall of gas onto a pre-existing SBH.
The bulk of the gas is believed to originate far from 
the nucleus and to be driven inward by gravitational torques
\citeaffixed{shlosman-90}{e.g.}.
The fact that quasar activity peaks at approximately
the same epoch as galaxy mergers is seen as
strong support for this picture, since mergers are
efficient at driving gas into a nucleus 
\citeaffixed{hk-00}{e.g.}.
Capture or disruption of stars by the SBH are now generally
believed to be
energetically insignificant, at least in a time-averaged
sense.

\end{itemize}

A  recurring theme of this article will be the
distinction between {\it collisionless}
nuclei -- which have central relaxation times longer
than $10^{10}$ yr -- and {\it collisional}  nuclei
in which $T_r\lap 10^{10}$ yr (\S 3).
As noted above, the majority of observed nuclei are
``collisionless'' in this sense.
The morphology of a collisionless nucleus
is constrained only by the requirement that the stellar
phase-space density $f$ satisfy Jeans's theorem, i.e.
that $f$ be constant along orbits, as they are defined
in the combined gravitational potential of the SBH
and the stars.
This weak condition is consistent with a wide
variety of possible equilibrium configurations (\S 4), 
including non-axisymmetric nuclei, 
and nuclei in which the majority of orbits are chaotic.
In a nucleus with $T_r\lap 10^{10}$ yr, on the other hand, 
the stellar distribution will have had time to evolve to a 
more strongly constrained, collisionally relaxed steady state
(\S 5).
For a single-mass population of stars
moving in the Keplerian potential of a black hole, the steady-state
density profile is $\rho(r) \approx \rho_0r^{-7/4}$, the so-called
Bahcall-Wolf (1976) solution.  
The distribution of stars near the Milky Way SBH appears
to be consistent with this collisionally-relaxed form 
\cite{schoedel-06} but there is no evidence for Bahcall-Wolf
cusps in any other galaxies.

The recent explosion of data from the Galactic center
has motivated a large and ever-expanding
number of theoretical studies of the central star cluster
and its evolution due to collisional processes
(as reviewed by \citename{alexander-05} \citeyear*{alexander-05}).
It is important to keep in mind that the Galactic
center is extreme, not only in the sense of harboring the nearest
SBH, but also in the sense of having the {\it smallest}
SBH with a well-determined mass \cite{FF-05}.
The bulge of the Milky Way is also one of the faintest 
systems known to contain a SBH, and its atypically high central 
density and short relaxation time are consistent with its
low luminosity given 
the parameter correlations defined by more luminous  systems (\S3).
As important as the Galactic center is, it is not typical
of the majority of nuclei known to contain SBHs.

The distinction between collisionless and collisional
nuclei is important in the context of star-SBH interactions.
The ``loss cone'' of a SBH, i.e. the set of orbits that intersect
its event horizon or tidal disruption sphere, is rapidly
depopulated in a spherical or axisymmetric nucleus, 
and continued feeding of stars to the SBH requires some mechanism
for loss-cone repopulation; typically this is assumed to be
gravitational encounters between stars.
Classical loss cone theory \cite{FR-76,LS-77,CK-78}
was worked out in the late 1970s
in  the context of massive black holes in globular 
clusters.
Globular  clusters are many relaxation times old, and loss cone
dynamics was approached from the point
of view that the distribution of stars 
would reach a steady state under the competing effects
of gravitational scattering and capture.
The same theory was later applied,  with only minor modifications,
to galactic  nuclei,
most of which however are far too young 
for this steady state to have been achieved.
In a collisionless nucleus, SBH feeding rates can be 
much higher than in a relaxed nucleus, e.g. if the nucleus
is triaxial and many of the orbits are ``centrophilic;''
or much lower, e.g. if orbits 
around the loss cone were substantially depeleted by a 
binary SBH before its coalescence.
Considerations like these have prompted a 
reinvestigation of the nuclear loss cone problem, 
as reviewed in \S 6.

In a collisionless nucleus, the distribution of
stars around the SBH should still reflect the
energy input from the binary SBH that preceded
the current, single hole.
(Some nuclei might  contain {\it un}coalesced binaries, 
particularly if they experienced a recent merger,
although the observational evidence for binary SBHs is still
largely circumstantial; see \citename{komossa-03}
\citeyear*{komossa-03}.)
The effect of binary SBHs on the morphology and kinematics
of galactic nuclei is discussed in \S 7.
An important success of the binary SBH model is its ability
to quantitatively explain the low central densities of bright
elliptical galaxies, i.e. the ``cusp-core dichotomy.''
Previously, this dichotomy had been widely interpreted to 
mean that dwarf and giant elliptical galaxies experienced very different
formation histories \citeaffixed{kormendy-85}{e.g.}.

The literature on galactic nuclei and SBHs is enormous and 
exponentially expanding, 
and the writing of this review article was greatly facilitated
by the recent appearance of a number of other 
articles that emphasize different aspects of this broad topic.
A comprehensive review of observational evidence for
SBHs is given by \citeasnoun{FF-05}. 
\citename{alexander-03} \citeyear{alexander-03,alexander-05}
reviews the physics of star-star
and star-SBH interactions, with an emphasis on dissipative
processes and on the nucleus of  the Milky Way;
as noted above, such processes are of less importance
in the nuclei of most galaxies known to contain SBHs.
\citename{komossa-02} \citeyear{komossa-02,komossa-03} 
reviews the observational evidence
for interaction of single and binary SBHs with stars
and gas in galactic nuclei. 
(For a more theoretically 
oriented review of binary SBHs, see \citename{living} \citeyear*{living}.)
While intermediate mass black holes (IMBHs) are still 
seen as speculative by many astrophysicists,
many of the models proposed for their formation are
closely similar to early models for the dissipative 
evolution of galactic nuclei, as discussed by 
\citeasnoun{MC-04} and \citeasnoun{marel-04}.
The prospect of detecting gravitational  waves from 
coalescing black holes in galactic nuclei is a prime
motivation behind much recent work on nuclear dynamics.
While no single article can do justice to this exciting topic,
the reviews by \citeasnoun{schutz-02} and
\citeasnoun{hughes-03} are excellent places to start.
Dark matter is almost certainly not a dominant
component of galactic nuclei, but its density there is
likely to be higher than elsewhere in galaxies,
making galactic nuclei ideal targets for so-called
``indirect-detection'' studies, as reviewed by
\citeasnoun{BM-05}.
Finally, some early review articles on the evolution 
of galactic nuclei, while outdated in some respects,
are still very much worth reading today, including those by 
\citeasnoun{spitzer-71}, \citeasnoun{saslaw-73},
\citeasnoun{rees-84}, and Gerhard (1992,1994).

\section{Characteristic Length and Time Scales}

The radius of the event horizon of a nonrotating hole 
of mass $\mh$ is
\beq
r_S = {2G\mh\over c^2} \approx 9.6 \times 10^{-6} {\rm pc}
\left({\mh\over 10^8\msun}\right).
\eeq
The dynamical influence of a SBH extends far beyond
$r_S$ however.
In a stellar nucleus with 1D velocity dispersion 
$\sigma$,
the SBH's ``influence radius'' is customarily defined as 
\begin{equation}
r_h = {G\mh\over\sigma^2} \approx 11\ {\rm pc} \left({M_\bullet\over 10^8M_\odot}\right) \left({\sigma\over 200\ {\rm km\ s}^{-1}}\right)^{-2}.
\label{eq:rh}
\end{equation}
In a singular isothermal sphere (SIS) nucleus,
$\rho=\sigma^2/2\pi Gr^2$, the velocity dispersion is
constant with radius in the absence of the SBH, 
and the stellar mass within $r_h$ is $2\mh$.
This is often taken as an alternate definition of $r_h$:
the influence radius is the
radius containing a stellar mass equal to $2\mh$, or
\beq
M_\star(r\le r_h) = 2\mh.
\label{eq:def_rh}
\eeq
The second definition of $r_h$ is generally to be 
preferred since $\sigma$ is a function of radius near the SBH,
and unless otherwise stated, this is the definition that
will be adopted below.
In the Milky Way, $r_h\approx 3$ pc according
to either definition.

The tidal disruption radius $r_t$ is the distance
from the SBH where tidal forces can pull a star apart.
Strictly, $r_h$ depends on the structure of the
star and on the shape of its orbit, but it is of order
\beq
r_t\approx r_\star \left({\mh\over m_\star}\right)^{1/3} \approx
1.0 \times 10^{-5} {\rm pc} \left({\mh\over 10^8\msun}\right)^{1/3}
\label{eq:rt1}
\eeq
where the latter expression assumes stars of Solar mass and radius.
In terms of the radius of the event horizon,
\beq
{r_t\over r_S} \approx \left({\mh\over 10^8\msun}\right)^{-2/3}
\eeq
so that SBHs more massive than $\sim 10^8\msun$ 
``swallow stars whole.''

The relaxation time is defined as the time for (mostly distant)
gravitational encounters between stars to establish a locally 
Maxwellian velocity distribution.
Assuming a homogenous isotropic distribution of equal-mass stars,
the relaxation time is approximately
\numparts
\begin{eqnarray}
T_r &\approx& {0.34\sigma^3\over G^2 \rho m_\star \ln\Lambda} \\
&\approx&
0.95\times 10^{10} {\rm yr} \left({\sigma\over 200\ {\rm km\ s}^{-1}}\right)^3 \left({\rho\over 10^6\msun {\rm pc}^{-3}}\right)^{-1} \left({m_\star\over \msun}\right)^{-1} \left({\ln\Lambda\over 15}\right)^{-1}
\label{eq:tr}
\end{eqnarray}
\endnumparts
\cite{spitzer-87}.
The Coulomb logarithm, $\ln\Lambda$, is a ``fudge factor'' that 
accounts approximately for the divergent total perturbing force 
in an infinite homogeneous medium.
Within the SBH's sphere of influence,
$N$-body experiments  \cite{preto-04} suggest that
\beq
\ln\Lambda\approx \ln(r_h\sigma^2/2Gm_\star) \approx
\ln\left(\mh/ 2m_\star\right) \approx \ln(N_\bullet/2)
\eeq
with $N_\bullet\equiv \mh/m_\star$ the number of stars whose
mass makes up $\mh$.
For $m_\star=\msun$ and $\mh=(0.1,1,10)\times 10^8\msun$,
$\ln\Lambda\approx (15,18,20)$.
In the gravitational field of a SBH, 
i.e. at $r\lap r_h$, encounters lead
to a steady-state distribution of orbital energies
in a time $\sim T_r$, as discussed in \S 5.

\begin{figure}
\centering
\includegraphics[scale=0.6]{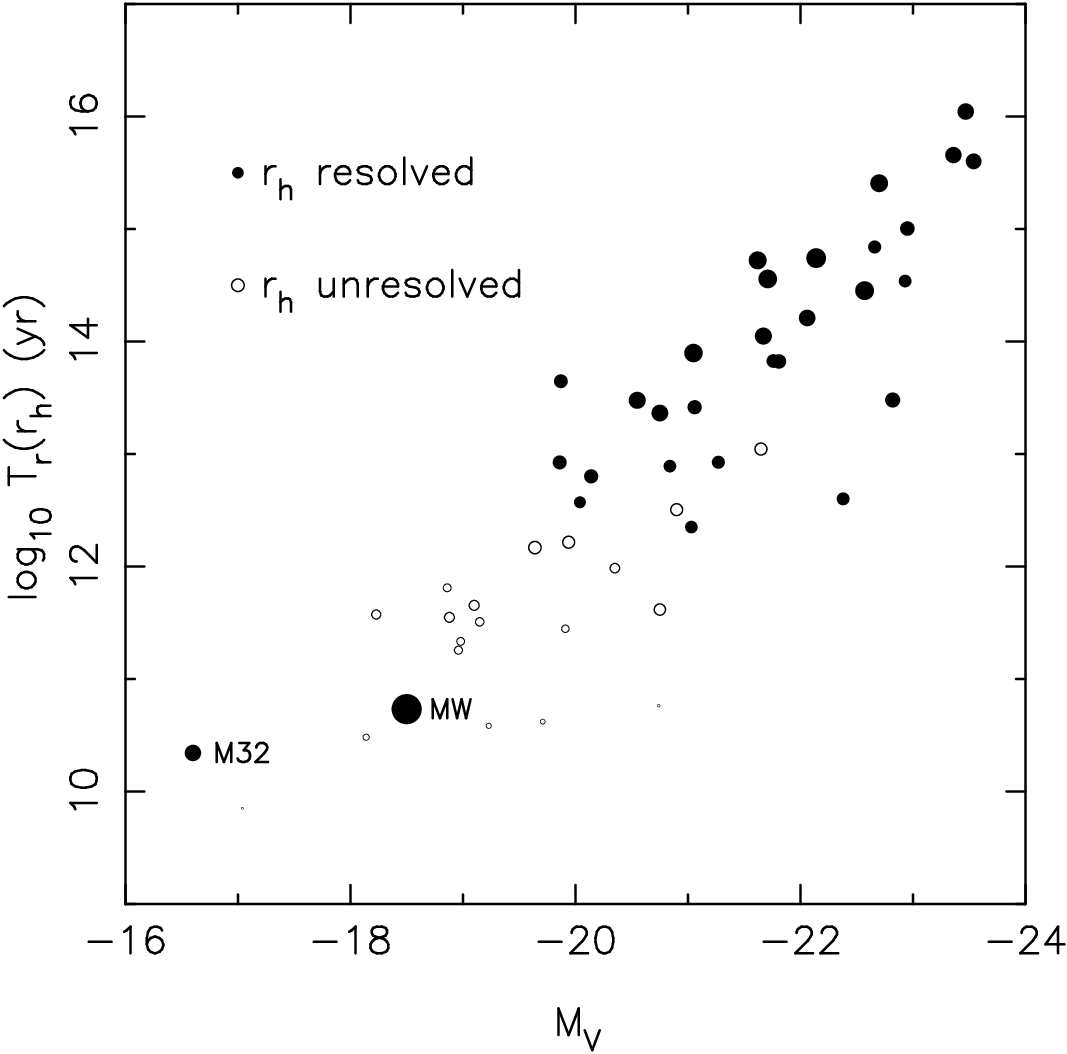}
\caption{Estimates of the relaxation time $T_r$ 
(Eq.~\ref{eq:tr}) at the 
  SBH's influence radius $r_h$ (Eq.~\ref{eq:def_rh}),
in the sample of early-type galaxies modelled by 
\citeasnoun{wang-04}.
SBH masses were computed from the $M_\bullet-\sigma$
relation,
except in the case of the Milky Way, for which
$M_\bullet=3.7\times 10^6M_\odot$ was assumed
\cite{ghez-05}.
The stellar mass was set equal to $0.7M_\odot$ when computing
$T_r$.
Horizonal axis is absolute visual  magnitude of the
galaxy or, in the case of the Milky Way, the stellar bulge.
The size of the symbols is proportional to
$\log_{10}(\theta_{r_h}/\theta_{obs})$, where $\theta_{r_h}$ is
the angular size of the black hole's influence radius and
$\theta_{obs}$ is the observational resolution.
Filled symbols have $\theta_{r_h}>\theta_{obs}$ 
($r_h$ resolved) and open circles have
$\theta_{r_h}<\theta_{obs}$ ($r_h$ unresolved).
Values of $T_r(r_h)$ in the unresolved galaxies should
be considered approximate.
In the Milky Way, the stellar density profile
{\it is} resolved at $r\ll r_h$ and $T_r$ is found
to drop below its value at $r_h$, to $\sim 4\times 10^9$ yr
at $\sim 0.1r_h\approx 0.3$ pc.
Time scales for physical collisions between stars are longer
than $T_r$ (Eq.~\ref{eq:tcoll2}).
\citeaffixed{MS-05}{From}
\label{fig:tr}
}
\end{figure}

Figure~\ref{fig:tr} shows estimates of $T_r$,
measured at $r=r_h$, in a sample of early-type galaxies
and bulges.
The relaxation time at $r_h$ is almost always in excess
of $10^{10}$ yr, although there is a clear trend with
luminosity, suggesting that $T_r(r_h)$
drops below $10^{10}$ yr for spheroids fainter than
absolute magnitude $M_V\approx -18$, roughly the luminosity
of the Milky Way bulge.
Furthermore, in the handful of Local Group
galaxies for which the SBH influence radius is well resolved,
$T_r$ continues to decrease inside of $r_h$:
in the case of the Milky Way, to $\sim 6\times 10^9$ yr
at $0.2 r_h$ and $\sim 3.5\times 10^9$ yr at $\sim 0.1 r_h$.
At radii $r\ll r_h$ where $\sigma^2\propto r^{-1}$,
the relaxation time varies as
$T_r\sim r^{\gamma-3/2}$ if $\rho\sim r^{-\gamma}$.
Severak Local Group galaxies are known to have $\gamma\approx 1.5$
at $r\lap r_h$ \cite{lauer-98} implying that $T_r$ is approximately
constant into the SBH in these galaxies.
Relaxation times in the nuclei of brighter galaxies
are probably always longer
than a Hubble time, and the stellar distribution at
$r\lap r_h$ in these galaxies should still reflect
the details of their formation.

In what follows, {\it collisional nuclei} are defined
as those which have $T_r\lap 10^{10}$ yr at $r<r_h$, while
{\it collisionless nuclei} have $T_r\gap 10^{10}$ yr.

A number of other physical processes have time scales
that are related to $T_r$.
The core-collapse time $T_{cc}$ is $\sim 10^2T_r$;
in a time of order $T_{cc}$, a nucleus {\it lacking} a SBH 
develops a high-density core and a power-law envelope,
$\rho\sim r^{-2.2}$
The core collapse time is interestingly short at the centers
of M33 and NGC205: both have $T_r\lap 10^8$ yr
\cite{hernquist-91,valluri-05}
and both lack dynamical signatures of a SBH.
However core collapse is probably unimportant in 
galaxies with SBHs, both because $T_r$ is $\gap 10^9$ yr,
and because the collapse would be inhibited by the presence of the hole.
Another relaxation-driven mechanism is 
diffusion of stars into the tidal disruption sphere of the SBH.
The rate is approximately 
$\dot N \approx N/\left[\ln(2/\theta_{lc})\right]T_r$,
where $\theta_{lc}\approx \sqrt{r_t/r_h}$ is the angular
size subtended by the sphere $r=r_t$ as seen from $r=r_h$
and $N$ is the number of stars within $r_h$
\cite{FR-76,LS-77}.

Physical collisions between stars take place in a time
$T_{coll}$ where
\numparts
\begin{eqnarray}
T_{coll} &=& 
\left[16\sqrt{\pi}n\sigma r_\star^2\left(1+\Theta\right)\right]^{-1} \\
&\approx& 1.1\times 10^{11} {\rm yr} 
\left({n\over 10^6\ {\rm pc}^{-3}}\right)^{-1} 
\left({\sigma\over 200\ {\rm km\ s}^{-1}}\right)^{-1} 
\left({r_\star\over R_\odot}\right)^{-2} 
\left({1+\Theta\over 3}\right)^{-1}
\end{eqnarray}
\endnumparts
and $\Theta$ is the Safronov number:
\numparts
\begin{eqnarray}
\Theta &=& {Gm_\star\over 2\sigma^2 r_\star} \\
&\approx& 2.38 \left({m_\star\over M_\odot}\right) 
\left({\sigma\over 200\ {\rm km\ s}^{-1}}\right)^{-2}
\left({r_\star\over R_\odot}\right)^{-1}.
\end{eqnarray}
\endnumparts
The ratio of the collision time to the relaxation time is
\beq
{T_{coll}\over T_r} \approx 0.8\ln\Lambda {\Theta^2\over 1+\Theta}
\approx 0.3 \left({r\over 0.01\ {\rm pc}}\right)^4;
\label{eq:tcoll2}
\eeq
the right-hand expression assumes Solar-type stars around
the Milky Way SBH, and shows that even in the very
dense environment of the Galactic center, 
physical collisions are significant only at very small radii,
$r\lap 0.02$ pc $\approx 10^{-2}r_h$ for Solar-type stars.
Stellar collisions may be responsible for the depletion
of luminous, late-type giant stars at the Galactic
center \cite{phinney-89,alexander-99,alexander-03}
or for the formation of the so-called S-stars
\cite{morris-93,genzel-03}.
There is little evidence that stellar collisions
have affected the stellar populations at the centers
of any of the other Local Group galaxies however
\citeaffixed{lauer-98}{e.g.}.
From the point of view of nuclear dynamics,
stellar collisions are probably always of minor importance
\citeaffixed{freitag-02}{e.g.}
and they will  not be discussed further in this article.

The spheroids that host SBHs are believed to have formed
via mergers, and in many cases, the merging galaxies would
have contained pre-existing SBHs.
The result is a binary SBH \cite{BBR-80}.
A number of additional length and time scales are
associated with the binaries.
Let $M_1$ and $M_2$ be the masses respectively of the
larger and smaller of the two SBHs, with
$q=M_2/M_1\le 1$ the mass ratio, $M_{12}=M_1+M_2$ the total mass,
and $\mu=M_1M_2/M_12$ the reduced mass.
Two SBHs form a gravitationally-bound pair
when their separation falls below the influence
radius defined by the larger hole.
Approximating the relative orbit as Keplerian,
i.e. ignoring the force perturbations from stars,
the binary's binding energy is
\beq
\left| E\right| = {GM_1M_2\over 2a} = {G\mu M_{12}\over 2a}
\eeq
with $a$ the semi-major axis,
and the orbital period is
\beq
P=2\pi\left({a^3\over GM_{12}}\right)^{1/2} = 
9.36\times 10^3\ {\rm yr} 
\left({M_{12}\over 10^8 \msun}\right)^{-1/2}
\left({a\over 1\ {\rm pc}}\right)^{3/2} .
\label{eq:period}
\eeq
The relative velocity of the two SBHs,
assuming a circular orbit, is
\beq
V_{bin} = \sqrt{GM_{12}\over a} = 658\ {\rm km\ s}^{-1} \left({M_{12}\over 10^8 M_\odot}\right)^{1/2} \left({a\over 1\ {\rm pc}}\right)^{-1/2}.
\eeq

A massive binary is called ``hard'' when its binding energy per unit
mass, $|E|/M_{12} = G\mu/2a$, exceeds $\sim\sigma^2$.
(The motivation for this definition is given in \S 7.)
A standard definition for the semi-major axis of a hard binary is
\beq
a_h = {G\mu\over 4\sigma^2} \approx 2.7\ {\rm pc} \left(1+q\right)^{-1} 
\left({M_2\over 10^8 M_\odot}\right) \left({\sigma\over 200\ {\rm km\ s}^{-1}}\right)^{-2}.
\label{eq:ah}
\eeq
Defining $r_h$ as $G\m12/\sigma^2$ then implies
\beq
a_h = {\mu\over 4\m12}r_h = {1\over 4} {q\over (1+q)^2} r_h.
\eeq
For an equal-mass binary, $a_h\approx 0.06 r_h$,
and for a more typical mass ratio of $q=0.1$,
$a_h\approx 0.02 r_h$.

If the binary's semi-major axis is small enough that its 
subsequent evolution is dominated by emission of gravitational radiation,
then $\dot {a} \propto -a^{-3}$ and
coalescence takes place in a time $t_{gr}$, where \cite{peters-64}
\begin{eqnarray}
t_{gr} &=& {5\over 256 F(e)}{c^5\over G^3} {a^4\over\mu \m12^2}, \nonumber \\
\label{eq:peters0}
F(e) &=& \left(1-e^2\right)^{7/2}\left(1+{73\over 24}e^2 + 
{37\over 96} e^4\right).
\end{eqnarray}
This can be written
\begin{eqnarray}
t_{gr} &=& {5\over 16^4 F(e)}{G\mu^3 c^5\over \sigma^8 \m12^2}
\left({a\over a_h}\right)^4 \nonumber \\
&\approx& {3.07\times 10^{8}{\rm yr}\over F(e)} {q^3\over (1+q)^6} 
\left({\m12\over 10^8\msun}\right) 
\left({\sigma\over 200\ {\rm km\ s}^{-1}}\right)^{-8}\left({a\over 10^{-2} a_h}\right)^4
\label{eq:peters}
\end{eqnarray}
with $e$ the orbital eccentricity.
Separations much less than $a_h$ are required in order for
emission of gravitational waves to induce coalescence
in $10^{10}$ yr.

Many of these expressions can be simplified by making use of 
the empirical correlations between SBH mass and galaxy properties.
The tightest of these is the $M-\sigma$ relation, which has
two extant forms, based either on the velocity dispersion $\sigma_c$
measured in an aperture centered on the nucleus (but large
compared with $r_h$) \cite{fm-00} 
or on the global rms stellar velocity \cite{gebhardt-00}.
The first form is the more relevant here since $\sigma$ as used
above is defined near the center of a galaxy;
the alternative form \cite{gebhardt-00} defines $\sigma$ as a 
mean value along a slit that extends over the entire half-light 
radius of the galaxy and can be strongly influenced by rotation,
inclination and other factors extraneous to nuclear dynamics.
(The \citename{gebhardt-00} form of the relation also exhibits 
substantially more scatter.)
In terms of $\sigma_c$, the best current determination
of the $M-\sigma$ relation is 
\beq
\left({\mh\over 10^8\msun}\right) 
= (1.66\pm 0.24) \left({\sigma_c\over 200\ {\rm km\ s}^{-1}}\right)^\alpha
\label{eq:ms}
\eeq
with $\alpha= 4.86\pm 0.43$ \cite{FF-05}.
The intrinsic scatter in this relation is consistent with zero
\cite{MF-01,FF-05}, although the number of galaxies with well-determined
SBH masses is still very small, of order a dozen or less 
\cite{valluri-04}.
The relations between $\mh$ and the mass or luminosity of the host spheroid
(elliptical galaxy or spiral galaxy bulge) appear to be less tight;
the mean ratio of $\mh$ to $M_{gal}$ is $\sim 1.3 \times 10^{-3}$
\cite{MF-01b,marconi-03}.

Using Equation~\ref{eq:ms}, one can write
\numparts
\begin{eqnarray}
r_h &\approx& 18\ {\rm pc} \left({\sigma_c\over 200\ {\rm km\ s}^{-1}}\right)^{2.86} \\
&\approx& 13\ {\rm pc} \left({M_\bullet\over 10^8M_\odot}\right)^{0.59}
\label{eq:rhnew} 
\end{eqnarray}
\endnumparts
and
\numparts
\begin{eqnarray}
a_h &\approx& 4.5\ {\rm pc} {q\over (1+q)^2}\left({\sigma_c\over 200\ {\rm km\ s}^{-1}}\right)^{2.86} \\
&\approx& 3.2\ {\rm pc} {q\over (1+q)^2}\left({M_\bullet\over 10^8M_\odot}\right)^{0.59}.
\end{eqnarray}
\endnumparts
Setting $M=\m12$, $F(e)=1$ in Equation~(\ref{eq:peters}) gives
\begin{eqnarray}
t_{gr} &\approx& 5.0\times 10^8 {\rm yr} {q^3\over (1+q)^6} 
\left({\sigma_c\over 200\ {\rm km\ s}^{-1}}\right)^{-3.14}\left({a\over 10^{-2}a_h}\right)^4 
\nonumber \\
&\approx& 7.1\times 10^8 {\rm yr} {q^3\over (1+q)^6} 
\left({M_{12}\over 10^8\msun }\right)^{-0.65}\left({a\over 10^{-2}a_h}\right)^4.
\label{eq:peters2}
\end{eqnarray}

\section{Structural Properties of Galaxies and Nuclei}

Knowledge of the distribution of mass and light near the
centers of galaxies is limited by the $\sim 0.1''$ angular
resolution of the Hubble Space Telescope
(although ground-based adaptive optics are starting to 
improve on this, e.g. \citename{melbourne-05} \citeyear*{melbourne-05},
\citename{davidge-05} \citeyear*{davidge-05}).
The corresponding linear scale is 
$\sim 5\ {\rm pc}(D/10{\rm Mpc})$;
at the distance of the Virgo cluster
($D\approx 16$ Mpc),
resolving $r_h$ is only possible in galaxies with
$\mh\gap 10^8\msun$ (Eq.~\ref{eq:rhnew}).
While a great deal is known about the 
morphology of bright ($M_B\lap -20$) 
galaxies on scales $r<r_h$,
the central structure of fainter spheroids
is poorly understood, and what little we know
comes almost entirely from Local Group galaxies
(including the bulge of the Milky Way).
Not surprisingly, parametric functions that were
developed to fit the luminosity profiles of
distant galaxies often fail to describe the
well-resolved centers of Local Group galaxies.
In this section, an attempt is made to synthesize
recent work on the light distributions in both
bright (distant) and fainter (mostly nearby) galaxies.

Among the various functional forms that have
been proposed to fit galaxy luminosity profiles,
the most generally successful have been those based on 
\possessivecite{sersic-68} law,
\beq
\ln I(R) = \ln I_e - b(n_S)\left[\left(R/R_e\right)^{1/n_S} - 1\right].
\label{eq:sersic}
\eeq
The constant $b$ is normally chosen such that $R_e$ is the projected
radius containing one-half of the total light.
The shape of the profile is then determined by $n_S$;
$n_S= 4$ is the \citeasnoun{devauc-48} law, which is a good
representation of bright elliptical galaxies,
while $n_S=1$ is the exponential law, which approximates
the luminosity profiles of dwarf galaxies \cite{binggeli-84}.
Deviations from the best-fitting S\'ersic law
are typically $0.05$ mag rms and, at least in the case
of galaxies outside the Local Group, the fits
are often good over the full observed range,
typically two to three decades in radius
\cite{gg-03,graham-03,trujillo-04}.
An alternative way to write Equation~(\ref{eq:sersic}) is
\beq
{d\ln I\over d\ln R} = -{b\over n_S}\left({R\over R_e}\right)^{1/n_S},
\label{eq:serslopes}
\eeq
i.e. the logarithmic slope varies as a power of the projected
radius, falling to zero at the center.
While there is no consensus on why the
S\'ersic law is such a good representation
of galactic spheroids, a possible hint comes from the
dark-matter halos produced in 
$N$-body simulations of hierarchical structure
formation, which have density profiles
that are also well described by Equation~(\ref{eq:serslopes})
\cite{navarro-04,merritt-05,graham-05}.
This functional form may be characteristic of systems
that form via chaotic, collisionless relaxation.
 
The S\'ersic index $n_S$ correlates reasonably well with
galaxy absolute magnitude \cite{gg-03}, 
and less well with other structural parameters like $R_e$
\cite{caon-93,gg-03}. 
In a recent comprehensive study \cite{acs6},
the Advanced Camera for Surveys on the Hubble Space Telescope
was used to measure luminosity profiles of 100 early-type
galaxies in the Virgo cluster.
The best-fitting relations $n_S(M_B)$ and $R_e(M_B)$
were found to be
\begin{eqnarray}
\log n_S &\approx& \log(3.89) - 0.10(M_B + 20), \\
\log R_e &\approx& \log(17.8) - 0.055(M_B+20)
\end{eqnarray}
where $R_e$ is in units of kpc.

In galaxies resolved on scales of order a few $r_h$
or better,
systematic deviations from the S\'ersic law often
appear near the center.
These deviations are of two kinds.
Galaxies fainter than $M_B\approx -20$ are generally
observed to have {\it higher} surface brightnesses
at small radii than predicted by S\'ersic's law.
These galaxies are sometimes called ``power-law''
galaxies since $I(R)$ can be reasonably well approximated
as a power law at $R\lap R_e$.
Galaxies brighter than $M_B\approx -20$
generally exhibit central {\it deficits} in the
intensity, inside of a ``break'' radius $R_b$
that is of order a few times $r_h$.
These are sometimes called ``core'' galaxies.

\begin{figure}[t]
\begin{center}
\includegraphics[width=.7\textwidth]{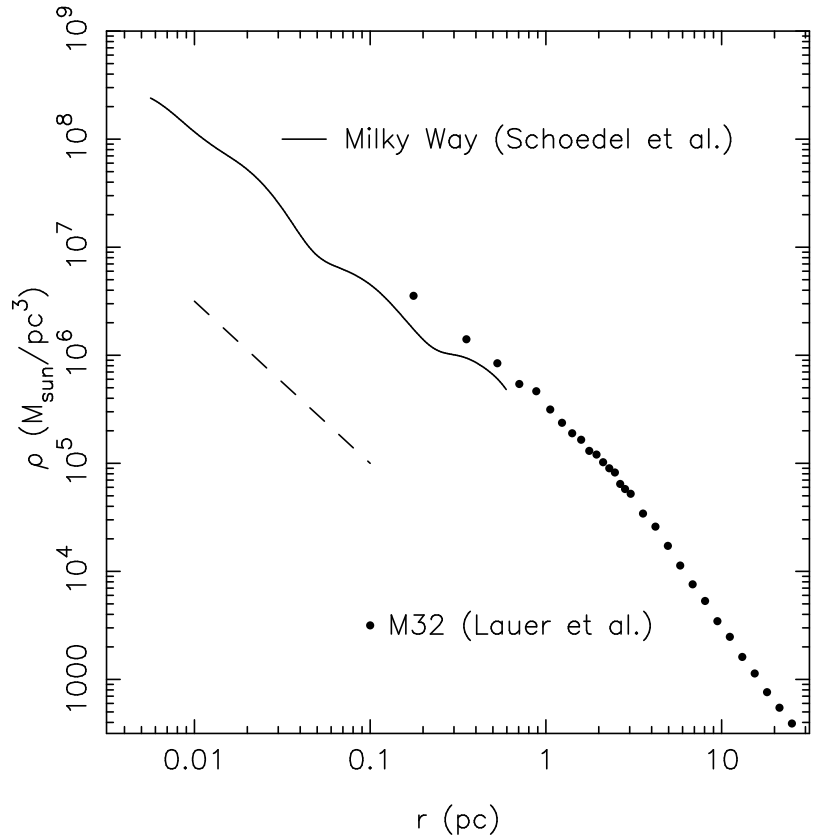}
\end{center}
\caption[]{Mass density profiles near
the centers of the Milky Way and M32
\cite{schoedel-06,lauer-98}.
Dashed line is $\rho \propto r^{-1.5}$.
Both galaxies contain SBHs with masses
$\sim 3\times 10^6\msun$ and with
influence radii $r_h\approx 3$ pc.
\label{fig:LG}}
\end{figure}

The surface brightness surpluses at small radii
are most clearly defined in Local Group galaxies like
the Milky Way, M31 and M32.
Each of these galaxies exhibits an approximately
power-law dependence of $I$ on $R$ into the
innermost resolved radius, $I\sim R^{-\Gamma}$,
implying a spatial density $\rho\sim r^{-\gamma}$,
$\gamma\approx \Gamma + 1$  (Figure~\ref{fig:LG}).
In the bulge of the Milky Way, the
stellar density has been derived from number counts 
that extend down to $\sim 0.005\ {\rm pc}\approx 10^{-3}r_h$;
the result is
\beq
\rho(r) \approx 1.8\times 10^5 M_\odot {\rm pc}^{-3}
\left({r\over 0.38{\rm pc}}\right)^{-\alpha}
\label{eq:rhoMW}
\eeq
with  $\alpha\approx 2.0$ at $r\ge 0.38$ pc 
and $\alpha\approx 1.4$ at $r<0.38$ pc
\cite{genzel-03,schoedel-06}.
M31 and M32 have central density profiles very similar
to that of the Milky Way,
both in slope and normalization \cite{lauer-98},
although $I(R)$ in both galaxies can not be measured on 
scales much smaller than $r_h$.
Deviations from a S\'ersic law in M32 appear inside
of $R\approx 10''\approx 40\ {\rm pc}\approx 20r_h$
\cite{tonry-84,graham-02}, which suggests that
the central surplus can not be ascribed solely to the
dynamical influence of the SBH.

\begin{figure}[t]
\begin{center}
\includegraphics[width=0.9\textwidth]{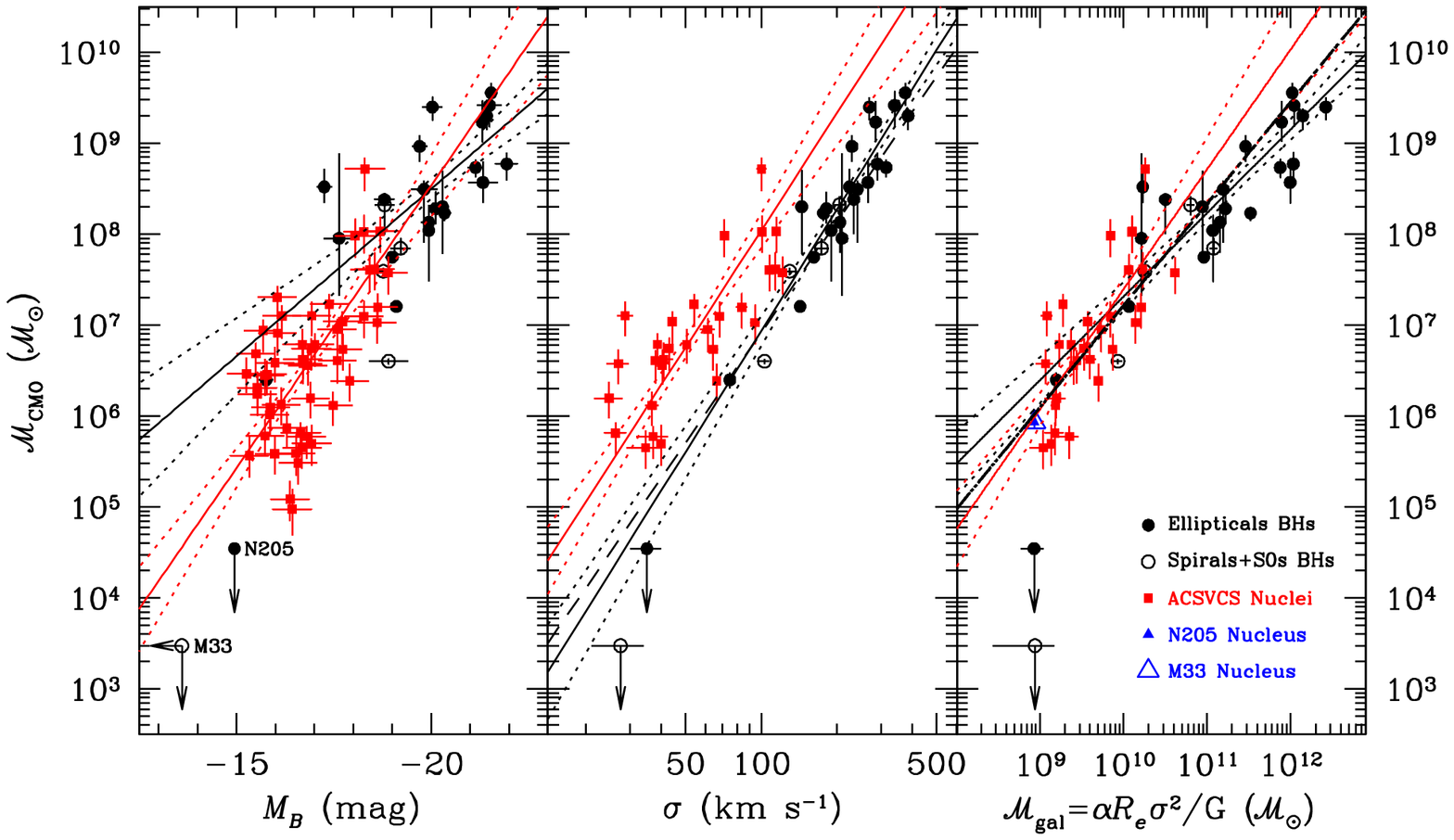}
\end{center}
\caption[]{
{\it (Left Panel)} Mass of the ``central massive object'' (CMO)
plotted against absolute blue magnitude of the host galaxy (or bulge
for spiral galaxies).
Stellar nuclei (defined as an excess with respect to the best-fitting
S\'ersic profile) are shown as red squares. 
SBHs in elliptical  and spiral
galaxies are  shown as filled and open circles respectively.  
Upper limits on the SBH mass are shown for NGC205 and M33.  
{\it (Middle Panel)} 
CMO mass as a function of velocity dispersion of the  host
galaxy.  
{\it (Right Panel)} CMO mass plotted against galaxy mass.
The solid red and black lines show the best-fit relations
fit to the nuclei and to the SBH samples respectively, 
with 1$\sigma$ confidence levels on the slope
shown by the dotted lines. 
In the middle panel, the dashed line is the $M_\bullet-\sigma$ 
relation.
In the right panel, 
the dashed line is the fit obtained for the combined nuclei+SBH sample.
(Adapted from \citename{CMO-06} \citeyear*{CMO-06}.)
\label{fig:CMO}}
\end{figure}

Inner light surpluses are also seen in galaxies beyond
the Local Group (e.g. \citename{acs8} 2006 and references
therein),
although in most cases the central regions are not
well enough resolved that it is possible to
determine the functional form of the deviation.
At the innermost resolved radii, most galaxies in
this class have $-2.5\lap d \log\rho/d\log r\lap -1.5$
\cite{gebhardt-96,acs6}.
A reasonable ansatz is that the inner, unresolved density 
profiles in these galaxies are similar to the power laws
 observed at the centers of Local Group galaxies.
However there is evidence that the central surpluses in 
spheroids fainter than $M_B\approx -18$ can be modelled
as distinct nuclei, i.e. as components that rise
suddenly above the best-fitting S\'ersic law at
some radius, then level out into a constant-density
core.
Such a profile is definitely seen in NGC205 \cite{valluri-05}; 
the upward inflection appears at $R\approx 3''$ and the core 
radius is $\sim 0.1''$.
Distinct nuclei are ubiquitous in spheroids with 
$M_B\approx -17$, gradually disappearing in spheroids
fainter than $M_B\approx -12$ \cite{vdb-86}.

There is no compelling explanation for the power-law density 
profiles observed at the centers of galaxies like the Milky Way and M32.
The scale-free nature of a power law 
suggests a gravitational origin, 
but no obvious dynamical mechanism suggests itself.
In the region dominated by the gravitational force
from the SBH, $r\lap 0.2 r_h$ say, processes like
adiabatic contraction \cite{young-80}
and collisional relaxation \cite{bw-76}
can produce power-law profiles, but these mechanisms
are ineffective at $r\gap r_h$.

Very recently,  a possible connection has been found between
the central luminosity excesses in low-luminosity galaxies
and the SBHs in brighter galaxies \cite{CMO-06,harris-06}.
If the ``nucleus'' is defined as the interior light
or mass in excess of the best-fitting S\'ersic profile,
then the ratio of ``nuclear'' mass to total galaxy mass
is found to be roughly equal to the ratio $\mh/M_{gal}$ in galaxies
with SBHs, or $\sim 10^{-3}$
(Fig.~\ref{fig:CMO}).
This has led to the suggestion
that galaxies always form a ``central massive object''
(CMO): either a SBH, or a compact stellar nucleus.
A possible objection to this appealing picture is that it fails
to describe some of the best-resolved galaxies, e.g.
M31, M32 and the Milky Way, all of which contain SBHs, as well as
``nuclei,'' i.e. central excesses with 
respect to the best-fitting S\'ersic law \citeaffixed{graham-02}{e.g.}.
It is possible that the ``nuclei'' in some of the
unresolved galaxies in Figure~\ref{fig:CMO} are similar to  the
power-law cusps observed in the well-resolved Local 
Group  galaxies, rather than distinct, compact cores
like in NGC 205.

\begin{figure}[t]
\begin{center}
\includegraphics[width=.70\textwidth]{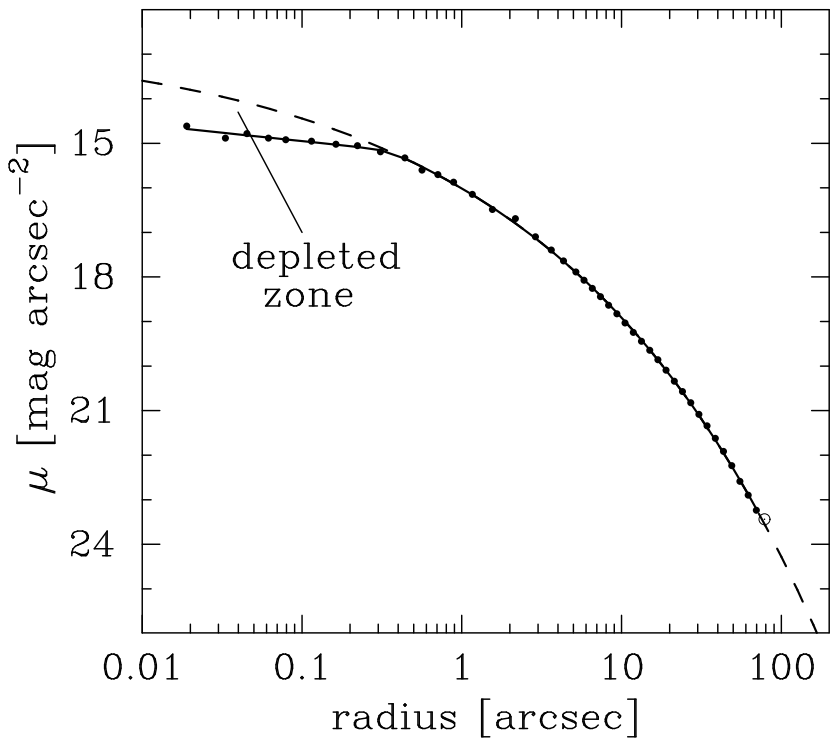}
\end{center}
\caption[]{Surface brightness profile in the $R$ band
of NGC 3348, a ``core'' galaxy.
The dashed line is the best-fitting S\'ersic model;
the observed profile (points, and solid line) falls
below this inside of a break radius $r_b\approx 0''.35$
\citeaffixed{graham-04}{From}}.
\label{fig:sersic}
\end{figure}

At the other extreme of luminosity, bright elliptical
galaxies generally show central {\it deficits} in $I(R)$,
or ``cores.''
The cores extend outward to a break radius $R_b$ 
of order a few times $r_h$. 
The profiles at $R\lap R_b$
are well described as power laws but with small slopes, 
$\Gamma\lap 0.3$, and furthemore the profiles often
show a distinct inflection, or change of slope, at $R_b$
\cite{lauer-95,trujillo-04,acs6}.
Such cores are apparent in elliptical galaxies brighter
than $M_V\approx -19.5$, but this number may be resolution-dependent
since fainter galaxies are mostly unresolved on scales of $r_h$.

One model, discussed in detail below, attributes the cores
to the dynamical influence of binary SBHs.
If this model is correct, the ``mass deficit'' --
the amount of mass that was removed in creating the core --
is a measure of the time-integrated effect of the
binary on the nucleus.
The mass deficit is defined \cite{milos-02}
as the difference in integrated mass 
between the deprojected density profile $\rho(r)$ and an
inward extrapolation of the outer, deprojected profile
$\rho_{fit}(r)$:
\begin{equation}
M_{def}\equiv 4\pi \int_0^{r_b} 
\left[\rho_{fit}(r) -\rho(r)\right] r^2 dr.
\label{eq:defmdef}
\end{equation}
Here, $r_b$ is the (spatial) radius where the 
luminosity profile departs from the fitted profile.
Figure~\ref{fig:sersic} illustrates the computation
of $M_{def}$ in NGC 3348.
Note that the pre-existing profile was assumed to be
a S\'ersic law;
under this assumption, the mass deficit is found to be
roughly equal to $\mh$, the mass currently in the SBH
\cite{graham-04}.
This is consistent with the mass displaced by an infalling
SBH (\S 7).

\begin{figure}[t]
\begin{center}
\includegraphics[width=.70\textwidth]{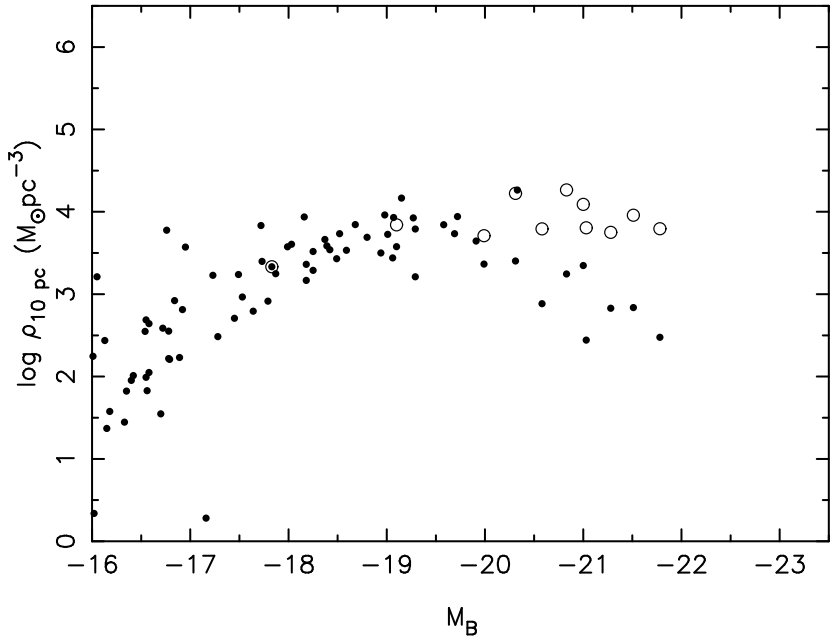}
\end{center}
\caption[]{Filled circles show the density at $r=10$ pc 
in a sample of early-type galaxies in Virgo
\cite{acs6}; $M_B$ is the absolute blue magnitude.
After increasing monotonically with luminosity over
$\sim 10$ magnitudes,
the central density begins to decline at 
$M_B\approx -20$,
roughly the magnitude at which galaxies begin to exhibit
central depletions, or  ``cores.''
Open circles show the density computed via an
inward extrapolation of a S\'ersic profile fit to 
the large-radius data,
as in Fig.~\ref{fig:sersic}.
The open circles presumably represent more closely
the density before
a binary SBH ``carved out'' a core; this density
follows the trend of increasing density with galaxy
luminosity exhibited by the fainter galaxies.
This argument, first made by \citeasnoun{JB-97}
and \citeasnoun{gg-03}, plausibly explains why
the centers of dwarf and giant galaxies appear
at first sight to define two distinct ``families'' 
\cite{kormendy-85}.}
\label{fig:rho10}
\end{figure}

The idea that the cores of bright elliptical galaxies are 
created during mergers provides a nice solution to a 
long-standing puzzle: why dwarf and giant elliptical 
galaxies appear to occupy two distinct families
in terms of their central properties \cite{JB-97,gg-03}.
Dwarf ellipticals define a continuous sequence
spanning ten magnitudes, such that the central
surface brightness increases linearly with 
absolute magnitude.
In galaxies brighter than $M_B\approx -20.5$, 
the central surface brightness {\it declines} 
with increasing luminosity, leading
\citeasnoun{kormendy-85} to suggest that
``dwarf elliptical galaxies are very different from
the sequence of giant ellipticals.''
But $M_B\approx -20.5$ is also roughly where
cores appear, and if the cores are ``removed''
as in Figure~\ref{fig:sersic} by fitting a smooth
profile to the surface brightness data at $R>R_b$,
one finds that bright ellipticals smoothly continue
the sequence defined by dwarf ellipticals of 
central density increasing with luminosity.
Figure~\ref{fig:rho10} illustrates this for a sample
of early-type galaxies in Virgo.
Even after this adjustment, 
the central densities of the ``core'' galaxies in 
Figure~\ref{fig:rho10} should probably be 
interpreted as lower limits,
since the pre-binary-SBH nuclei might
have had central excesses like those observed 
in many fainter galaxies.

Our current understanding of nuclear cusps and cores
presents an interesting contrast to the earlier view, 
in which cores were seen as
generic \citeaffixed{tremaine-97}{e.g.}
and the presence of power-law inner profiles
was attributed to some secondary process like
adiabatic contraction driven by growth of the SBH
\citeaffixed{marel-99}{e.g.}.
Nowadays, power-law profiles are deemed
``natural'' and special explanations are sought
for cores.
Preconceptions aside, the existence of a 
core begs the question: What determines the core radius?  
A reasonable answer in the case of galactic nuclei, 
as discussed in \S7,
is that the size of the core is determined
by the radius at formation of the binary SBH
that preceded the current, single SBH.
(A similar explanation appears to work for cores
of globular  clusters; see \citename{merritt-04b} \citeyear*{merritt-04b}.)
No comparably compelling explanation currently exists
either for the inner power-law profiles or 
the compact nuclei observed in fainter elliptical 
galaxies, although it is intriguing that the latter
appear to be present only in galaxies that lack a SBH.

\section{Collisionless Equilibria}

Collisionless nuclei have relaxation times greater
than the age of the universe (\S2),
so the distribution of stars around the SBH
still reflects to some extent the details of the nuclear 
formation process, including possibly the effects
of binary SBHs (\S7).
While little can be said {\it ab initio} about the 
expected form of $\rho({\bf r})$,
a collisionless steady state must satisfy
the coupled equations
\numparts
\begin{eqnarray}
\rho({\bf r}) &=& \int\int\int f(E,I_2,I_3) d^3v, \label{eq:jeans1}\\
\nabla^2\Phi &=& 4\pi G\rho({\bf r}). \label{eq:jeans2}
\end{eqnarray}
\endnumparts
Here $f$ is the number density of stars in phase space;
$\Phi({\bf r})$ is the gravitational potential,
which includes contributions both from the SBH and 
from the stars; 
$E=v^2/2 + \Phi({\bf r})$ is the orbital
energy per unit mass; and $I_2$ and $I_3$ are
additional isolating integrals of the motion in
$\Phi$, if they exist.
For instance, in an axisymmetric nucleus, $I_2=J_z$,
the angular momentum about the symmetry axis.
Jeans's theorem states that a steady-state $f$ must 
be expressible in terms of the isolating integrals
of the motion in the potential $\Phi$,
or equivalently that $f$ must be independent of phase
on every invariant torus.

A useful way to think about the self-consistency
problem is to view each set of integral values
$(E,I_2,I_3)$ as defining a single, time-averaged orbital density;
the total density $\rho({\bf r})$ must then be
representable as a superposition of orbits
with non-negative weights \cite{schwarz-79,voort-84}.
Roughly speaking, self-consistency requires that
there exist at least as many distinct orbit families 
as there are dimensions in $\rho$; for instance,
a triaxial mass distribution requires three
isolating integrals \cite{schwarz-81}.  
This condition places almost no restrictions on the
form of $\rho$ in spherical or axisymmetric geometries
but can be an important constraint in non-axisymmetric nuclei
(see \citename{merritt-99} \citeyear*{merritt-99} for a review).

\subsection{Spherical Nuclei}

The most general form of $f$ that preserves spherical
symmetry is $f=f(E,L^2)$ with ${\bf L}={\bf r}\times {\bf v}$
the angular momentum per unit mass.
If the velocity distribution is assumed to be isotropic, $f=f(E)$,
$f$ is determined uniquely by $\rho$ and $\Phi$ via 
\Eref{eq:jeans1}.
For example, if the stellar density follows a power law
near the SBH, $\rho(r) = \rho_0(r/r_0)^{-\gamma}$, 
and assuming that $r\ll r_h$ so that the contribution to $\Phi$ 
from the stars can be ignored, then 
\begin{equation}
f(E) = \frac{3-\gamma}{8} \, \sqrt{\frac{2}{\pi^5}} \,
\frac{\Gamma(\gamma+1)}{\Gamma(\gamma-\case{1}{2})} \,
\frac{\mh}{m_\star} \,
\frac{\phi_0^{3/2}}{(G\mh)^3} \,
\left(\frac{|E|}{\phi_0}\right)^{\gamma-3/2}, \, \gamma> 1/2.
\label{eq:fofe}
\end{equation}
Here $m_\star$ is the stellar mass,
$\phi_0 = G \mh/r_h$, and $r_h$ is the radius containing
a mass in stars equal to twice $\mh$ (\S2).

For $\gamma\le 1/2$, \Eref{eq:fofe} states that $f(E)$ is 
undefined; the reason is that the low-$L$ orbits at each
$E$ force the density to increase faster than $r^{-1/2}$
toward the center.
Achieving a steady state in this case requires a depopulation of 
the eccentric orbits, i.e. a
velocity ellipsoid that is biased toward circular orbits.
The amplitude of the anisotropy required can be estimated
by repeating the derivation above after setting 
$f=f(E,L^2)=KL^{-2\beta}|E|^q$,
which imposes a constant degree of anisotropy,
$\sigma_t^2/\sigma_r^2=1-\beta$;
here $\sigma_t$ and $\sigma_r$ are the velocity dispersions perpendicular
and parallel to the radius vector.
The result is
\begin{equation}
f(E,L^2) = \frac{3-\gamma}{2^{3-\beta}} \, \sqrt{\frac{2}{\pi^5}} \,
{\Gamma(\gamma+1-2\beta)\over \Gamma(1-\beta)\Gamma(\gamma-\case{1}{2}-\beta)} \,
\frac{\mh}{m_\star} \,
\frac{\phi_0^{3/2}}{(G\mh)^3} \,
\left(\frac{L^2}{L_0^2}\right)^{-\beta} \,
\left(\frac{|E|}{\phi_0}\right)^{\gamma-3/2-\beta}
\label{eq:fofel}
\end{equation}
where $L_0^2=G\mh r_h$.
In this more general case, a non-negative
$f$ implies $\beta<\gamma-1/2$,
i.e. $\sigma_t^2/\sigma_r^2 > 3/2 - \gamma$.

\begin{figure}
\begin{center}
\includegraphics[width=0.9\textwidth]{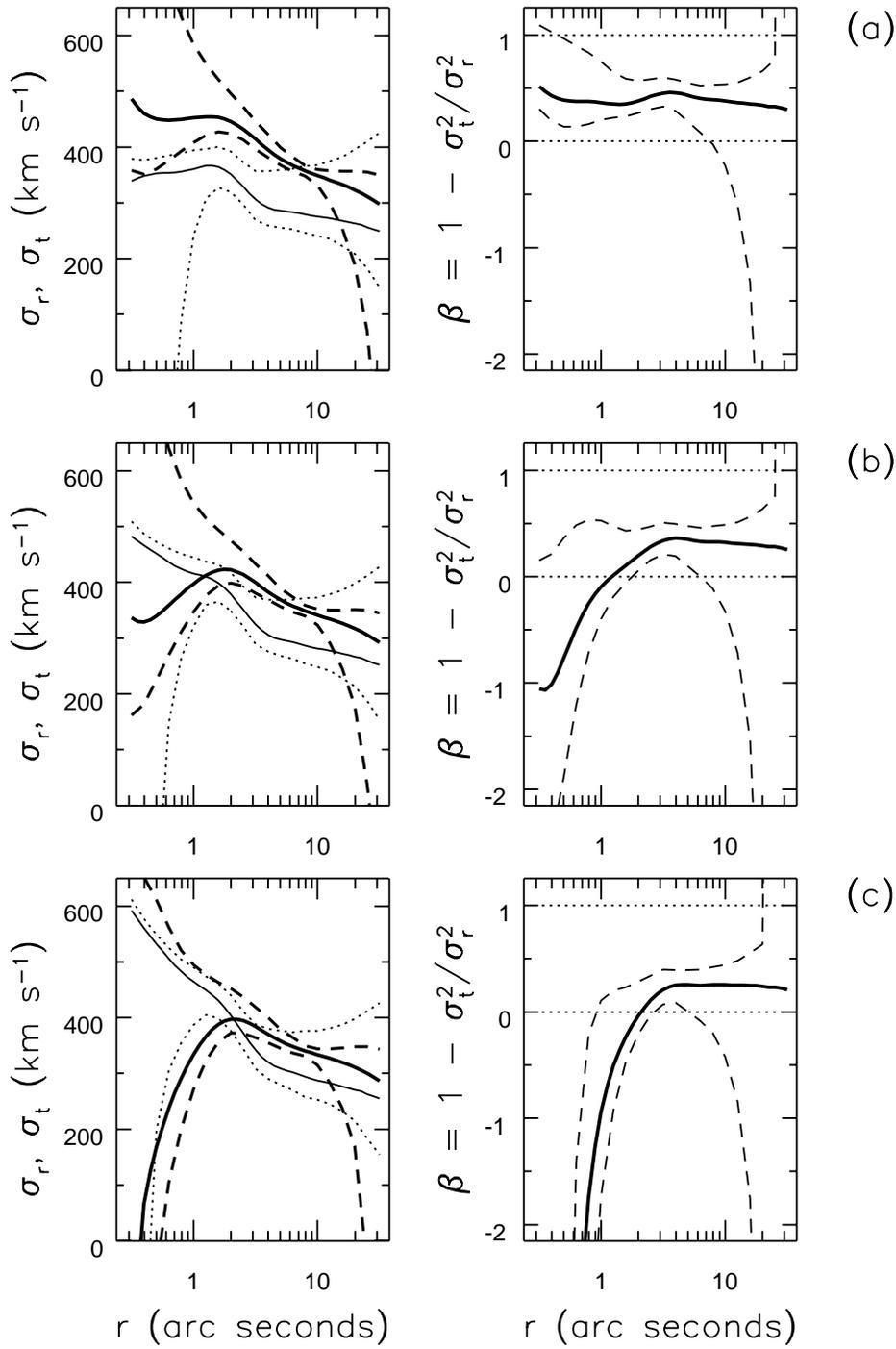}
\end{center}
\caption[]{Velocity dispersions and velocity anisotropy in
a spherical model of the center of M87, a ``core'' galaxy,
under three different assumptions about the mass of the SBH: 
(a) $\mh=1.0\times 10^9\msun$, $r_h\approx 0.3''$;
(b) $\mh=2.4\times 10^9\msun$, $r_h\approx 0.8''$;
(c) $\mh=3.8\times 10^9\msun$, $r_h\approx 1.3''$.
The individual components $\sigma_r,\sigma_t$ of the velocity
dispersion tensor were derived from the observed
run of line-of-sight velocity dispersion with radius,
under the assumption that the mass-to-light ratio
of the stars is constant.
Models with $\mh\gap 1\times 10^9\msun$ are 
characterized by tangential anisotropy,
$\sigma_t>\sigma_r$, at $r\lap r_h$, 
a consequence of the very flat stellar density profile
and the presence of a central mass.
The best current estimate of $\mh$ is $3.6\pm 1\times 10^9\msun$
\cite{macchetto-97}.
(From \citeasnoun{merritt-97}.)
}
\label{fig:m87}
\end{figure}

This result is relevant to the ``core'' galaxies (\S3), 
some of which have essentially flat or even
centrally-decreasing densities within $r_h$ \citeaffixed{lauer-02}{e.g.}.
The velocity distribution of stars near the SBHs in these galaxies
must be biased toward circular orbits.
The effect has probably been seen in M87
\citeaffixed{merritt-97,cappellari-05}{Figure~\ref{fig:m87};}.
An example of a formation mechanism that produces an anisotropic
core is ejection of stars on radial orbits by a binary SBH (\S7).

Even in spherical nuclei  without  cores, Jeans's theorem
permits large anisotropies, both toward circular orbits
($\sigma_t>\sigma_r$) and radial orbits ($\sigma_r>\sigma_t$)
\cite{merritt-85,dejonghe-89}.
Circularly-biased velocity distributions are generically stable
on dynamical time scales \cite{barnes-86,dejonghe-88};
very radial velocity distributions, like that of a
proposed, SBH-free model for the nucleus of M87 \cite{mamon-82,newton-84}, 
are unstable to nonspherical
modes that convert the nucleus into a triaxial spheroid
\cite{merritt-87}. 
However it is hard to see how strong radial aniostropies
would develop in the first place.
A moderate radial anisotropy has been claimed for stars 
within $\sim 0.1$ pc of the Milky Way SBH based on proper
motion measurements \cite{schoedel-03}.

Figure~\ref{fig:m87} illustrates an important point about
anisotropic models for galactic nuclei.
In general, one expects a {\it degeneracy} in the inferred 
gravitational potential as a result of the nonuniqueness
of $f$.
For any choice of $\Phi(r)$, there are many 2D functions $f(E,L^2)$ that can 
precisely reproduce the 1D function $\rho(r)$.
The same turns out to be true if additional moments of
the velocity distribution function are measured, e.g.
the line-of-sight velocity dispersion: there will generally
exist a range of functions $\Phi(r)$ such that a
non-negative $f(E,L^2)$ can be found which reproduces
a finite set of observed moments exactly.
In practice, the range of allowed $\Phi$'s is often extremely
wide \citeaffixed{dejonghe-92,merritt-93}{e.g.}.
This means that one can not expect to infer a unique 
value for $\mh$, the mass of the SBH, in spherical nuclei
unless substantially more data are available than the low-order velocity
moments; for instance, each of the three models illustrated
in Figure~\ref{fig:m87} makes identical predictions about
the observed velocity dispersion profile. 
A similar indeterminacy afflicts axisymmetric models, as discussed
below.

Anisotropy in the stellar velocity distribution at $r<r_h$ 
has consequences for the rate of interaction of stars with the SBH 
(\S7).
It also complicates inferences, based on the observed kinematics,
about the mass of the SBH, as discussed in more detail below.

\subsection{Axisymmetric Nuclei}

Axisymmetric nuclei have a density $\rho=\rho(\varpi,z)$
where $z$ is parallel to the symmetry axis and $\varpi^2=x^2+y^2$. 
The axisymmetric analog of an isotropic spherical model
is a so-called ``two-integral'' model, 
$f=f(E,L_z)$ with $L_z$ the component of the angular
momentum about  the symmetry axis.
In two-integral models, the functional form of $f$ is determined
uniquely once $\rho(\varpi,z)$ and $\Phi(\varpi,z)$ are specified;
the only freedom that remains is the choice of which sign
to attach to $L_z$ for each orbit, i.e. the degree of streaming
about the symmetry axis
\cite{lynden-62,dejonghe-86,hunter-93}.
Since $f$ depends on $v_\varpi$ and $v_z$ through the symmetric combination
$v_r^2+v_\varpi^2$, the velocity distribution at every position
is forced to be the same with respect to $\varpi$ and $z$, 
e.g. $\sigma_\varpi=\sigma_z$ (sometimes called ``isotropy'').
As the flattening of the mass distribution increases, 
the tensor virial theorem demands that $\sigma_z$ decrease, 
and so $\sigma_\varpi$ falls as well,
i.e. the velocity distribution becomes more biassed toward
circular motions; in the disk limit, two-integral models
contain only circular orbits.

Two-integral models are consistent with
virtually every oblate-spheroidal mass distribution, i.e.
the $f$ inferred from $\rho$ and $\Phi$ 
is nonnegative at every $(E,L_z)$.
However prolate mass models tend to require negative
$f$'s unless they are nearly spherical
\cite{batsleer-93,dejonghe-86} or have unrealistic
isodensity contours \cite{jiang-02}, implying that
prolate or barlike nuclei are dependent on a third integral.

The ease with which two-integral models can be constructed
has made them popular for modelling the central parts of galaxies
\citeaffixed{bdi-90,kuijken-95,verolme-02}{e.g.},
and the same approach is sometimes used when estimating the mass of the SBH
from kinematical data
\citeaffixed{marel-94,dehnen-95,magorrian-98}{e.g.}.
However it is dangerous to base estimates of $\mh$ on two-integral modelling,
since one has no freedom to adjust $f$ once the galaxy mass model
has been specified; the kinematical data are not used at all
aside from determining the fraction of orbits
that rotate in the two directions about the symmetry axis.
The best-fit $\mh$ is determined by the choice mandated for $f$
\cite{marel-99b}.
Indeed most of the putative SBH detections based on two-integral models 
\citeaffixed{magorrian-98}{e.g.} are now believed to be spurious 
since they were based on data that failed to resolve the
SBH's sphere of influence, sometimes by as much as two orders
of magnitude \cite{MF-01c}.

Numerical integrations reveal that the majority of orbits in
realistic axisymmetric potentials are regular, i.e.
they respect a third isolating integral $I_3$
\citeaffixed{richstone-82,evans-94}{e.g.}.
The main family of regular orbits are the tubes, which
fill a torus-shaped region having the same symmetries
as the potential.
Varying $I_3$ at fixed $E$ and $L_z$ is roughly equivalent
to varying the extent of the orbit in the $z$ direction.
Since the self-consistency problem (\ref{eq:jeans1},
\ref{eq:jeans2}) generally has a solution for the restricted
form $f=f(E,L_z)$, 
allowing the additional freedom of a third integral
typically results in a large degeneracy of solutions, 
just as in the spherical geometry with $f=f(E,L)$.
This freedom is exploited by modellers to construct
self-consistent models which reproduce not only 
$\rho(\varpi,z)$, but also whatever additional
kinematical information is available, e.g. the rotation
curve near SBH \cite{marel-98,bower-01,gebhardt-03,verolme-02,valluri-05}.

\begin{figure}
\begin{center}
\includegraphics[width=0.75\textwidth,angle=0.]{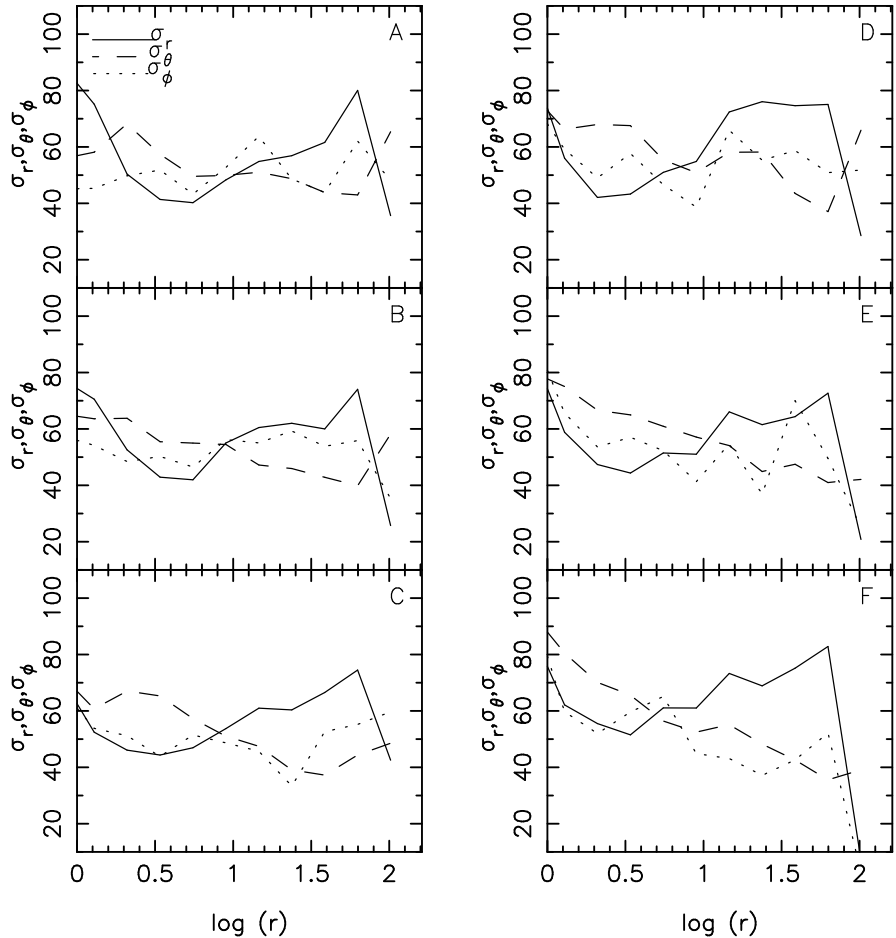}
\caption{Internal kinematics of six, three-integral
axisymmetric models of the galaxy M32.
Each panel shows the three velocity dispersions
$\sigma_r, \sigma_\phi, \sigma_\theta$ versus distance
along the major axis in seconds of arc
($1'' \approx 3$ pc).
The assumed value of $\mh$ increases from
$1.4\times 10^6\msun$ (Model A; $r_h\approx 0.4''$) to 
$4.8\times 10^6\msun$ (Model F; $r_h\approx 1''$).  
Each of these models provides an equally good fit to 
the HST/FOS and ground-based kinematical data for M32.
Models with small $\mh$ have radially-biased velocity ellipsoids
($\sigma_r>\sigma_\phi,\sigma_\theta$) at $r\lap r_h$, 
while models with large $\mh$ have tangentially-biased velocities;
in other words, the indeterminacy in $\mh$ implies
an indeterminacy also in the nuclear kinematics
 \cite{valluri-04}.
\label{fig:m32}
}
\end{center}
\end{figure}

Originally it was hoped that kinematical information
could eliminate the degeneracy inherent in three-integral
modelling and permit well-constrained estimates of $\mh$
in nearby galaxies.
Instead, it was found that in many galaxies,
variations in the assumed value of $\mh$ could be
compensated for by variations in $f(E,L_z,I_3)$
without changing the predicted line-of-sight kinematics at all; 
indeed models with $\mh=0$ often provide fits to the nuclear
kinematics that are of precisely
the same quality as models with a putative ``best-fit'' value
of $\mh$
(e.g. in NGC 4342, \citename{cretton-99} \citeyear*{cretton-99};
in NGC 3379, \citename{gebhardt-00} \citeyear*{gebhardt-00}).
This degeneracy is similar to that which characterizes
the spherical geometry (Figure~\ref{fig:m87}):
the extra dimensionality in the data (1D$\rightarrow$2D)
is matched (crudely speaking)
by the extra freedom in $f$ ($f(E,L^2)\rightarrow f(E,L_z,I_3)$).
The degeneracy is illustrated in Figure~\ref{fig:m32} for
the galaxy M32, one of the best observed and best resolved
of the SBH candidate galaxies; each of the models whose properties
are illustrated there is an equally good fit
to the kinematical data even though the assumed value of $\mh$ 
varies from $1.4$ to $4.8\times 10^6\msun$ \cite{valluri-04}.

While M32 is currently the only galaxy for which such 
comprehensive
modelling has been carried out, it is likely
that $\mh$ as derived from stellar kinematics
in other galaxies 
is comparably degenerate, if not more so, since many
of these galaxies were observed at lower effective resolutions
than M32. 
Indeed a conservative, but justifiable, view is that
no best-fit value of $\mh$ has been derived from
stellar kinematics in {\it any} galaxy aside from the Milky
Way.
Even in the case of the Milky Way, it is instructive to recall
that until about 2004, stellar-kinematical estimates of $\mh$ 
varied by roughly a factor of two,
from $\sim 1.8\times 10^6\msun$ \cite{chakra-01}
to $\sim 3.3\times 10^6\msun$ \cite{genzel-00},
in spite of the availability of velocity data that were 
resolved into a distance of $\sim 10^{-3} r_h$ from the SBH.
It was only after the orbits of individual stars, some 
with pericenter distances smaller than $10^{-4}r_h$, had
been traced that the degeneracy was removed; the best
current determination is $\mh=3.7\pm 0.2\times 10^6\msun$ 
\cite{ghez-05}.
No external galaxy has stellar kinematical data of quality 
remotely comparable with that of the Galactic center,
even {\it ca} 2000,
and so a factor of four degeneracy in $\mh$
in a galaxy like M32 (the nearest external galaxy,
which is resolved on a scale of $\sim 0.1 r_h$) 
is not surprising.

It is common practice to ``remove'' the degeneracy in
spherical or axisymmetric modelling by imposing
ad hoc regularization on the solutions, e.g.
``maximum entropy'' \cite{RT-88,gebhardt-03}.
The regularization constraint has the effect of
singling out a single solution $(\mh, f)$ as
``most probable,'' even (or especially) in cases
where the data are of insufficient quality to
select a best-fit model.

The indeterminacy in stellar-dynamical estimates of $\mh$ 
is currently the biggest impediment to refining 
our understanding of SBH demographics,
and it also severely limits the inferences that
can be drawn about the stellar kinematics 
of nuclei (e.g. Figs.~\ref{fig:m87},~\ref{fig:m32}).
In collisional nuclei like that of M32, 
the indeterminacy might be reduced by 
requiring the stellar distribution function to
be collisionally relaxed in the sense defined
below (\S 5).
However the majority of galaxies with claimed
SBH detections have collisionless nuclei, and
the indeterminacy in stellar-dynamical estimates of $\mh$
in these galaxies will probably persist for the forseeable future.

\subsection{Nonaxisymmetric Nuclei}

The degeneracy inherent in axisymmetric models
becomes even more marked in triaxial models, 
and this fact, plus the complexity of dealing with
an additional degree of freedom, has kept most galaxy modellers
from venturing beyond the axisymmetric paradigm.
But there are compelling reasons for considering non-axisymmetric
models of galactic nuclei.
Imaging of the centers of galaxies reveals a wealth of 
features in the stellar distribution on scales $r\approx r_h$
that are not consistent with axisymmetry,
including bars, bars-within-bars and nuclear spirals
\citeaffixed{shaw-95,peng-02,erwin-02}{e.g.}.
Even if these features are transient, they may be
present for a significant part of a galaxy's lifetime.
Since orbits in non-axisymmetric potentials do
not conserve any component of the angular momentum,
``centrophilic'' orbits like the boxes \cite{schwarz-79} are allowed,
which can pass arbitrarily close to the center after a finite time.
This can imply chaos in the motion \cite{gerhard-85}, 
as well as much greater rates of interaction of stars with the SBH
\cite{norman-83}.

Triaxiality is usually defined via the index $T$ where
\beq
T \equiv {a^2-b^2\over a^2-c^2}
\eeq
and $(a,b,c)$ are the scale lengths of the long ($x$), 
intermediate ($y$),
and short ($z$) axes respectively.
Oblate spheroids have $T=0$, prolate spheroids have $T=1$,
and $T=1/2$ is the ``maximally triaxial'' case.

\begin{figure}
\begin{center}
\includegraphics[width=.90\textwidth]{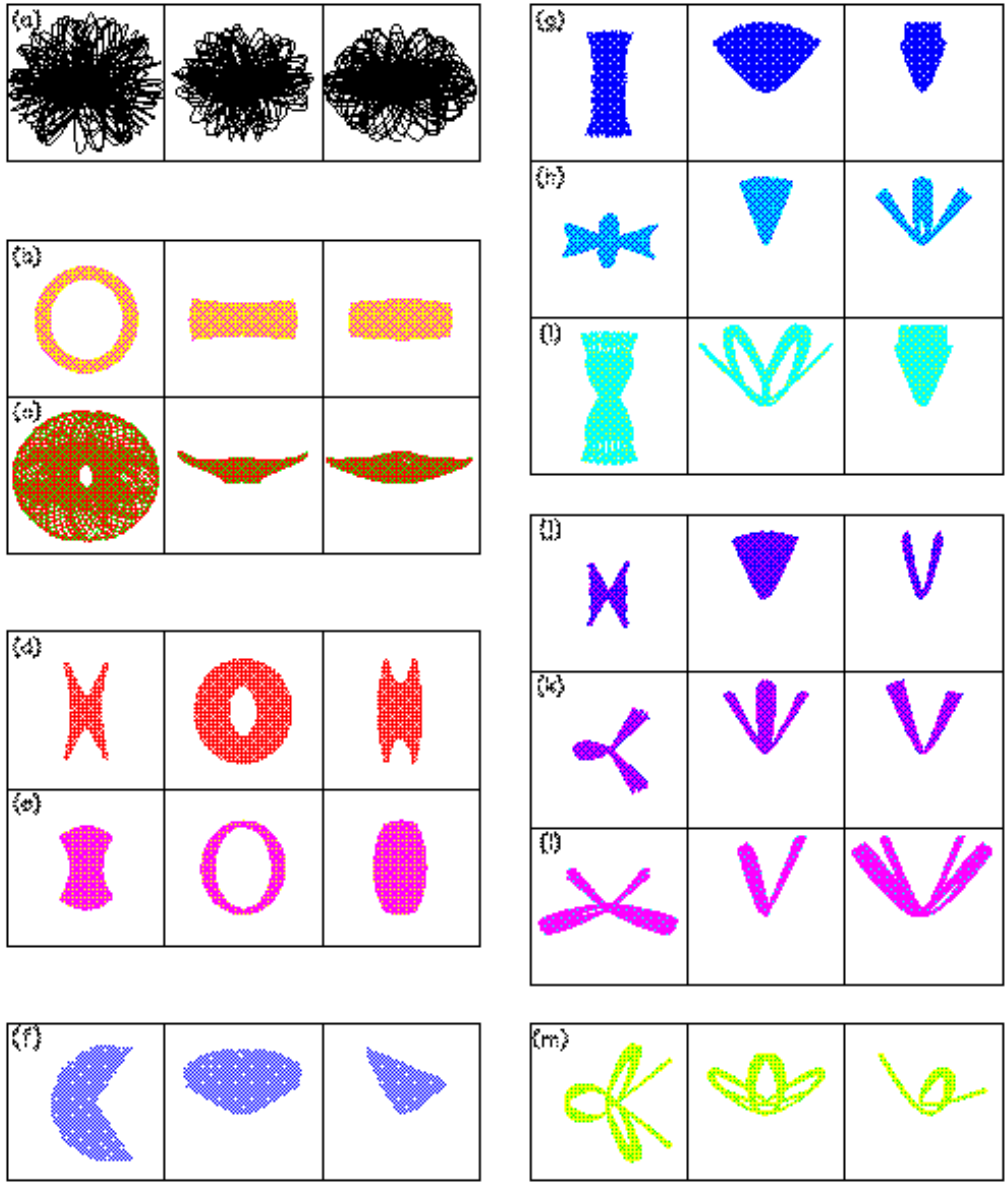}
\end{center}
\caption[]{
Major families of orbits in triaxial black-hole nuclei.
Each set of three frames shows, from left to right,
projections onto the ($x,y$), ($y,z$) and ($x,z$) planes.
(a) Stochastic orbit.
(b) Short-axis tube orbit.
(c) Saucer orbit, a resonant short-axis tube.
(d) Inner long-axis tube orbit.
(e) Outer long-axis tube orbit.
(f) ($1,-2,1)$ resonant orbit.
(g) Pyramid orbit.
(h) ($3,0,-4$) resonant pyramid orbit.
(i) ($0,6,-5$) resonant pyramid orbit.
(j) Banana orbit.
(k) $2:3:4$ resonant banana orbit.
(l) $3:4:6$ resonant banana orbit.
(m) $6:7:8$ resonant orbit.
(From \citeasnoun{poon-01}.)
}
\label{fig:triorbits}
\end{figure}

In a triaxial nucleus containing a SBH,
the character of the orbits depends on the distance from the center.
(1) In the region near the SBH, $r\lap r_h$,
the motion is nearly Keplerian and the forces
from the stars constitute a perturbation,
causing orbits to precess typically without destroying
their integrability \cite{sridhar-99,sambhus-00}.
Nearly-circular orbits are converted into tubes
similar to the tube orbits in axisymmetric potentials.
Nearly-radial orbits are converted into ``pyramids,''
Keplerian ellipses with one focus lying near the 
SBH and which precess in $x$ and $y$ 
\citeaffixed{poon-01}{Figure~\ref{fig:triorbits};}.
A symmetric pair of pyramid orbits oriented above
and below the $x-y$ plane looks similar to a 
classical box orbit \cite{schwarz-79},
although with the opposite orientation, i.e. along the
short axis of the figure, making it
less useful for reconstructing the stellar density.
A number of other, minor orbit families 
associated with resonances can be identified (Figure~\ref{fig:triorbits}).
(2) At intermediate radii, 
the SBH acts as a scattering center, 
rendering almost all of the boxlike orbits chaotic \cite{gerhard-85}.
This ``zone of chaos'' extends outward from a few
times $r_h$ to a radius where the enclosed stellar mass is
roughly $10^2$ times the mass of the SBH 
\cite{valluri-98,laskar-98,poon-01,kalap-04,kalap-05}.
Integrable tube orbits continue to exist at these energies,
as do resonant families like the $2:1$ ``banana'' orbits
that avoid the center \cite{escude-89}.
(3) At still larger energies, the phase space
is a complex mixture of chaotic and regular orbits,
including resonant boxlike orbits that remain stable 
by avoiding the center \cite{mv-99}.

Little is apparently known about the influence of figure rotation 
on the structure of orbits in triaxial black-hole nuclei.
One study \cite{valluri-99} found that figure rotation tends to
increase the degree of orbital chaos, apparently because
the Coriolis forces broaden orbits that would otherwise
be thin, driving them into the destabilizing center.

The importance of these various orbit families is
a function of how useful they are for solving the
triaxial self-consistency problem.
Following M. Schwarzschild's \citeyear{schwarz-79,schwarz-93}
pioneering work, it was generally assumed that
triaxial nuclei would need to be supported by the regular
(nonchaotic) orbit families like the pyramids and bananas
\cite{kuijken-93,syer-98,zhao-99,jalali-02}.
The first full self-consistency studies of triaxial nuclei 
\cite{poon-02,poon-04} revealed that this was only partly correct.
When only regular orbits were included in the orbital libraries,
solutions were found in the nearly oblate and ``maximally'' triaxial
geometries; the dominant orbits were the tubes circulating
around the  short axis, and the pyramid orbits.
These models had power-law ($\rho\propto r^{-1}, r^{-2}$)
radial density profiles and extended outward to $\sim 5r_h$.
But when chaotic orbits were also included, at least $40\%$,
and as much as $75\%$, of the mass was found to be
assigned to these orbits.
$N$-body integrations of these self-consistent models
confirmed their stability, at least for several
crossing times.
Highly prolate models could not be constructed, with
or without chaotic orbits.

\begin{figure}
\begin{center}
\includegraphics[width=.90\textwidth]{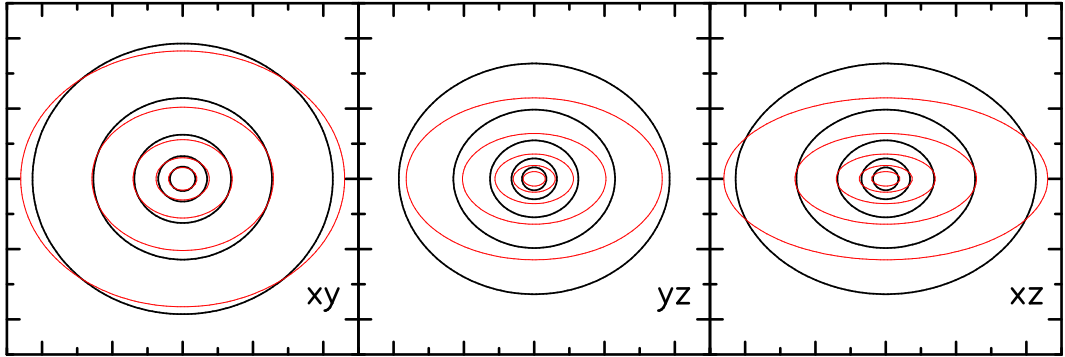}
\end{center}
\caption[]{Principal-plane cuts throught the equipotential (black/thick) 
and equidensity (red/thin) surfaces of a triaxial nucleus containing
a SBH.
The density falls off as $r^{-3/2}$ and the density axis
ratios are $c/a=0.5$, $b/a=0.7$.
The plots extend to $\sim 5$ times the SBH's influence radius $r_h$
in each direction.
The equipotential surfaces are only moderately rounder than
the equidensity surfaces.
Chaotic orbits, which fill equipotential surfaces, can therefore
be useful building blocks in the triaxial self-consistency
problem.
}
\label{fig:equipot}
\end{figure}

This work showed for the first time that chaotic orbits
could be major components of galactic nuclei.
In retrospect, this need not have been surprising.
Chaotic orbits fill a volume defined by an equipotential
surface.
Far from the center of a galaxy, the equipotential surfaces 
are nearly spherical, and so chaotic orbits at large energies
are not very useful for reconstructing an elongated
triaxial figure.
Near the center of a galaxy with a power-law density
profile, however, equipotentials are only slightly rounder
than equidensities (Figure~\ref{fig:equipot}).
Add to this the fact that many of the regular orbit families
in triaxial nuclei tend to be oriented counter to the
figure \cite{poon-01}, and it follows that self-consistent solutions
(if they exist) will draw heavily from the chaotic orbits.

Jeans's theorem in its usual form 
\citeaffixed{bt-87}{e.g. Equation~\ref{eq:jeans1};}
seems to exclude chaotic orbits.
However a little thought confirms that chaotic orbits
are perfectly acceptable components of steady-state 
galaxies, as long as they are populated with a uniform
phase-space density throughout the accessible part
of phase space \cite{kandrup-98};
indeed chaos is considered almost a {\it requirement} for
defining a steady state in statistical mechanics
\citeaffixed{sinai-63}{e.g.}.
In triaxial black-hole nuclei, the time scale for
achieving a uniform population of chaotic phase space
-- the ``chaotic mixing'' time -- would be very short, 
of order a few crossing times
\cite{mv-96,valluri-00,kandrup-02,kandrup-03}.

\begin{figure}
\begin{minipage}[b]{0.5\linewidth} 
\centering
\includegraphics[width=8.5cm]{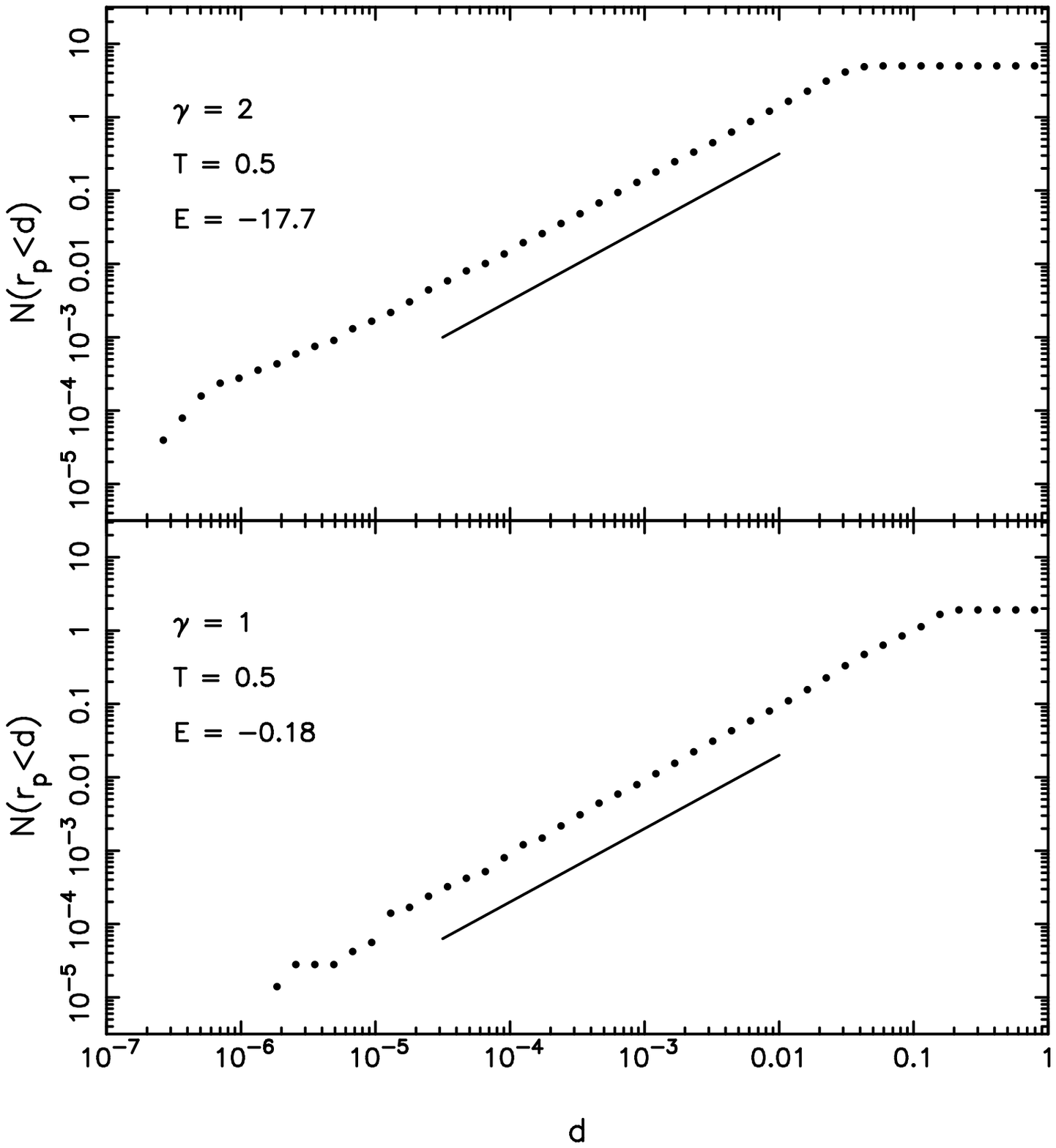}
\end{minipage}
\hspace{0.5cm} 
\begin{minipage}[b]{0.5\linewidth}
\centering
\includegraphics[width=8.5cm]{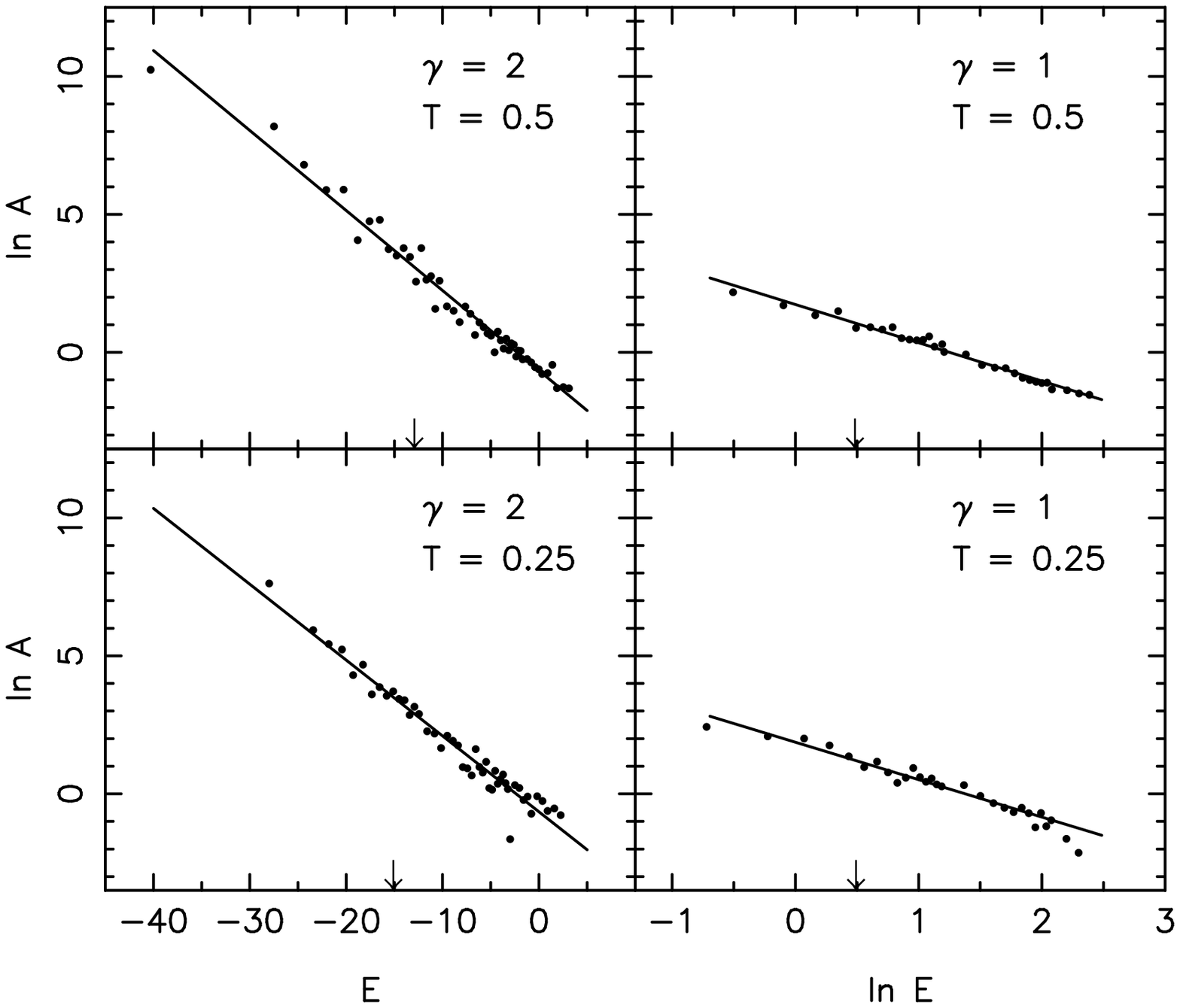}
\end{minipage}
\caption{{\it Left panel:}
Distribution of close approaches of a chaotic orbit to the
central SBH in two triaxial nucleus models.
Vertical axis shows the cumulative number of approaches
having pericenter distances less than $d$; the
SBH's tidal disruption radius is of order $10^{-7}$ in these units.
The radial  density profile is $\rho\propto r^{-\gamma}$
with $\gamma=2$ (top panel) and $\gamma=1$ (bottom panel).
Both orbits have energies $\sim E_h\equiv \Phi(r_h)$.
The solid lines have unit slope.
{\it Right panel:} The function $A(E)$ (Equation~\ref{eq:aofe}) that describes 
the cumulative rate of pericenter passages in four triaxial 
nucleus models.
The triaxiality index is $T\equiv (a^2-b^2)/(a^2-c^2)$;
$T=0.25$ is a nearly-oblate model and $T=0.5$ is a
``maximally triaxial'' model.
Points are from integrations of chaotic orbits; lines show fits;
arrows indicate $E_h$. \citeaffixed{mp-04}{From}}
\label{fig:nofd}
\end{figure}

Chaotic orbits have well-defined statistical properties.
Let $N_E(r_p<d)$ be the number of pericenter passages
per unit time for a chaotic orbit of energy $E$, 
such that the distance from the SBH at pericenter $r_p$
is less than $d$.
Numerical integrations show (e.g. Figure~\ref{fig:nofd})
that 
\beq
N_E(r_p<d) \approx A(E)\times d ,
\label{eq:aofe}
\eeq
i.e. the number of close encounters with the SBH
is an approximately linear function of the encounter distance.
The function $A(E)$, in a singular isothermal sphere
($\rho\propto r^{-2}$) nucleus, is roughly
\beq
A(E) \approx {\sigma^5\over G^2\mh^2} e^{-\left(E-E_h\right)/\sigma^2}
\label{eq:aofesis}
\eeq
and in a triaxial nucleus with $\rho\propto r^{-1}$,
\beq
A(E) \approx \sqrt{G\mh\over r_h^5}\left({E\over E_h}\right)^{-1.4}
\eeq
with $E_h\equiv\Phi(r_h)$ \cite{mp-04}.
These relations, together with estimates of the fraction
of chaotic orbits at each energy (e.g. from the self-consistent
solutions), can be used to estimate the total rate at which
stars are ``scattered'' by the triaxial potential into the
SBH, or into its tidal disruption sphere (\S6).
Here we emphasize that this mechanism of feeding stars to the SBH
is collisionless, i.e. independent of gravitational encounters.

Poon \& Merritt's (2004) study is still the only attempt
to solve the self-consistency problem for triaxial, black-hole
nuclei on scales $r\lap r_h$.
There is however a large body of work addressing the 
{\it large-scale} effects of central mass concentrations on triaxial
spheroids or bars
\citeaffixed{norman-85,hasan-93,dubinski-94,mq-98,holley-02,kalap-04,kalap-05,athan-05}{e.g.}.
These studies generally proceed by first constructing
an $N$-body model for the bar or triaxial spheroid, then a compact
mass is inserted or grown at the center and the model
is integrated forward.
Typically the model isophotes evolve toward rounder and/or
more axisymmetric shapes on scales $\gap r_h$;
the evolution can be very striking when the mass of the central
object exceeds $\sim 1$\% of the total galaxy mass
(compared with $\sim 0.1$\% for real SBHs)
and when the figure is elongated.
When the mass of the SBH is smaller, $\mh\lap 10^{-3}M_{gal}$,
evolution is often still observed but  the final shape
can still be non-axisymmetric.
The sudden evolution for large $\mh$ is likely due to conversion of the
box orbits to chaotic orbits and the ensuing chaotic mixing.
It is currently unclear whether Nature would select
stable triaxial configurations for galactic nuclei
like those constructed by \citeasnoun{poon-04},
or whether the presence of a SBH would mitigate against
such equilibria, as it seems to do on larger scales.

\citeasnoun{statler-04} used the measured streaming velocities
of stars in NGC 4365 to model its intrinsic shape; they concluded
that the galaxy was strongly triaxial, $T\approx 0.45$, and
fairly elongated, $c/a\approx 0.6$.
This result implies that SBH's of mass $\mh\approx 10^{-3}M_{gal}$ 
do not impose large-scale axisymmetry in galaxies.
The velocity data on which this conclusion was based
were only resolved on scales $r\gap 10 r_h$ however
and so do not constrain the triaxiality of the nucleus.

Due to the near-Keplerian nature of the potential at $r<r_h$,
orbits like the pyramids are not centered on the SBH.
\citeasnoun{sridhar-99} noted that off-center orbits can persist even
in nuclei where the SBH itself is offset from the center
of the stellar spheroid; furthermore they identified one family of
loop orbits for which the offset was in the same direction as that of
the spheroid.
\citeasnoun{salow-01} and \citeasnoun{sambhus-02} used this
result to construct
self-consistent, planar, lopsided models for the nucleus
of M31, and
\citeasnoun{jacobs-01} showed via $N$-body simulations that the
lopsided models could be relatively long-lived.
 
\subsection{The ``Adiabatic Growth'' Model}

It is tempting to try to derive the distribution
of stars around a SBH from first principles \cite{peebles-72}.
For instance, if the stellar velocity distribution
is assumed to be Maxwellian, $f(v)\propto e^{-v^2/2\sigma^2}$,
with constant $\sigma$,
then Jeans's theorem implies $f(E)\propto e^{-E}\propto e^{-[v^2/2+\Phi(r)]/\sigma^2}$ and \Eref{eq:jeans1} gives for the stellar density near the SBH
\beq
\rho(r) \propto \int_0^{\sqrt{2G\mh/r}} e^{-[v^2/2+\Phi(r)]/\sigma^2}v^2dv 
\propto e^{GM_\bullet/\sigma^2r}, \ \ \ \ r\ll G\mh/\sigma^2 .
\label{eq:max}
\eeq
This expression implies an exponentially divergent stellar mass within $r$.
It has other unphysical features as well: for instance,
the fact that the velocity dispersion $\sigma$ is constant implies that
typical kinetic energies near the SBH are much smaller
than binding energies, hence most stars must be near their apocenters.

A slightly more sophisticated approach is to start with
an ``isothermal'' nucleus, $f\propto e^{-E}$, {\it without}
a central mass, then increase the value of $\mh$ from
zero and ask what happens to $f$ and $\rho$.
The initial model has a constant-density core;
as the black hole grows, stars are pulled in and the density
increases.
In the limit that the rate of change of $\mh$ is slow compared
with orbital periods, this ``adiabatic growth'' model yields
for the final stellar distribution at $r\ll r_h$
\beq
f(E) = {\rm const.},\ \ \ \ \rho(r) \propto \left({r\over r_h}\right)^{-3/2}
\label{eq:PY}
\eeq
\cite{peebles-72b,young-80}.
This model and variations (nonspherical or rotating
nuclei, etc.) has been very widely investigated
(see Merritt 2004 for a comprehensive review).

\begin{figure}
\begin{center}
\includegraphics[width=0.5\textwidth,angle=-90.]{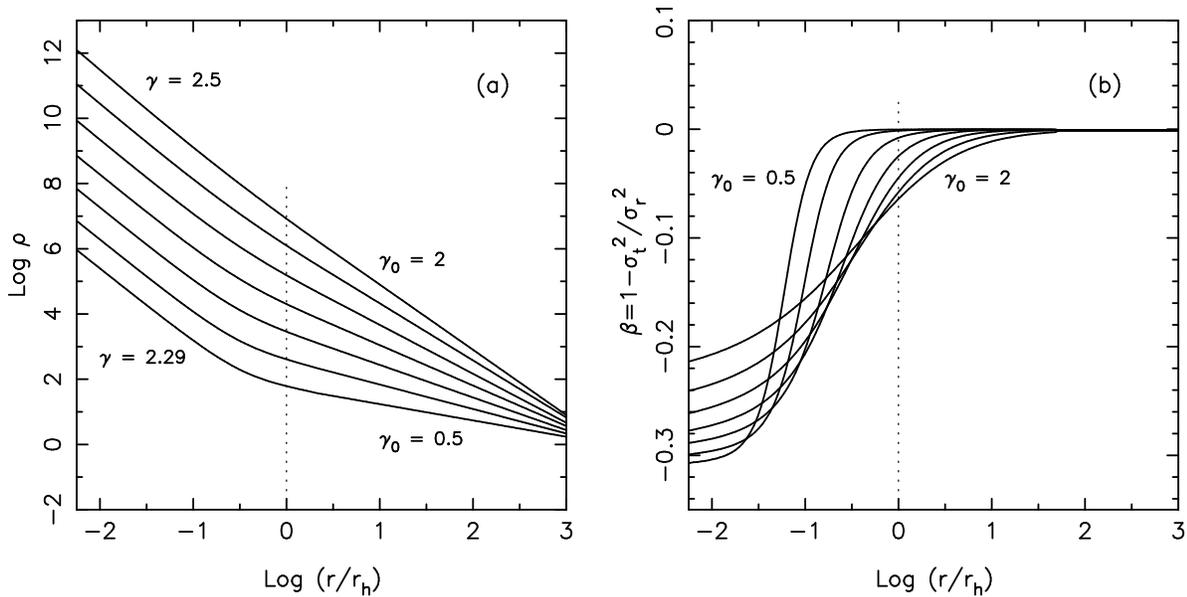}
\caption{
Influence of the adiabatic growth of a black hole on
its nuclear environment in a spherical, isotropic galaxy.
(a) Density profiles after growth of the black hole.
Initial profiles were power laws, $\rho_i\propto r^{-\gamma_0}$,
with $\gamma_0$ increasing upwards in steps of $0.25$.
The radial scale is normalized to $r_h$ as defined in the initial
galaxy (Eq.~\ref{eq:def_rh}).
The slope of the final profile at $r< r_h$ is 
almost independent of the initial slope.
(b) Velocity anisotropies after growth of the black hole.
A slight bias toward circular motions appears at $r< r_h$.
\label{fig:adiabat}}
\end{center}
\end{figure}

At first sight, \Eref{eq:PY} is very promising: as Figure~\ref{fig:LG} shows, 
the stellar densities near the centers of the Local Group
galaxies M32 and the Milky Way increase as
$\rho\sim r^{-3/2}$ at $r\lap r_h$.
But there is a consensus that the
adiabatic growth model is probably {\it not} relevant to the
structure of galaxy cores, for the following reasons.
\begin{itemize}
\item Both galaxies in Figure~\ref{fig:LG} have collisional
nuclei, i.e. the relaxation time at $r<r_h$ is shorter than the age
of the galaxy.
The nuclear density profiles in these galaxies must have
been strongly modified by energy exchange between stars (\S5).
\item The postulates of the adiabatic growth model --
that the SBH grew in mass at a fixed location via spherically-symmetric
accretion --
are extremely unlikely; almost all models for growth of SBHs
invoke strong departures from spherical
symmetry in order to remove angular momentum from the infalling gas
\citeaffixed{shlosman-90}{e.g.}.
\item Density profiles in the Milky Way, M32 and many other
galaxies are steep power laws, $\rho\sim r^{-2}$, 
at $r\gap r_h$.
If a SBH grows at the center of a galaxy
with a pre-existing power-law cusp,
the profile at $r\lap r_h$ is an even steeper power law:
an initial profile $\rho\propto r^{-\gamma_0}$
becomes $\rho\propto r^{-\gamma}, \ \gamma=2 + (4-\gamma_0)^{-1}>2$
(Figure~\ref{fig:adiabat}).
Such steep profiles are not observed (although they might
be present, but unresolved, in some galaxies).
\item The adiabatic growth model can not explain the
flat inner density profiles of the ``core'' galaxies
(\S3) without ad-hoc assumptions
\citeaffixed{marel-99b}{e.g.}.
The existence of the cores is now generally attributed
to ejection of stars by a binary SBH following
a galaxy merger (\S7); this process would likely
have taken place at some point during the formation
of virtually every spheroid, destroying an adiabatic
cusp even if it had been present.
\end{itemize}

Recently there has been a revival of interest in the
adiabatic growth model in the context of the {\it dark matter}
distribution at the center of the Milky Way and other
galaxies \cite{gondolo-99}.
Some of  the objections just raised to the adiabatic growth 
model for stellar nuclei do not apply to dark matter
\cite{BM-05}.

\section{Collisional Equilibria}

Most galaxies with well-determined SBH masses have central
relaxation times much longer than $10^{10}$ yr
(Figure~\ref{fig:tr}): their nuclei are ``collisionless.''
In a collisionless nucleus, the distribution of stars 
near the SBH will reflect the details of the nuclear
formation process, and many steady-state configurations
are possible, as discussed in \S4.
But Figure~\ref{fig:tr} reveals a clear trend of $T_r(r_h)$ with
luminosity, such that galactic spheroids fainter than 
$M_V\approx -18$  have relaxation times at $r_h$
shorter than $10^{10}$ yr.
The Milky Way nucleus has $T_r(r_h)\approx 5\times 10^{10}$ yr
but the steep density profile implies shorter relaxation times
at smaller radii:
$\sim 6\times 10^9$ yr at $0.2r_h$ ($0.6$pc)
and $\sim 3.5\times 10^9$ yr at $0.1r_h$ ($0.3$ pc)
(assuming Solar-mass stars).
The nucleus of the Milky Way is therefore ``collisional''
in the sense defined in \S2.
Three other Local Group galaxies, M32, M33 and NGC 205,
also have collisional nuclei \cite{lauer-98,hernquist-91,valluri-05}
although M32 is the only one of these to exhibit 
dynamical evidence for a SBH \cite{mfj-01,valluri-05}.

Beyond the Local Group,
essentially all of the galaxies
for which the SBH's influence radius is spatially resolved
are ``core'' galaxies (\S3)
with low nuclear densities and long relaxation times.
Still, it is reasonable to suppose that collisional nuclei
are present in at least some galaxies 
with spheroid luminosities
below the value at which the cores appear, $M_V\approx -20$.
Furthermore the nuclei of the ``core'' galaxies
may have been much denser before the cores were
created by binary SBHs (\S7).

In a collisional nucleus, an approximately steady-state
distribution of stars is set up around the SBH in a time
$\sim T_r(r_h)$.
In the case of a single stellar mass, the steady-state density
is $\rho\propto r^{-7/4}$; if there is
a mass spectrum, the heavier stars will concentrate to the center
(``mass segregation'').
``Core collapse,'' the runaway increase in density
that occurs in isolated stellar systems
after $\sim 10^2T_r$, does not occur in nuclei
containing SBHs, because
the time required is too long and because
the presence of the SBH inhibits the runaway \cite{marchant-80}.

Continued loss of stars to the SBH implies that no precisely
steady-state equilibrium can exist; for instance, 
the nucleus will slowly expand due to the effective
heat input as stars are destroyed
\cite{shapiro-77}.
This and related effects are discussed in more detail in \S 6.

\subsection{The Bahcall-Wolf Solution}

Gravitational encounters drive the local velocity
distribution toward a Maxwellian,
but a Maxwellian velocity
distribution implies an exponentially divergent mass near
the SBH (Equation \ref{eq:max}).
The existence of a region close to the hole where stars
are captured or destroyed prevents the nucleus from
reaching thermal equilibrium.
The density must drop to zero on orbits
that intersect the SBH's event horizon at $r=r_S$, 
or that pass within the tidal disruption sphere at $r_t$;
the latter radius is most relevant since
galaxies with collisional nuclei probably always have 
$\mh\lap 10^8M_\odot$ (\S 2).
Ignoring for the moment a possible dependence of the 
stellar phase space density on orbital angular momentum $L$,
and assuming spherical symmetry,
the evolution of $f$ can be approximated via the isotropic 
orbit-averaged Fokker Planck equation,
\numparts
\begin{eqnarray}
& & 4\pi^2p(E){\partial f\over\partial t} = -{\partial F_E\over\partial E}
 , 
\label{eq:fpa}
 \\
& & F_E(E,t) = -D_{EE}{\partial f\over\partial E} - D_Ef
\label{eq:fpb}
\end{eqnarray}
\endnumparts
\cite{spitzer-87}.
\noindent Here $p(E)$ is a phase-space volume element,
$p(E)=2^{-3/2}\pi G^3\mh^3|E|^{-5/2}$ near the SBH,
$F_E$ is the flux of stars in energy space, and $D_E$ and $D_{EE}$ are
diffusion coefficients that describe the effects of small-angle 
scattering:
\begin{eqnarray}
D_{EE}(E) &=& 64\pi^4G^2m^2\ln\Lambda
\left[q(E) \int_{-\infty}^E dE'f(E') + 
\int_E^0 dE' q(E')f(E')\right], \nonumber \\
D_E(E) &=& -64\pi^4G^2m^2\ln\Lambda\int_E^0 dE'p(E')f(E')
\end{eqnarray}
\noindent with $\ln\Lambda\approx \ln(\mh/m)$ the Coulomb logarithm
and $q(E)=(2^{1/2}\pi/6)G^3\mh^3|E|^{-3/2}$.
The boundary conditions are $f(E_t)=0$ and $f(0)=f_0$;
$E_t=G\mh/r_t$ is the energy at which stars are lost to the SBH 
and $f_0$ is the phase space density at $E=0$. 
(The inner boundary condition is only approximate since
the condition for a star to pass inside $r_t$ is angular-momentum
dependent, as discussed in more detail below.
The outer boundary condition is also approximate since
a non-zero $f$ at large radii implies a population of
stars whose  contribution to the potential has been
ignored.)

\begin{figure}
\begin{center}
\includegraphics[width=0.5\textwidth,angle=-90.]{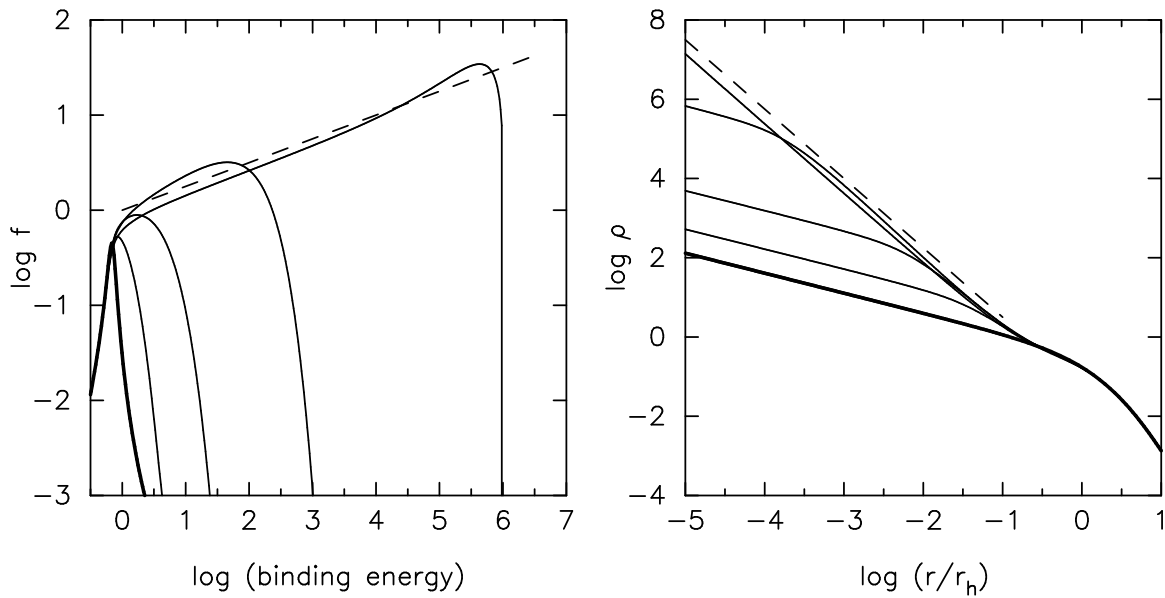}
\caption{
Evolution of the stellar distribution around a SBH due
to energy exchange between stars.
These curves were computed from the isotropic, orbit-averaged
Fokker-Planck equation (Equations~\ref{eq:fpa},~\ref{eq:fpb}) 
with boundary condition $f=0$ at $\log |E| = 6$.
Left panel: phase-space density $f$; right panel: 
configuration-space density $\rho$.
The initial distribution (shown in bold) had $\rho\propto r^{-0.5}$
near the SBH;
thin curves show $f$ and $\rho$ at times of $(0.2,0.4,0.6,1.0)$
in units of the relaxation time at the SBH's initial influence
radius $r_h$.
Dashed lines show the ``zero-flux'' solution
$f\propto |E|^{1/4}$, $\rho\propto r^{-7/4}$.
The steady-state density is well approximated by the
zero-flux solution at $r\lap 0.2 r_h$.
  \label{fig:fp}}
\end{center}
\end{figure}

\citeasnoun{bw-76} first presented numerical solutions
to Equations (\ref{eq:fpa}, \ref{eq:fpb}).
They found that a steady state is reached after roughly one relaxation 
time at $r_h$.
If $|E_t|\gg G\mh/r_h$, i.e. if the disruption radius $r_t$
is much smaller than $r_h$ (which is the case in real nuclei),
the steady-state solution is close to a power law,
\beq
f(E) = f_0|E|^{1/4},\ \ \ \ \rho(r) = \rho_0 r^{-7/4}, \ \ \ \ 
|E| \ll |E_t|,\ \ \ \ r_t \ll r .
\label{eq:zflux}
\eeq
Equation (\ref{eq:zflux}) is a ``zero-flux'' solution, 
i.e. it implies $F_E=0$.
(An ``isothermal'' distribution, $f\sim e^{E/\sigma_0^2}$,
also implies zero flux but is unphysical for the reasons 
discussed above.)

Figure~\ref{fig:fp} illustrates the evolution of $f(E,t)$ in the
case of a cluster that extends beyond the SBH's influence
radius.
The Bahcall-Wolf cusp rises above the pre-existing density
inside a radius $\sim 0.2r_h$.

In the numerical solutions, the steady-state flux is found
to be small but non-zero, of order
\beq
F(E) \approx {n(r_t)r_t^3\over T_r(r_t)} \propto r_t.
\label{eq:fluxe}
\eeq
In other words, the flux is determined by the rate
at which stars can diffuse into the disruption sphere at $r_t$.
This flux is ``small'' in the sense that the one-way
flux of stars in or out through a surface
at $r_t\ll r \lap r_h$ is much greater;
except near $r_t$, the inward and outward fluxes almost cancel.
Thus, the steady-state flux is limited by the ``bottleneck''
at $r=r_t$.
As $r_t$ is reduced, the flux approaches
zero and the numerical solution approaches the power-law
form of Equation~(\ref{eq:zflux}).

At the center of the Milky Way, the flux implied by the 
Bahcall-Wolf solution would only be of order $\sim 10^{-12}$ stars yr$^{-1}$.
\citeasnoun{FR-76} pointed out that the actual  
loss rate to a black  hole would be dominated by changes in
angular momentum, not energy, implying a much
higher flux.
(This is discussed in more detail in \S 6.)
\citeasnoun{bw-77} included these loss-cone
effects heuristically, by finding steady-state solutions of
the modified equation
\beq
4\pi^2p(E){\partial f\over\partial t} = -{\partial F_E\over\partial E} - \rho_{lc}(E,t)
\label{eq:fpmod}
\eeq
where the term $\rho_{lc}$ is an approximate representation of
the true loss rate into the SBH \cite{LS-77}.
They found that the addition of the loss term had only a small
effect on the  steady-state form of $f(E)$ and $\rho(r)$
even though it substantially increased the implied
loss rate.

Solutions of the full, anisotropic Fokker-Planck equation,
including a careful treatment of loss cone dynamics as well as
physical collisions between stars,
were first presented by \citeasnoun{CK-78}.
(The parameters in this study were chosen to mimic the distribution
of stars around a black hole in a globular cluster.)
The logarithmic derivative of the steady-state density was found to
be $d\ln\rho/d\ln r\approx -1.65$ for $r$ in the range
$10^{-3}<r/r_h<10^{-1}$, compared with the Bahcall-Wolf 
value of $-1.75$.
The velocity anisotropy was found to be close to zero
for $r\gap 10^{-3} r_h$.
The Bahcall-Wolf solution has been verified in a number of
other studies based on fluid \cite{amaro-04}
or Monte-Carlo \cite{marchant-80,duncan-83,freitag-02}
approximations to the Fokker-Planck equation.

\begin{figure}
\begin{center}
\includegraphics[width=0.65\textwidth,angle=0.]{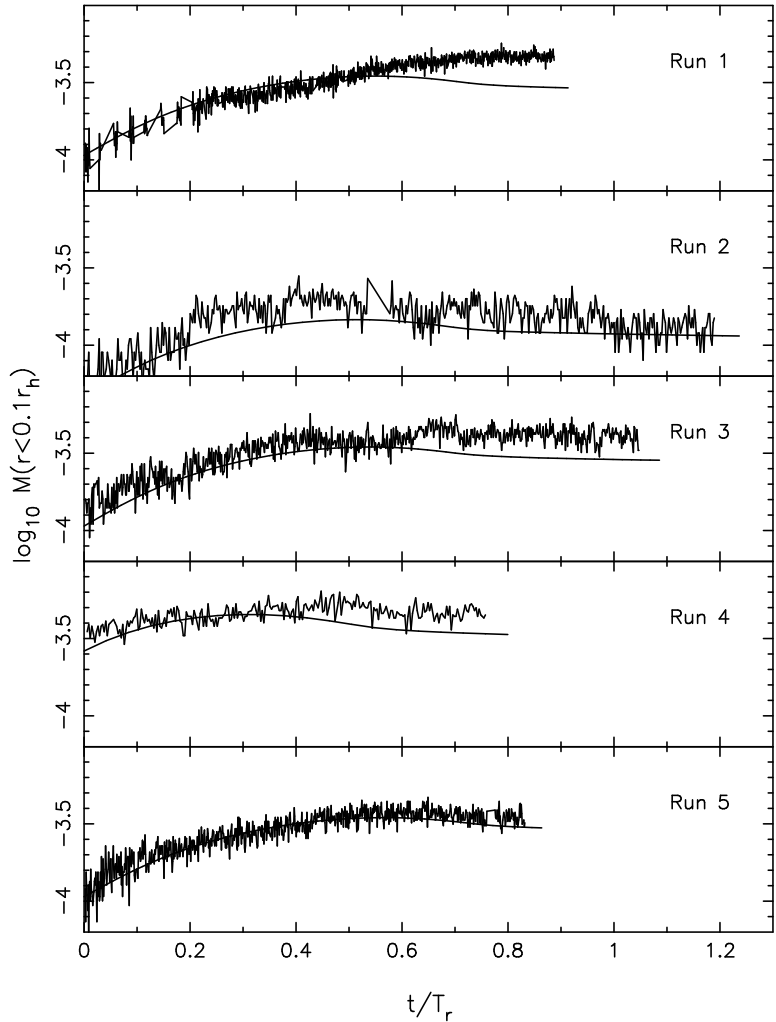}
\caption
{Evolution of the mass in stars within a distance
$0.1 r_h$ from a massive body (``black hole'') 
in a set of $N$-body simulations with different initial
conditions and different particle numbers.
A chain-regularization algorithm was used to handle
close encounters between stars and the black hole.
Smooth curves are solutions to the Fokker-Planck equation 
(\ref{eq:fpa},\ref{eq:fpb}).
\citeaffixed{preto-04}{From}
\label{fig:preto}
}
\end{center}
\end{figure}

Most recently, advances in computer hardware
\citeaffixed{namura-03}{e.g.}
and software \citeaffixed{mikkola-90,mikkola-93}{e.g.}
have made it possible to test the Bahcall-Wolf solution
via direct $N$-body integrations, avoiding the approximations
of the Fokker-Planck formalism \cite{preto-04,baumgardt-04a,MS-05}.
Figure~\ref{fig:preto} shows a set of $N$-body simulations 
of collisional cusp growth compared with the predictions of the 
isotropic Fokker-Planck equation.
In these simulations, there was no loss of stars and
so a true ``zero-flux'' equilibrium was established.

The Milky Way is probably the only galaxy in which
a Bahcall-Wolf cusp could currently be detected: its central
relaxation time is shorter than  $10^{10}$ yr and
its nuclear density profile is resolved on scales
$\ll r_h$.
Figure~\ref{fig:LG} suggests that the bright stars
have $\rho\sim r^{-1.5}$
at $10^{-3}\lap r/r_h \lap 10^{-1}$.
This is slightly shallower than the Bahcall-Wolf 
``zero-flux'' prediction but probably consistent given
the uncertainties in the observed profile \cite{schoedel-06}.
The presence of a mass spectrum also implies a smaller slope
(\S\ref{sec:mm}).
In addition, the time required to reach a steady state
at the Galactic center may be $\gap 10^{10}$ yr
\cite{MS-05}.

\subsection{Multi-Mass Equilibria}
\label{sec:mm}

Galaxies contain stars with a range of masses.
The initial mass function, i.e. the distribution
of masses at the time of formation, is believed
to be roughly a power-law, 
$n(m)\propto m^{-\alpha},\alpha\approx 2$
\cite{salpeter-55,miller-79},
but the mass function changes with time 
as stars lose mass  and as new stars are formed.
Mass functions near the centers of galaxies
are difficult to constrain observationally;
this is true even at the Galactic center
due to crowding and obscuration.
Evolutionary models assuming a constant rate of star
formation \citeaffixed{alexander-05}{e.g.} 
suggest that $\sim 75\%$ of the mass after 10 Gyr would be in 
the form of ``live'' stars with $\langle m\rangle\approx 0.5\msun$,
$\sim 20\%$ in white dwarves with $0.6\msun\le m\le 1.1\msun$,
and a few percent in neutron stars ($m=1.4\msun$)
and black holes ($m\approx 10\msun$).
At the center of  the Milky Way 
there is also known to be a population of more massive stars,
$3\msun\lap m\lap 15\msun$, inside $\sim 10^{-2} r_h$,
which probably could not have formed {\it in situ}
due to tidal stresses from the SBH \cite{ghez-03,eisenhauer-05}.

Exchange of energy between stars with different masses
tends to establish local equipartition of kinetic energy,
leading to spatial segregation \cite{SS-75}:
the more massive stars congregate closer to the center.

\begin{figure}
\begin{center}
\includegraphics[width=0.45\textwidth,angle=-90.]{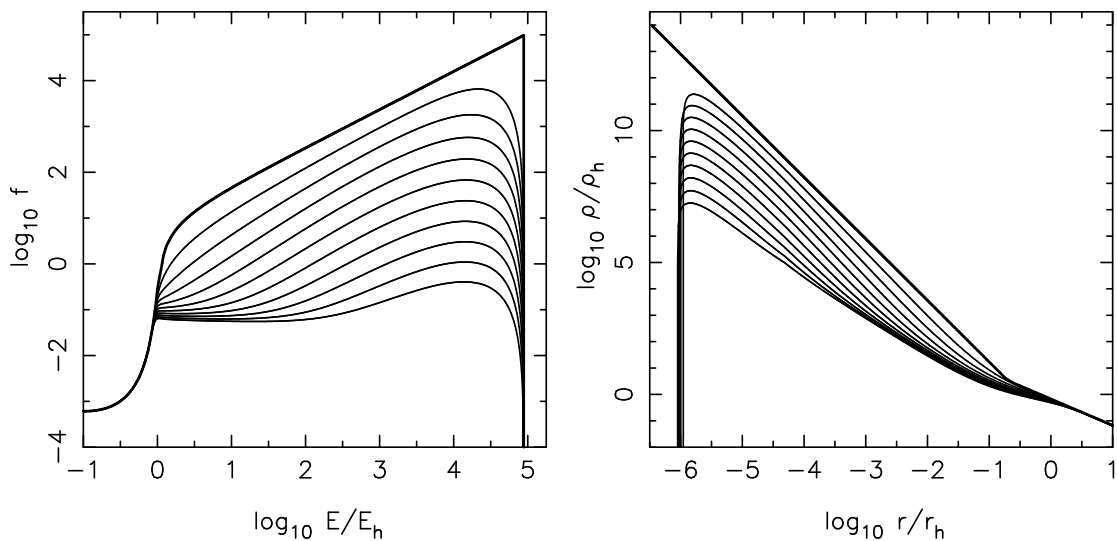}
\caption{
Evolution of a population of low-mass objects due to
heating from a dominant, high-mass population and
to scattering into the SBH.
$f(E,t)$ and $\rho(r,t)$ are the phase-space and 
configuration-space densities respectively of the low-mass population.
Times shown are 2, 4, .... 20 in units of the time
$T_2$ defined in Eq.~\ref{eq:t1t2}.
The space density evolves from $\rho\sim r^{-2.3}$
to $\sim r^{-1.5}$.
(Adapted from \citename{merritt-04} \citeyear*{merritt-04})
  \label{fig:dm}}
\end{center}
\end{figure}

The mass-segregation time scale can be estimated by
considering two mass groups $m_1$ and $m_2$.
Assuming Maxwellian velocity distributions,
the mean rate of change of kinetic energy
for stars of mass $m_1$ due  to encounters with stars of mass $m_2$ is
\begin{equation}
\frac{d\epsilon_1}{dt}
=\frac{8\left(6\pi\right)^{1/2} G^2m_1\rho_2\ln\Lambda}{\left(v_{1,rms}^2 + 
v_{2,rms}^2\right)^{3/2}} (\epsilon_2 - \epsilon_1)
\label{eq:equipart}
\end{equation}
where $\epsilon_i={1\over 2}m_i v_{i,rms}^2$;
the same expression with indices interchanged gives $d\epsilon_2/dt$ \cite{spitzer-87}.
Suppose that $m_2\ll m_1$ and assume $v_{1,rms}\approx v_{2,rms}$,
appropriate shortly after the nucleus forms.
Then
\begin{equation}
T_{1} \equiv \left| {1\over\epsilon_1} {d\epsilon_1\over dt}\right|^{-1} 
= {0.0814 v_{rms}^3\over G^2m_1\rho_2\ln\Lambda}, \ \ \ \ 
T_{2} \equiv \left| {1\over\epsilon_2} {d\epsilon_2\over dt}\right|^{-1} 
= {\rho_2\over \rho_1} T_1.
\label{eq:t1t2}
\end{equation}
Consider first the case that the heavier stars dominate the total 
density and have a relaxation time $T_{r,1}$. 
Then $T_2\approx T_{r,1}$ and the light stars reach equipartition
on the same time scale that the heavy stars establish 
a collisional steady-state.
On the other hand, if the light stars dominate,
with relaxation time $T_{r,2}$,
then $T_1\approx (m_2/m_1)T_{r,2}$, and the heavy stars 
lose energy to the light stars very rapidly compared with $T_{r,2}$.
The first case describes a nucleus containing stars
and particle dark matter, while the latter describes a nucleus
containing stars and a population of massive remnants.
In either case, the time for the sub-dominant component 
to reach equipartition with the dominant component is
of order $T_r$ for the dominant component or less.

The single-mass Fokker-Planck equation can be generalized
to the multi-mass case by defining $f(E,m,t)dm$ as the number density of
stars in phase space with masses in the range $m$ to $m+dm$.
Then \citeaffixed{merritt-83}{e.g.}
\begin{eqnarray}
& & 4\pi^2p(E){\partial f\over\partial t} = -{\partial F_E\over\partial E}
 , 
\nonumber \\
& & F_E(E,m,t) = -D_{EE}{\partial f\over\partial E} - mD_Ef,
\label{eq:fpmm}
\end{eqnarray}
with diffusion coefficients
\begin{eqnarray}
D_{EE}(E) &=& 64\pi^4G^2\ln\Lambda
\left[q(E) \int_{-\infty}^E dE'h(E') + 
\int_E^0 dE' q(E')h(E')\right], \nonumber \\
D_E(E) &=& -64\pi^4G^2\ln\Lambda\int_E^0 dE'p(E')g(E')
\end{eqnarray}
and $g$ and $h$ are moments over mass of $f$:
\begin{eqnarray}
g(E,t) &=& \int_0^\infty f(E,m,t) m\ dm, \nonumber \\
h(E,t) &=& \int_0^\infty f(E,m,t) m^2\ dm. 
\end{eqnarray}

Considering again the case of a nucleus
containing stars with just two masses,
$m_1\gg m_2$,
the evolution equation for the lighter component
may be found by taking the first moment over mass of \Eref{eq:fpmm},
restricting the integral to the mass range which characterizes
the lighter stars.
The result is
\numparts
\begin{eqnarray}
& & 4\pi^2p(E){\partial g_2\over\partial t} = {\partial \over\partial E}
\left(D_{EE}{\partial g_2\over\partial E}\right), \\
D_{EE}(E) &=& 64\pi^4G^2\ln\Lambda
\left[q(E) \int_{-\infty}^E dE'h_1(E') + 
\int_E^0 dE' q(E')h_1(E')\right]
\end{eqnarray}
\endnumparts
with $g_2$ the phase-space mass density of the lighter stars.
(These equations apply also to the more general case of 
a distribution of light-star masses; 
$g_2$ is then the total mass density of the lighter population.)
The steady-state solution is
obtained by setting $\partial g_2/\partial E=0$,
yielding a density profile for the light component 
$\rho_2\propto r^{-3/2}$
(cf. Equation~\ref{eq:fofe}), {\it independent} of 
the $f$ that describes the heavier stars.
Figure~\ref{fig:dm} illustrates the evolution toward
this  state.
If the heavier objects dominate the density, then their 
steady-state density is $\rho\propto r^{-7/4}$;
in other words, the lighter component is less centrally
concentrated than the heavier component.
Note however the difference in density slopes is fairly small
even in this extreme case.

\citeasnoun{bw-77} solved the coupled Fokker-Planck
equations for systems containing stars of two
masses around a SBH.
They assumed that the two populations had similar densities at $r\gap r_h$.
The heavier component was found to always attain a steady-state
density with slope close to the single-mass value $-7/4$,
while the lighter component was less centrally condensed,
with index
\begin{equation}
-{d\ln\rho_2\over d\ln r} \approx \left({m_2\over 4m_1} + {3\over 2}\right).
\label{eq:mm}
\end{equation}
This expression gives the expected results 
$\rho_2\propto r^{-7/4}$ when $m_2=m_1$
and $\rho_2\propto r^{-3/2}$ when $m_2\ll m_1$.

\citeasnoun{MCD-91} incorporated realistic mass
spectra into their time-dependent solutions of
the isotropic Fokker-Planck equation; they also
included physical collisions as well as mass loss from stars.
\citename{MCD-91} found that the more massive component 
always attained a power-law slope close to $-7/4$, 
the least massive component had
a slope near $-3/2$, and intermediate mass groups
scaled approximately as \Eref{eq:mm}.
\citename{baumgardt-04b} \citeyear{baumgardt-04b,baumgardt-05}
carried out $N$-body simulations of star clusters containing 
a massive black hole and a range of stellar masses; 
they observed mass segregation but did not compare their 
results in detail with the Bahcall-Wolf formula.
Baumgardt et al. found that the density profile
of the ``bright stars'' in their simulations
(i.e. the particles representing giants)
was very flat after a relaxation time, 
with no hint of a power-law cusp.
This puzzling result appears not to have been 
satisfactorily explained.
Most recently, \citeasnoun{freitag-06} presented
Monte-Carlo models of the evolution of the Galactic
center star cluster including a \citeasnoun{kroupa-93}
mass function.
The heaviest mass group in their simulations
consisted of the $\sim 10\msun$ stellar black  holes;
in models designed to mimic the Galactic center
cluster, they found that the black holes
segregated to the center in $\sim 5$ Gyr,
after which loss of stars into the SBH
drove a general expansion (\S6.5).
The stellar black holes were found to dominate
the mass density within $\sim 0.2$ pc $\approx 0.1r_h$
of the center.
The density profile of the black holes was
found to be ``compatible'' (modulo noise)
with the Bahcall-Wolf single mass form, $d\log\rho/d\log r = -7/4$,
while the main-sequence stars near the center had
$-d\log\rho/d\log r\approx 1.3-1.4$.
\citename{freitag-06}'s simulations were targeted
toward nuclei with SBHs in the mass range
$10^5\msun \le\mh\le 10^7\msun$, even though
there is no firm evidence for SBHs with masses
below $\sim 10^{6.5}\msun$, since these
would be the SBHs of most interest to gravitational-wave
physicists.
Their results suggest that mass segregation is likely to
be of marginal importance in the majority of nuclei
with confirmed SBHs due to the long relaxation times
(\S 2).

\subsection{Cusp Regeneration}

\begin{figure}
\begin{center}
\includegraphics[width=0.5\textwidth,angle=0.]{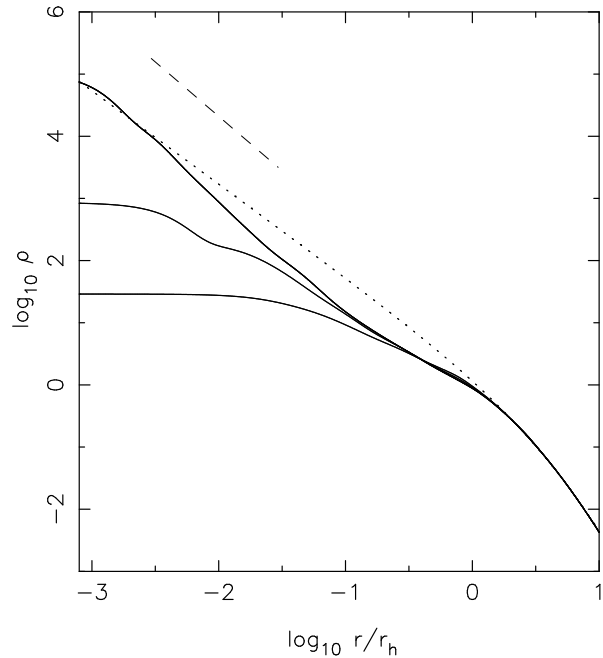}
\caption
{Regeneration of a Bahcall-Wolf cusp around a black
hole after destruction by a massive binary.
Dotted line is the initial density profile of
an $N$-body model containing a massive central
particle (the ``black hole'') of mass $0.01$,
in  units where the total galaxy mass is one.
Lower solid line shows the density after infall of
a second ``black hole'' of mass $0.005$ has 
destroyed the cusp.
Upper solid lines show the evolving density after
the two black hole particles were combined into one,
simulating coalescence.
The final time would  be rougly $10^{10}$ yr if scaled
to the Galactic center.
The dashed line has logarithmic slope of $-7/4$.
(Adapted from \citename{MS-05} \citeyear*{MS-05}.)
\label{fig:regen}
}
\end{center}
\end{figure}

As discussed below (\S 7), binary SBHs formed during
galaxy mergers are efficient at destroying dense nuclei.
Essentially all stellar spheroids are believed to have experienced
such events in the past, and evidence for the
``scouring'' effect of binary SBHs is seen in the
flat, central density profiles or ``mass deficits''
of bright elliptical galaxies (Figure~\ref{fig:sersic}).
An important question is whether the existence
of dense cusps at the centers of galaxies like the
Milky Way and M32 implies that no binary SBH was
ever present, or whether a collisional cusp could have
spontaneously regenerated after being  destroyed.
\citeasnoun{MS-05} used $N$-body integrations
to simulate cusp destruction by a binary SBH,
then combined the two ``black holes''  into one
and continued the integrations until a Bahcall-Wolf
cusp had formed around the single massive particle.
Figure~\ref{fig:regen} shows the results of one
such integration for mass ratio $q=0.5$ and initial
density profile $\rho\sim r^{-1.5}$.
Cusp regeneration was found to require roughly
one relaxation time as measured at the SBH's
influence radius $r_h$; for reasonable mass ratios
($M_2/M_1\lap 0.5$), the binary hardly
affects the density at this radius (Fig.~\ref{fig:regen}).
Growth of the cusp is preceded by a stage
in which the stellar velocity dispersion evolves toward
isotropy and away from the tangentially-anisotropic
state induced by the binary.
When scaled to the Galactic center,
these experiments suggest that a dense cusp could
have been regenerated in $10^{10}$ yr, although
evolution toward the steady-state profile might
still be occurring.

As discussed above (\S3), ``mass deficits,''
or cores, are observed to disappear in galaxies
fainter than $M_V\approx -19.5$.
This might be due in part to cusp regeneration
in these galaxies, although central
relaxation times are almost always too long for this
explanation to be convincing (Fig.~\ref{fig:tr}).
An alternative explanation is suggested by
Figure~\ref{fig:tr}: galaxies fainter
than $M_V\approx -19.5$ are mostly unresolved
on scales of $r_h$, which is also approximately
the size of a core created by a binary SBH.
The lack of mass deficits in galaxies with
$M_V\gap -19.5$ probably just  reflects a failure
to resolve the cores in these galaxies.

\section{Loss-Cone Dynamics}

As discussed in \S5.1, the existence of a region $r\le r_t$ close
to the SBH where stars are captured or destroyed can have
a significant influence on the steady-state distribution
of stars even at radii $\gg r_t$, since it precludes
the formation of an ``isothermal'' distribution of velocites.
Loss of stars is also important because of its observational
consequences: tidally disrupted stars are expected
to produce X- and UV radiation with luminosities
of $\sim 10^{44}$ erg s$^{-1}$, potentially outshining
their host galaxies for a period of days or weeks
\cite{rees-90,kobayashi-04,khoklov-96,kochanek-94}.
A handful of X-ray flaring events have been observed
that have the expected signature 
\cite{komossa-02,komossa-04,halpern-04},
and the number of detections is crudely consistent
with theoretical estimates of the event rate 
\cite{donley-02,wang-04}.
Tidal flaring events may dominate the $X$-ray 
luminosity function of AGN at $L_X\lap 10^{44}$ erg s$^{-1}$
\cite{MMH-06}.

Compact objects (neutron stars or stellar-mass black holes)
can remain intact at much smaller distances from the SBH;
these objects would emit gravitational waves at potentially
observable amplitudes before spiralling in, and may 
dominate the event rate for low-frequency gravitational
wave interferometers like LISA \cite{sr-97,gair-04,hopman-06}.
(The other major categories of low-frequency gravitational
wave events are compact, stellar binaries and binary
SBHs; the latter are discussed in \S7.)

A central SBH acts like a sink, removing stars on ``loss-cone''
orbits, i.e. orbits that intersect the capture or disruption
sphere $r=r_t$.
In a spherical galaxy,
this removal is complete after just one galaxy crossing time,
and continued supply of stars to the SBH requires some
mechanism for loss-cone repopulation.
The  most widely discussed mechanism is gravitational
encounters, which drive a diffusion in energy ($E$) and
angular momentum ($L$).
The latter dominates the loss rate \cite{FR-76,LS-77}.
Roughly speaking, many of the stars within the SBH's
influence radius will be deflected into $r_t$ in one
relaxation time, i.e. the loss rate is roughly
$(M_\bullet/m_\star)T_r(r_h)^{-1}$.
In a collisional nucleus (\S2) with $M_\bullet\approx 10^6M_\odot$, 
this is $\sim 10^6/(10^{10}$ yr$)\approx 10^{-4}$ yr$^{-1}$.

Classical loss cone theory \cite{bw-76,LS-77,ipser-78,CK-78}
was directed toward understanding
the observable consequences of massive black holes 
at the centers of globular clusters.
Globular clusters are many relaxation times old, 
and this assumption was built into the theory, 
by requiring the stellar phase space density $f$
near the black hole to have reached an approximately
steady state under the influence of gravitational
encounters.
One consequence of the much longer
($\gap 10^{10}$ yr) relaxation times in galactic nuclei
is that the stellar density profile near
the SBH need not have the collisionally-relaxed, 
Bahcall-Wolf form (\S5.1).
Another is that nuclei can be strongly nonspherical
or non-axisymmetric,
which permits the existence of centrophilic (box or chaotic)
orbits (\S4.3); stars on centrophilic orbits can easily dominate 
the loss rate.
In addition, depending on the details of the nuclear
formation process, the stellar phase-space density
in a galactic nucleus
can depart strongly from its steady-state form near
the loss cone, even in a precisely spherical galaxy.
For instance, if the current SBH was preceded by a 
binary SBH, stars on orbits with pericenter distances
$\sim a_h \gg r_t$ will have been ejected by the binary,
and until these orbits are repopulated (on a time scale $\sim T_r$), 
the loss rate to the SBH can be much smaller than in
the steady state.
Other initial conditions, e.g. an ``adiabatic'' cusp around
the SBH (\S4.4),
could result in higher loss rates than 
predicted by the steady-state theory.

\subsection{Classical Loss-Cone Theory}

Stars are swallowed or tidally disrupted after coming within 
a certain distance of the hole.
The tidal disruption radius is
\beq
r_t=\left(\eta^2\frac{M_\bullet}{m_*}\right)^{1/3} r_* .
\label{eq:rt}
\eeq       
Here $m_*$ and $r_*$ are the stellar mass and 
radius, and $\eta$ is a form factor of order unity;
$\eta=0.844$ for an $n=3$ polytrope.
Tidal disruption occurs outside of the hole's event
horizon for a solar-type star when $\mh\gap 10^8\msun$.
In what follows, $r_t$ will be used indiscriminately 
to denote the radius of capture or disruption, 
and stars will be assumed to vanish instantaneously
when $r<r_t$.
Galaxies with the highest feeding rates are expected 
to be those with the smallest SBHs, 
$\mh\lap 10^7\msun$, and in these galaxies $r_t>r_{S}$.

In a spherical galaxy, an orbit that just grazes
the tidal disruption sphere has angular momentum
\beq
L_{lc}^2 = 2r_t^2\left[E-\Phi(r_t)\right]
\approx 2G\mh r_t;
\eeq
the latter expression assumes $|E|\ll G\mh/r_t$, i.e.
that stars are on nearly-radial orbits with apocenters 
much larger than $r_t$.

An upper limit to the consumption rate in a spherical galaxy
comes from assuming that stars are instantaneously replaced
on their original orbits after being consumed by the hole.
In this ``full loss cone'' model,
stars are consumed at a constant rate
\beq
\dot N^{\rm FLC} \approx \int {N_{lc}(E)\over P(E)} dE =
\int F(E)^{FLC} dE
\eeq
where $P(E)$ is 
the period of a nearly-radial orbit of energy $E$ and
$N_{lc}(E)dE$ is the number of stars at energies 
$E$ to $E+dE$ on orbits with $L\le L_{lc}$.
In a spherical galaxy, 
\beq
N(E,L)dEdL = 8\pi^2 L f(E,L) P(E,L)dEdL
\eeq 
\cite{spitzer-87} and
\beq
\dot N^{\rm FLC} \approx 4\pi^2 \int f(E) L_{lc}^2(E)dE 
\approx
8\pi^2 r_t^2 \int f(E) (E-\Phi_t) dE
\eeq
where $\Phi_t\equiv \Phi(r_t)$ and the stellar velocity 
distribution has been assumed to be isotropic.
Using Equation~(\ref{eq:fofe}) for $f(E)$ near the SBH
in a power-law ($\rho\propto r^{-\gamma}$) nucleus, 
this expression can be evaluated:
\beq
\dot N^{FLC} \approx (3-\gamma) \sqrt{2\over\pi} {\Gamma(\gamma+1)\over\Gamma(\gamma+3/2)}
\left({\mh\over m_\star}\right) \left({r_t\over r_h}\right)^{3/2-\gamma}
\left({G\mh\over r_h^3}\right)^{1/2}.
\label{eq:flcrate}
\eeq
For $\gamma=3/2$, which is close to the value observed near the
Milky Way SBH (Fig.~\ref{fig:LG}, the time $T_\bullet$ required to consume
$N_\bullet=\mh/m_\star$ stars is independent of $r_t$: 
\beq
T_\bullet\approx {2\sqrt{2}\over 9}\left({G\mh\over r_h^3}\right)^{-1/2} \approx 0.3 {r_h\over \sigma},
\eeq
i.e. the hole consumes its mass in stars in roughly one crossing time at $r_h$.

Early discussions of SBH feeding 
\citeaffixed{hills-75,ozernoy-76,shields-78}{e.g.}
sometimes assumed a full loss cone;
for instance, in the ``black tide'' model
for quasar fueling \cite{ysw-77,young-77b},
gas from tidally disrupted stars radiates
as it spirals into the SBH, and the
the hole shines as a quasar until its
mass reaches $\sim 10^8\msun$ at which point
stars are swallowed whole and the quasar fades.
A number of recent studies \cite{ZHR-02,escude-05,HB-05,HB-06} 
have revived the full loss cone idea
in the context of single or binary SBHs.
At least in spherical or axisymmetric galaxies,
feeding rates as large as (\ref{eq:flcrate})
would only persist for single orbital periods,
long enough for stars on loss-cone orbits to
be consumed; presumably these orbits would have
been depleted already during the chaotic events associated
with SBH formation.
Various justifications have been invoked for assuming
that $\dot N$ remains of order $\dot N^{FLC}$ for
much longer times; perhaps the most
plausible is that the nucleus is
strongly non-axisymmetric and contains centrophilic orbits
(\S6.4).
Under the full-loss-cone assumption,
feeding rates are high enough that
the mass accumulated in $10$ Gyr can be sufficient
to reproduce observed SBH masses \cite{ZHR-02},
although presumably only a fraction of the mass liberated
when $r_t>r_S$ would find its way into the hole.

A more common assumption is that loss cone orbits
were depleted at some early time, and that 
continued supply of stars to the SBH is limited
by the rate at which these orbits can be re-populated.
The most widely discussed mechanism for orbital
repopulation is two-body relaxation.
Consider again a spherical galaxy.
Star-star encounters induce changes in orbital
energy and angular momentum.
The former were considered above (\S 5)
and shown to imply very low consumption rates.
Much larger rates are implied by changes
in orbital angular momentum \cite{FR-76,LS-77,young-77b}.
In a time $T_r(E)$, the two-body relaxation time
for orbits of energy $E$,
the typical change in $L^2$ is of order
$\sim L_c^2(E)\approx G^2\mh^2/(2|E|)$, 
the squared angular momentum of a circular orbit.
In a single orbital period $P\ll T_r$, 
the rms angular momentum change is roughly
\beq
(\delta L)^2 \approx (P/T_r)L_c^2.
\eeq
Near the hole, orbital periods are short, and
a star will complete many orbits
before being deflected in.
Far from the hole, on the other hand,
orbital periods are long enough
that a star can be deflected in and out of the loss
cone in a single orbital period.
For stars in this outer (``pinhole'') region, consumption by the hole
has almost no effect on the orbital population, and the feeding
rate is equal  to the full loss cone rate defined above.
The energy separating the inner, diffusive region
from the outer, full-loss-cone region is $E_{crit}$,
defined as the energy such that
\beq
\delta L \approx L_{lc}.
\eeq
A more quantitive definition of $E_{crit}$ is in terms of 
the orbit-averaged quantity $q(E)$: $q(E_{crit})=1$, 
where
\beq
q(E)\equiv \frac{1}{R_{lc}(E)} \oint 
\frac{dr}{v_r}
\lim_{R\rightarrow 0}
\frac{\langle(\Delta R)^2\rangle}{2R} = {P(E){\bar\mu}(E)\over R_{lc}(E)};
\label{eq:defq}
\eeq
here $R\equiv L^2/L_c(E)^2$, a dimensionless
angular momentum variable, and $\langle(\Delta R)^2\rangle$
is defined in the usual way as the sum, over a unit
interval of time, of $(\Delta R)^2$ due to encounters.
As the second expression in Equation~(\ref{eq:defq}) shows,
$q(E)$ is the ratio of the orbital period to the 
orbit-averaged time $({\bar\mu}/R_{lc})^{-1}$
for angular momentum to change by $L_{lc}$,
with ${\bar\mu}(E)$ the
orbit-averaged diffusion coefficient.
In terms of $q$, the ``diffusive'' loss cone regime
has $q(E)<1$ and the ``pinhole'' regime has $q(E)>1$.

\begin{figure}
\begin{center}
\includegraphics[width=2.5in,angle=0.]{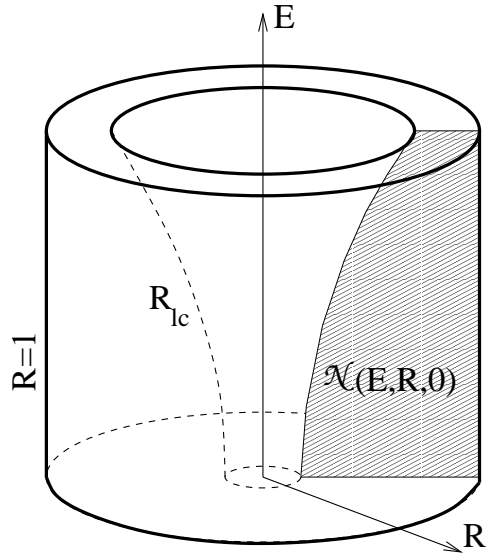}
\caption
{Schematic representation of the loss cone.  
\label{fig:cylinder}
}
\end{center}
\end{figure}

Typically, the total consumption rate is dominated
by stars with $q(E)\lap 1$, i.e. by stars 
that diffuse gradually into the loss cone.
For $q\ll 1$, there is almost no
dependence of density on orbital phase since
changes in $L$ take much longer than an orbital period.
In this regime,
is appropriate to describe the diffusion in terms
of the Fokker-Planck equation \cite{LS-77}.
Ignoring changes in $E$, 
the Fokker-Planck equation describing diffusion in
angular momentum is
\beq
\frac{\partial N}{\partial t}=
\frac{1}{2}\frac{\partial}{\partial R}
\left[\langle\left(\Delta R\right)^2
\rangle\frac{\partial N}{\partial R}\right] 
\eeq
where the dependence of $N$ on $E$ is understood. 
Taking the limit $R\rightarrow 0$
and averaging over one orbital period,
this becomes
\beq
\label{eq:fpnr}
\frac{\partial N}{\partial t} = {\bar\mu} \frac{\partial }{\partial R}
\left(R \frac{\partial N}{\partial R}\right)
\eeq
where ${\bar\mu}(E)$ is the orbit-averaged diffusion
coefficient defined above.
Equation~(\ref{eq:fpnr}) has the same form as the heat
conduction equation in cylindrical
coordinates, as can be seen by changing variables 
to $j=R^{1/2}$:
\beq
\label{eq:diffusion}
\frac{\partial N}{\partial t} = 
\frac{{\bar\mu}}{4j}\frac{\partial }{\partial j}
\left(j \frac{\partial N}{\partial j}\right) ;
\eeq
$j$ plays the role of a radial variable and ${\bar\mu}/4$ is the 
diffusivity.
Figure~\ref{fig:cylinder} illustrates the geometry.
The loss cone is sometimes called a ``loss cylinder''
by virtue of this analogy with the heat conduction problem.

The steady-state solution to Equation~(\ref{eq:fpnr}) is
\beq
N(R;E)=
\frac{\ln(R/R_{lc})}{\ln(1/R_{lc})-1} {\bar N}(E)
\eeq
where
\beq
{\bar N}(E)=\int_{R_{lc}}^{1} N(E,R) dR 
\eeq
or, in terms of the phase space density $f$,
\beq
\label{eq:soltimeind}
f(R;E)= \frac{\ln(R/R_{lc})}{\ln(1/R_{lc})-1} \bar f(E) .
\label{eq:fer}
\eeq
The phase space density tends logarithmically to zero
at the loss cone boundary $R=R_{lc}$.
The implied (number) flux of stars into the loss cone is 
\numparts
\begin{eqnarray}
F(E)dE &=& {4\pi^2 P(E)L_{c}^2(E) {\bar\mu}(E) {\bar f}(E) 
\over \ln(1/R_{lc})-1} dE \\
&\approx& {{\bar N}(E){\bar\mu}(E)\over \ln(1/R_{lc})} dE
\label{eq:fluxone}
\end{eqnarray}
\endnumparts
where the latter expression assumes $L_{lc}\ll L_c$
and $P(E,R)\approx P(E)$.
This is lower than the full-loss-cone feeding rate
by a factor $\sim q /\ln(1/R_{lc})\approx q/\ln(G\mh/4|E|r_t)$.
Equation~(\ref{eq:fluxone}) implies
that a large fraction of the stars in the diffusive
regime will migrate into the hole in one relaxation time;
the consumption rate depends only logarithmically
on the ``size'' $r_t$ of the tidal disruption sphere
\cite{FR-76,LS-77}.
In the 1D (energy-dependent) problem, the 
consumption rate is much smaller, scaling as $r_t^{-1}$
(\S5).

This analysis breaks down where $q\gap 1$, since
changes in $L$ over one orbital period are of order
$L$, and the separation of time scales that is implicit
in the orbit-averaged treatment does not apply.
\citeasnoun{CK-78} developed an approximate scheme
to deal with this situation.
In the presence of relaxation,
some stars will enter and exit the loss cone
$R\le R_{lc}$ in one orbital period, thus evading
disruption.
\citeasnoun{CK-78} wrote the $r$-dependent
Fokker-Planck equation for stars near the loss cone
and computed how the density changes {\it along}
orbits due to the competing effects of capture
at $r<r_t$ and relaxation-driven repopulation
at $r>r_t$.
They found that the $L$-dependence of the steady-state
phase-space density near the loss could be be approximated
by replacing $R_{lc}$ in Equation~(\ref{eq:fer})
by $R_0$, where
\beq
R_0(E) = R_{lc}(E) \times 
\cases{ \exp(-q), & $q(E) > 1$ \cr
  \exp(-0.186 q -0.824 \sqrt{q}), & $q(E) < 1$} .
\label{eq:r0}
\eeq
Near the hole, $q\ll 1$ and $R_0\approx R_{lc}$:
relaxation effects are small and the phase-space
density falls to zero just at the loss-cone boundary.
Far from the hole, relaxation dominates, and
$f$ only falls to zero for orbits with $L\ll L_{lc}$.
\citeasnoun{CK-78} then computed the flux by 
assuming that Equations~(\ref{eq:soltimeind}) 
and~(\ref{eq:fluxone}) still applied, even
in the ``pinhole'' regime where the diffusion 
approximation breaks down;
this asssumption gives the reasonable result
that the flux approaches the full loss cone value
far from the hole.

\citeasnoun{CK-78} included this ``boundary layer''
prescription into a fully  time-dependent calculation
of the evolution of $f(E,R,t)$ for stars around 
a black hole.
They were primarily interested in the case of
a $\sim 10^3\msun$ black hole in a globular cluster, and so 
their solution was required to match on to a 
constant-density (isothermal) core at large radii.
The contribution of the stars to the gravitational
potential was ignored.
\citeasnoun{CK-78} also assumed that the star cluster
was many relaxation times old, and so they evolved their
time-dependent equations until a steady state was reached.
The stellar
density profile evolved approximately to the
Bahcall-Wolf form, $\rho\sim r^{-7/4}$, near the hole
and the feeding rate attained a constant value.
This value was substantially larger than had been found
by earlier authors who adopted more approximate
descriptions of the loss-cone boundary, 
by factors ranging from $\sim 2$
\cite{SM-78} to $\sim 15$ \cite{LS-77},
showing the sensitive dependence of the feeding
rate on the assumed form of $f$ near the loss cone.

\subsection{Application to Galactic Nuclei}

Loss cone theory as developed in the classic papers of
\citeasnoun{FR-76}, \citeasnoun{LS-77} and
\citeasnoun{CK-78} can be used to estimate
feeding rates for SBHs in galactic nuclei.
A number of possible complications arise however:
\begin{itemize}
\item Much of the total consumption occurs from orbits
that extend beyond $\sim r_h$, hence the contribution
of the stars to the gravitational potential 
can not be ignored.
\item Galactic nuclei are often much less than
one relaxation time old (\S2).
This means that the stellar density near the SBH
need not have the Bahcall-Wolf (1976) steady-state form.
In the most luminous galaxies, the nuclear density profile
is in fact known to be much flatter than $\rho\sim r^{-7/4}$ 
at $r\lap r_h$ (\S3).
\item The dependence of $f$ on $L$ near the
loss cone boundary can also be very different from
its steady-state form, regardles of the form of $\rho(r)$.
For instance, if the current SBH was preceded
by a massive binary, almost all low-angular-momentum
stars will have been ejected by the binary.
\item Galactic nuclei need not be spherical
or even axisymmetric.
In a triaxial nucleus containing centrophilic
orbits, the mass in stars on orbits that intersect
the SBH's capture sphere can be enormous,
much greater than than $\mh$, so that the
loss cone is never fully depleted.
Even in a spherical nucleus, 
the velocity distribution can be anisotropic (\S4).
\item Galactic nuclei sometimes undergo catastrophic
changes, due to galaxy mergers, infall of star
clusters or black holes, star formation,
etc. all of which can substantially affect the
feeding rate on both the short and long terms.
\end{itemize}

\noindent
Starting with \citeasnoun{MCD-91},
a standard approach has been to consider only the
first two of these complications when computing
feeding rates of SBHs in galactic nuclei, 
i.e. to allow the density
profile around the SBH to have whatever form is
implied by the observations (which however often
do not resolve $r_h$) and to include the contribution
of the stars to the gravitational potential
when computing the angular-momentum diffusion
coefficient
\citeaffixed{MCD-91,SU-99,mt-99,wang-04}{e.g.}.
The energy dependence of the stellar distribution
function is then fixed by $\rho(r)$ and by the
assumed (or measured) value of $\mh$; the $L$-dependence of
$f$ near the loss cone is taken from the steady-state
theory.

\begin{figure}
\begin{center}
\includegraphics[width=0.65\textwidth,angle=0.]{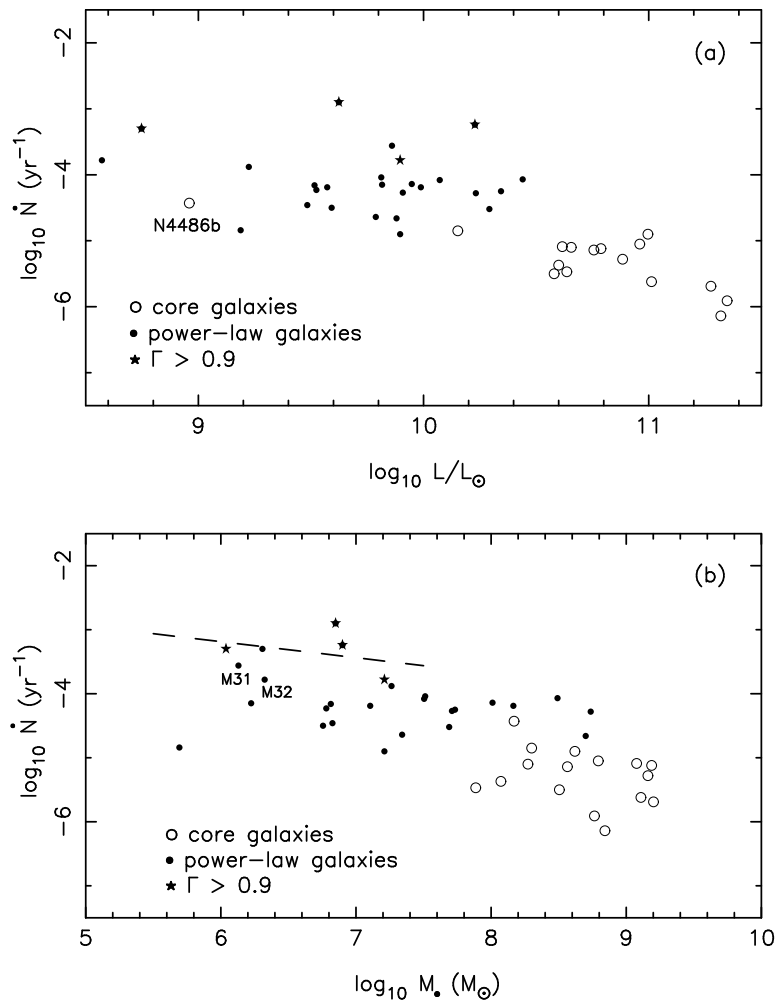}
\caption
{Consumption rate as a function of galaxy luminosity (a) 
and SBH mass (b) for a sample of elliptical galaxies.
Plotted is the rate at which stars (assumed to have the mass
and radius of the Sun) are scattered into
a radius $r_t$ (Eq.~\ref{eq:rt}); note that for the
most massive SBHs plotted here, $r_t\lap r_S$ and stars
would not be tidally disrupted before falling into the hole.
The dashed line in (b) is the relation defined by the singular
isothermal sphere, Eq.~\ref{eq:SIS};
it is a good fit to the galaxies plotted with stars,
which have central density profiles with
$\rho\sim r^{-2}$ at $r\gap r_h$.
(From~\citename{wang-04} \citeyear*{wang-04}.)
\label{fig:wang5}
}
\end{center}
\end{figure}

Figure~\ref{fig:wang5} shows tidal disruption rates
computed in this way for a sample of early-type galaxies
\cite{wang-04}.
There is a weak net dependence of $\dot N$ on $\mh$,
in the sense that nuclei with smaller SBHs tend to have
higher feeding rates.
In the smallest galaxies with well-determined SBH
masses, e.g. M32, $\dot N$ is predicted to 
exceed $10^{-4}$ yr$^{-1}$.
A similar rate would be predicted for the Milky Way,
which is almost an exact copy of M32 in terms of
its central density profile and SBH mass.
It follows that, at the Galactic center, 
the most recent tidal disruption event should have occurred
just a few thousand years ago.

\begin{figure}[htb]
\begin{minipage}[t]{.45\textwidth}
\includegraphics[width=\textwidth]{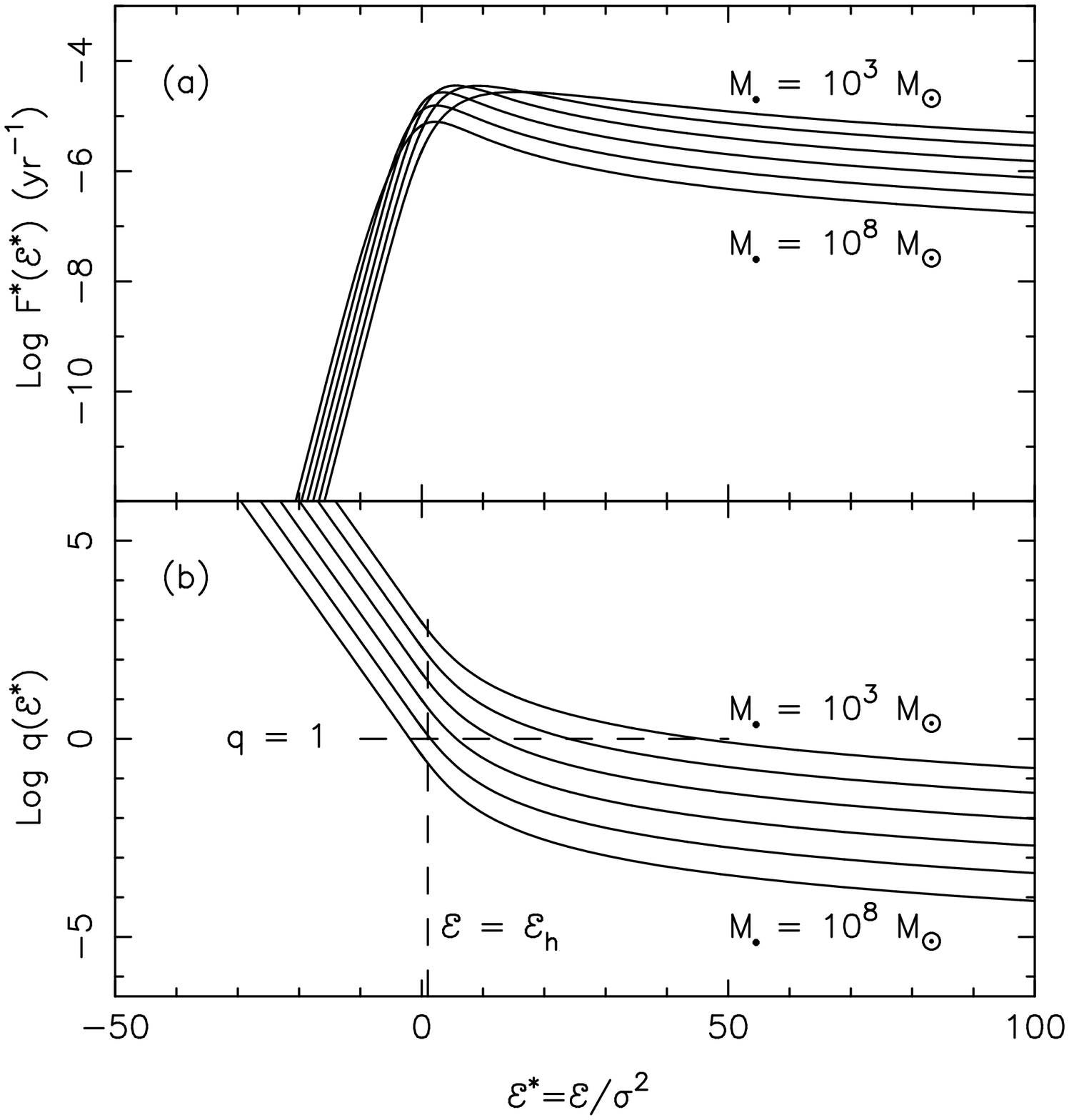}
\caption
{Energy-dependent loss cone flux $F(E)$ (Eq.~\ref{eq:fluxone}) 
and $q(E)$ (Eq.~\ref{eq:defq}) for a SBH in a singular isothermal sphere
nucleus, $\rho\propto  r^{-2}$.
The $\mh-\sigma$ relation was used to relate $\sigma$
to $\mh$.
(From~\citename{wang-04} \citeyear*{wang-04}.)
}
\label{fig:wang8}
\end{minipage}
\hfil
\begin{minipage}[t]{.45\textwidth}
\includegraphics[width=\textwidth]{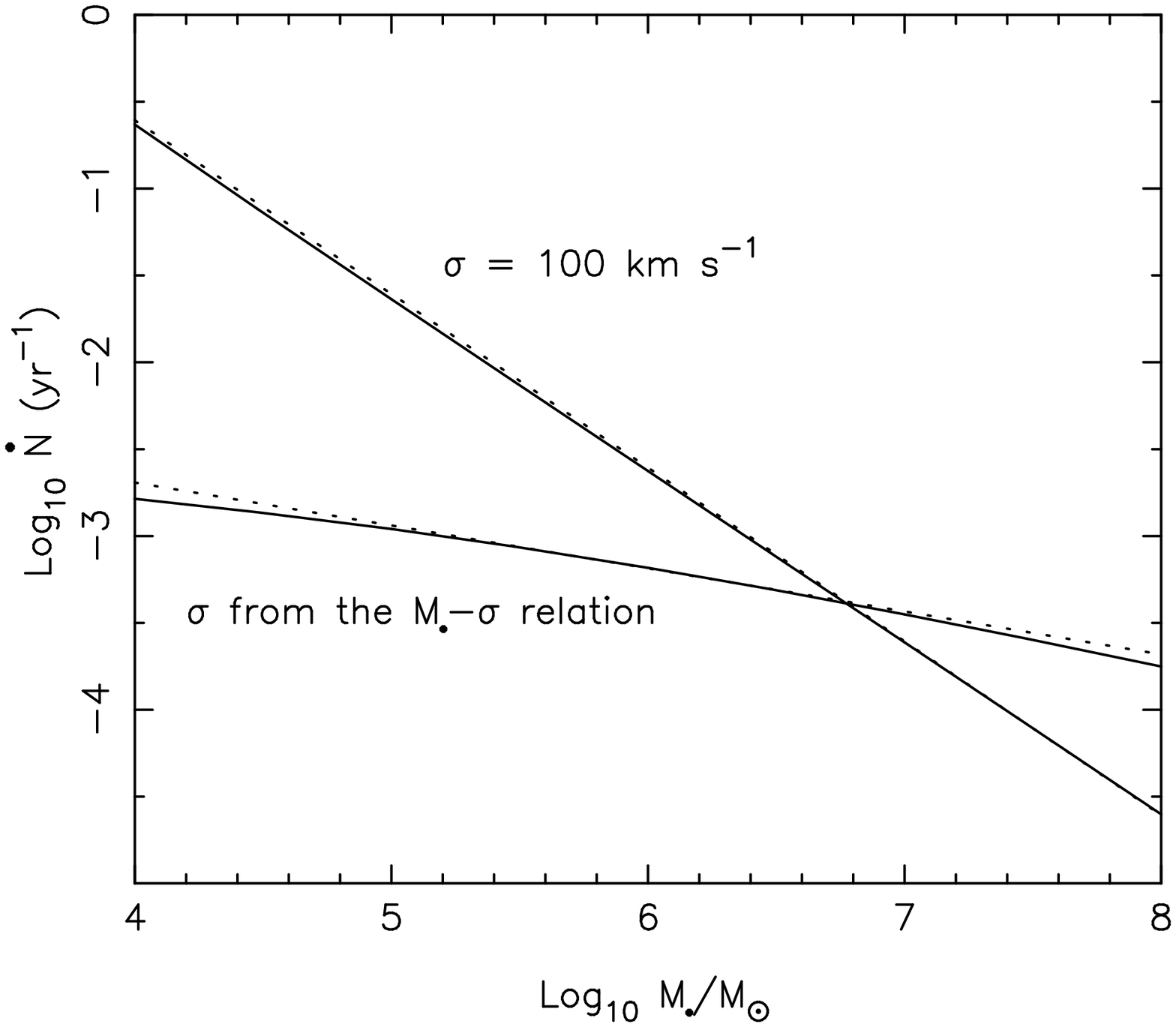}
\caption
{Consumption rate as a function of $\mh$ in singular isothermal
sphere nuclei, for two assumptions about $\sigma(\mh)$.
(From~\citename{wang-04} \citeyear*{wang-04}.)
}
\label{fig:wang9}
\end{minipage}
\end{figure}

The galaxies in Figure~\ref{fig:wang5} with the highest
feeding rates have density profiles $\rho\sim r^{-2}$ at
$r\gap r_h$.
In the case of the Milky Way bulge, which is the only
spheroid in this category near enough to be well resolved
on scales $\ll r_h$, we know that $\rho(r)$ flattens
from $\sim r^{-2}$ to $\sim r^{-1.5}$ at $r<r_h$
\cite{genzel-03}.
However much of the loss cone flux comes from stars
with apocenters $\gap r_h$, and so it is reasonable
to approximate $\rho(r)$ as a singular isothermal
sphere (SIS), $\rho\propto r^{-2}$, in these low-luminosity
galaxies.
Figures~\ref{fig:wang8} and \ref{fig:wang9}
show the predicted loss rates in SIS nuclei.
The total $\dot N$ depends
on the two parameters $(\mh/m_\star,r_h/r_t)$;
adopting (\ref{eq:rt}) for $r_t$, the second
of these parameters can be written
\numparts
\begin{eqnarray}
{r_h\over r_t} &=& {2\Theta\over\eta^{2/3}}\left({\mh\over m_*}\right)^{2/3} \\
&=& 21.5 \left({\mh\over m_*}\right)^{2/3}\left({\sigma\over 100\ \mathrm{km\ s}^{-1}}\right)^{-2} \left({m_\star\over\msun}\right) \left({r_\star\over R_\odot}\right)^{-1}
\label{eq:rbrt}
\end{eqnarray}
\endnumparts
with $\Theta=Gm_\star/2\sigma^2 r_\star$ the Safronov number;
$\eta$ in (\ref{eq:rt}) has been set to $0.844$.
In Figure~\ref{fig:wang9}, two different assumptions
have been made about the relation between $\mh$ and
$\sigma$: constant $\sigma$, and the $\mh-\sigma$ relation.
The stellar mass and radius were set equal to the Solar values.
In both cases, the results are well fit by the simple
expression
\beq
\dot N \approx 7.1\times 10^{-4} {\rm yr}^{-1} 
\left({\sigma\over 70\ {\rm km\ s}^{-1}}\right)^{7/2} 
\left({\mh\over 10^6\msun}\right)^{-1}.
\label{eq:SIS}
\eeq
The normalization constant in equation (\ref{eq:SIS}) was chosen
to reproduce $\dot N$ exactly for $\sigma=70$ km s$^{-1}$, 
$\mh=10^6\msun$;
the consumption rate 
scales as $m_\star^{-1/3} r_\star^{1/4}$ for non-Solar stars.
Equation~(\ref{eq:SIS}), combined with the $M_\bullet-\sigma$ relation,
implies $\dot N \sim \mh^{-0.25}$.
Equation~\ref{eq:SIS} is consistent with the rates computed directly from
the observed luminosity profiles in the  fainter galaxies
(Fig.~\ref{fig:wang5}).

Tidal disruption rates as high as $\sim 10^{-4}$ yr$^{-1}$
in nuclei with $\mh\approx 10^6\msun$ 
imply a liberated mass  of $\sim \mh$ after $10$ Gyr.
This is not necessarily a problem since only a fraction of
the gas removed from stars is expected to find its way into the hole
\cite{rees-90}.
Nevertheless, the high values of $\dot N$ in low-luminosity
galaxies suggest that matter tidally liberated
from stars might contribute substantially to SBH
masses in these galaxies.

The faintest systems in which there is solid kinematical evidence
for SBHs are M32 and the bulge of the Milky Way
($L\approx 10^9L_\odot, \mh\approx 10^{6.5}\msun$).
However there is strong circumstantial evidence for massive
black holes in fainter systems 
\citeaffixed{FH-03}{e.g.}
and for intermediate mass black holes
(IMBHs) in starburst galaxies and star clusters
\citeaffixed{marel-04}{e.g.}.
If nuclear black holes are common in the so-called
dE (dwarf elliptical) galaxies and in the bulges of late-type
spiral galaxies,
these systems would dominate the total tidal flaring rate, 
due both to their large numbers and to their high individual event rates.
A simple calculation \cite{wang-04} suggests that 
the tidal flaring rate due to dwarf galaxies in
the Virgo cluster alone would be of the order of
$10^{-1}$ yr$^{-1}$.
Nondetection of flares after a few years of monitoring
would  argue against the existence of IMBHs in dwarf galaxies.

\subsection{Time-Dependent Loss Cone Dynamics}

The majority of galaxies with detected SBHs have
collisionless nuclei (\S 2).
In these galaxies, the assumption that the stellar
phase space density has reached an approximate steady
state under the influence of gravitational encounters
breaks down.
One consequence is that the stellar density near the SBH
need not have the Bahcall-Wolf $r^{-7/4}$ form.
As discussed above, it is straightforward to modify
the classical loss-cone treatment for arbitrary $\rho(r)$.
But the fact that galactic nuclei are not collisionally
relaxed also has implications for the more detailed form
of the phase space density near the loss cone boundary.
For instance, in a nucleus that once contained
a binary SBH, stars on orbits such that
$L\lap L_{bin}= (2GM_{12}a_h)^{1/2}$ will
have been ejected, where $M_{12}$ is the binary
mass and $a_h\approx G\mu/4\sigma^2$ is the
``hard'' binary separation (Eq.~\ref{eq:ah}).
Since $a_h\gg r_t$, there will be a gap in angular
momentum space around the single SBH that subsequently
forms, corresponding to stars with $L\lap L_{bin}$ that
were ejected by the binary.
Before the single SBH can begin to consume stars at the
steady-state rate computed above,
this gap needs to be refilled.
At the other extreme, one can imagine a nucleus
that formed in such a way that the gradients in
$f$ near the loss cone boundary are much greater than
their steady state values, implying larger feeding rates.

Equation~(\ref{eq:fpnr}) implies a characteristic time
to set up a steady-state
distribution in angular momentum near the loss cone:
\beq
t_L\approx {R\over {\bar \mu}} \approx {L^2\over L_c^2} T_r.
\eeq
Setting $L^2\approx 2G\mh a_h$ -- the appropriate
value for a loss cone that was emptied by a binary SBH --
and $L_c^2\approx G\mh r_h$, 
appropriate for stars at a distance $\sim r_h$
from the hole,
\beq
{t_L\over T_r(r_h)} \approx {a_h\over r_h} \approx {M_2\over M_1} 
\eeq
with $M_2/M_1\le 1$ the mass ratio of the binary that
created the gap.
Since $T_r(r_h)$ can be much greater than $10^{10}$ yr
in the bright elliptical galaxies that show evidence
of cusp destruction (Fig.~\ref{fig:tr}),
even large mass ratio binaries can open up phase-space
gaps that would not be refilled in a galaxy's lifetime,
implying much lower rates of SBH feeding than in the
steady-state theory.

\begin{figure}[htb]
\begin{minipage}[t]{.45\textwidth}
\includegraphics[width=\textwidth]{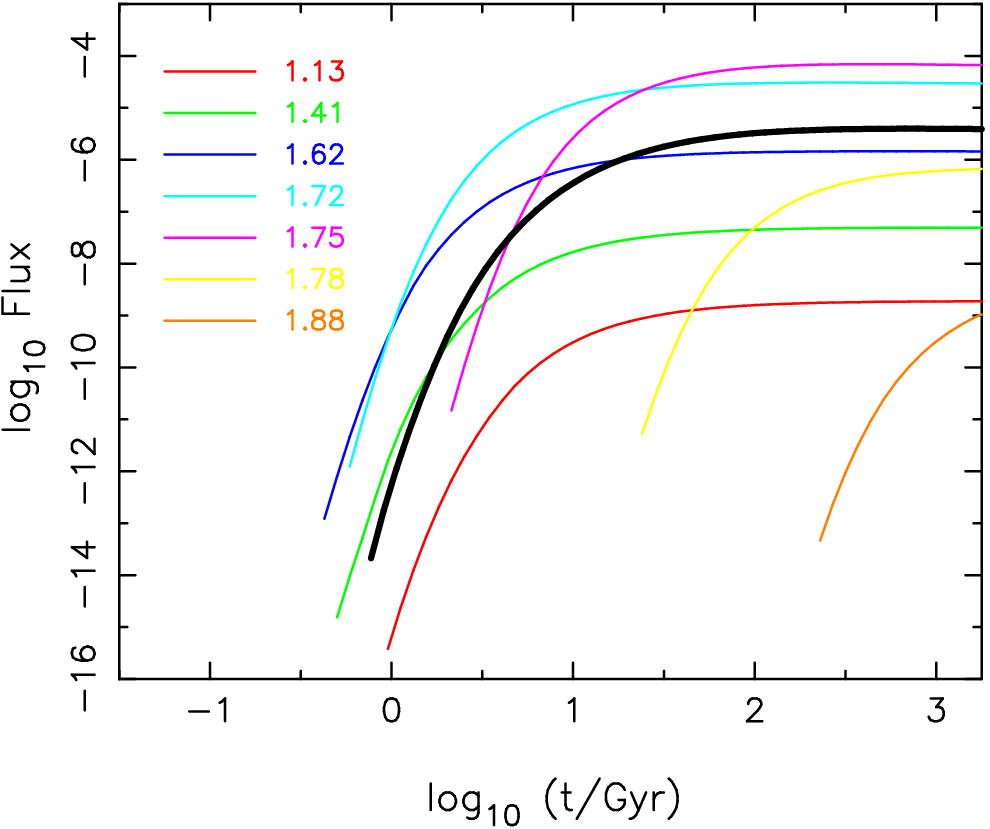}
\caption
{
Dependence of the loss-cone flux on energy and time in the
galaxy NGC 4168 assuming a binary mass ratio of $q=0.1$.
Curves are labelled by their binding energy; the value of the gravitational
potential at $r=r_h$ in this galaxy is $1.76$.
Thick black curve is the total flux, in units of stars per year.
(From~\citename{wang-05} \citeyear*{wang-05}.)
}
\label{fig:mw2}
\end{minipage}
\hfil
\begin{minipage}[t]{.45\textwidth}
\includegraphics[width=\textwidth]{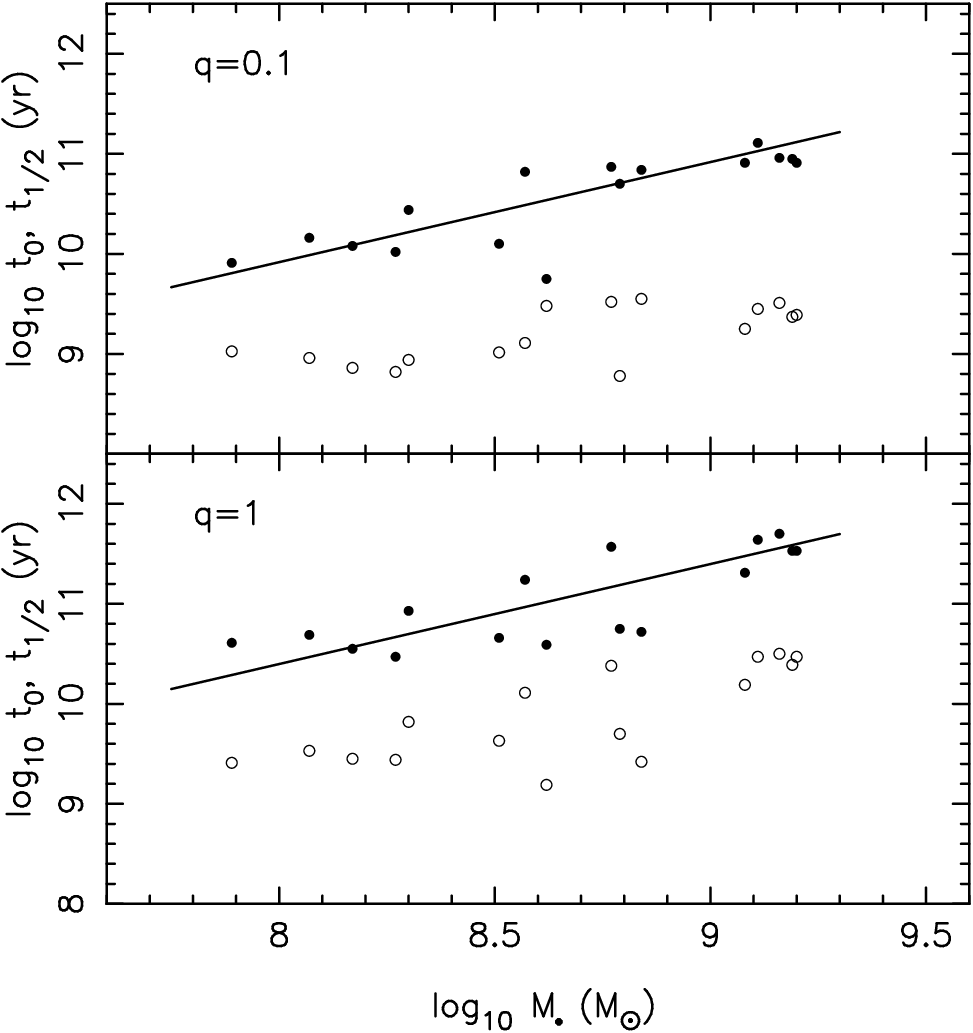}
\caption
{Two characteristic times associated with loss-cone
refilling in a sample of elliptical galaxies,
assuming that a phase-space gap was created by
a binary SBH with mass ratio $q$.
$t_0$ is the elapsed time before the first star is 
scattered into the single, coalesced hole
and $t_{1/2}$ is the time for the loss-cone flux to reach
$1/2$ of its steady-state value.
Solid lines are the approximate fitting function,
Eq.~\ref{eq:fit}.
(From~\citename{wang-05} \citeyear*{wang-05}.)
}
\label{fig:mw3}
\end{minipage}
\end{figure}

Figure~\ref{fig:mw2} illustrates this for a particular
giant elliptical galaxy, NGC 4168,
assuming a binary mass ratio $M_2/M_1=0.1$
\cite{wang-05}.
Equation~(\ref{eq:fpnr}) was integrated forward
assuming an initially sharp phase-space cutoff at
$L=L_{lc}$.
Stars with energies near $\Phi(r_h)$ are the first to
be scattered into the hole;
the total flux  reaches $1\%$, $10\%$, $50\%$ and $90\%$
of its steady state value in a time of $4.5, 9.8, 17$ and
$97$ Gyr.
Figure~\ref{fig:mw3} shows the results of a similar
calculation for each of the ``core'' galaxies from 
Figure~\ref{fig:wang8};
the time for the total flux to reach $1/2$ of its steady-state
value is roughly
\beq
{t_{1/2}\over 10^{11}{\rm yr}} \approx
{q\over (1+q)^2}{M_\bullet\over 10^8\msun}
\label{eq:fit}
\eeq
with $q\le 1$ the mass ratio of the pre-existing binary.
While highly idealized, calculations like these
demonstrate how different the feeding rates in collisionless
nuclei can be from the predictions of steady-state theory.
Hopefully, more progress on this important
problem can be expected in the near future.

\subsection{Nonaxisymmetric Nuclei}

A qualitatively different kind of SBH feeding can occur
in nonaxisymmetric (triaxial or barlike) nuclei.
Orbits in nonaxisymmetric potentials do not conserve
any component of the angular momentum and certain orbits,
the so-called centrophilic orbits, have ``filled centers'':
they pass arbitrarily close to  the potential center
after a sufficiently long time
\cite{norman-83,gerhard-85}.
In the presence of a massive central object like a SBH,
centrophilic orbits tend to be unstable (chaotic)
\citeaffixed{mv-99}{e.g.}
and this fact was long taken to imply that triaxiality
could not be maintained in galaxies containing
SBHs.
However as discussed in \S4, recent work suggests that
nuclei can remain stably triaxial even when most of their
stars are on chaotic orbits.
Even if long-lived triaxial equilibria are not possible,
the high frequency of barlike distortions observed at the centers
of galaxies suggests that transient departures from
axisymmetry are common.

The mass associated with stars on centrophilic orbits
in triaxial galaxies can easily be $\gg \mh$, 
greatly exceeding than the mass on loss-cone orbits 
in the spherical or axisymmetric geometries.
A lower limit on the feeding rate in the triaxial
geometry comes from ignoring collisional loss cone
refilling and simply counting the rate at which stars
on centrophilic orbits pass within a distance
$r_t$ from the center as they move along  their  orbits.
For a single orbit, this rate is $\sim A(E)r_t$
(\S4.3) where the function $A(E)$ can be determined
by numerical integrations \cite{mp-04,HB-06}.
If $N_c(E)dE$ is the number of stars on centrophilic 
orbits in the energy range $E$ to $E+dE$, then the
loss rate to the SBH (ignoring orbital depletion) is
\begin{equation}
\dot N \approx r_t\int A(E)N_c(E)dE.
\end{equation}
In a SIS ($\rho\sim r^{-2}$) nucleus,
if the fraction $f_c(E)$ of chaotic orbits at
energy $E$ is initially independent of $E$, the accumulated
mass after time $t$ is
\beq
\Delta M \approx
1 \times 10^8 \msun f_c \left({r_t\over r_s}\right)^{1/2}\left({\sigma\over 200\ {\rm km s}^{-1}}\right)^{5/2} 
\left({\mh\over 10^8\msun}\right)^{1/2}
\left({t\over 10^{10}\ {\rm yr}}\right)^{1/2}
\label{eq:dm}
\eeq
\cite{mp-04};
the $t^{1/2}$ dependence reflects the reduction in
the feeding rate as centrophilic orbits are depleted.
Even for modest values of $f_c$ ($\sim 0.1$),
this collisionless mechanism can supply stars to
the SBH at higher rates than collisional loss-cone
repopulation, particularly in galaxies with
$M_\bullet\gap 10^7 M_\odot$ in which 
relaxation times are very long.
Indeed after $10^{10}$ yr, Equation~(\ref{eq:dm})
predicts an accumulated mass 
\beq
{\mh\over 10^8\msun} \approx f_c 
\left({\sigma\over 200\ {\rm km\ s}^{-1}}\right)^5.
\label{eq:ms2}
\eeq
This is remarkably similar in form to the $\mh-\sigma$ relation,
and even the normalization is of the right order if 
$f_c\approx 1$.

Hills pointed out already in 1975 that 
full-loss-cone feeding rates in spherical
galaxies could grow $\sim 10^8\msun$ holes
in $10^{10}$ yr,
and \citeasnoun{ZHR-02} noted that the accumulated
mass would satisfy a relation like (\ref{eq:ms2})
between $\mh$ and $\sigma$.
Although there are many problems with this simple
idea -- for instance, when $\mh\lap 10^8\msun$,
most of the mass liberated from tidally disrupted stars 
would be lost from the nucleus -- the idea 
that supply rates can plausibly approach the
(spherical) full-loss-cone rate in a triaxial nucleus
appears to be quite solid.
An important next step will be to extend the
\citeasnoun{poon-04} self-consistency studies of 
triaxial black-hole nuclei to full galaxy models,
using both orbital-superposition and $N$-body techniques
to ensure that the models are long-lived.

\begin{figure}
\begin{center}
\includegraphics[width=0.5\textwidth,angle=-90.]{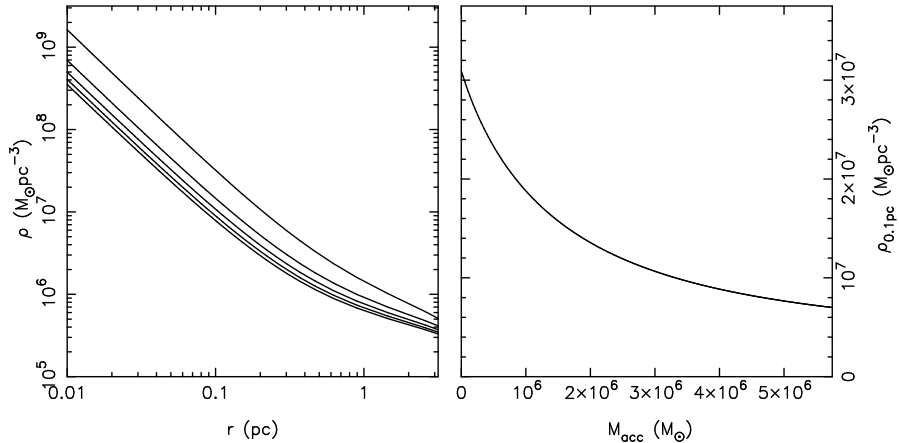}
\caption
{Black-hole-driven expansion of a nucleus, as computed via
the Fokker-Planck equation.
In this calculation, the total gravitational potential
(galaxy + black hole) was assumed fixed; this is a good
approximation since the potential in the region of changing
density is dominated by the hole, and at large radii the
evolution times are very long.
Stars were assumed to vanish instantaneously after being
scattered into the tidal destruction sphere.
The left panel shows density profiles at constant time intervals
after a Bahcall-Wolf cusp has been established;
the density normalization was fixed by requiring the
final (lowest) curve to have roughly the same density
as the nucleus of M32 at 0.1 pc, and the other
parameters in the model (e.g. black hole mass) also mimic M32;
the final influence radius is $r_h\approx 3$ pc.
The right panel shows the evolution of the density at
$0.1$ pc as a function of $M_{acc}$, 
the accumulated mass in tidally-disrupted stars.
As scaled to M32, the final time is roughly $2\times 10^{10}$ yr.
This plot suggests that the densities of collisional
nuclei like those of M32 and the Milky Way were once higher, 
by factors of $\sim $ a few, than at present.
\label{fig:m32expand}
}
\end{center}
\end{figure}

\subsection{Black-Hole-Driven Expansion}

Steady-state solutions like the ones described above
can only be approximate descriptions of nuclei,
since the supply of stars in a galaxy is finite,
and destruction of stars by the SBH will eventually 
cause the stellar density to drop.
In the equilibrium models of 
Bahcall \& Wolf (1976), Cohn \& Kulsrud (1978) and others,
this effect is absent since the stellar distribution
function is fixed far from the hole,
enforcing an inward flux of stars 
precisely large enough to replace the stars 
being destroyed or consumed by the hole.
In reality, the relaxation time beyond 
a certain radius would be so long that the
encounter-driven flux of stars could not 
compensate for losses near the hole, forcing
the density to drop.
Expansion occurs for two reasons: 
(1) Stars are physically destroyed, reducing their numbers.
(2) Disrupted stars are those most tightly bound to the
hole, on loss-cone orbits, and to achieve such an orbit
a star must have given up energy to other stars.
In effect, the black hole acts as a 
heat source
\cite{shapiro-77,dokuchaev-89},
in much the same way that hard binary stars inject energy into
a post-core-collapse globular cluster and cause it to re-expand
\cite{henon-61,henon-65}.

A very simple model that produces self-similar
expansion of a black-hole nucleus can be constructed
by simply changing the outer boundary condition in the
Bahcall-Wolf (1976) problem (\S5.1) 
from $f(0)=f_0$ to $f(0)=0$.
Equations~(\ref{eq:fpa},\ref{eq:fpb}) then describe the evolution
of a finite cluster of stars as they diffuse to lower
energies and are consumed by the hole.
One finds that the evolution after $\sim$one relaxation
time can be described as 
$\rho(r,t) = \rho_c(t)\rho^*(r)$, with
$\rho^*(r)$ slightly steeper than the $\rho\sim r^{-7/4}$
Bahcall-Wolf form; the normalization drops
off as $\rho_c\propto t^{-1}$ at late times 
(Merritt, unpublished).
Figure~\ref{fig:m32expand} shows the results of
a slightly more realistic calculation in a model
designed to mimic M32.
After reaching approximately the Bahcall-Wolf
steady-state form, the density drops in amplitude
with roughly fixed slope for $r\lap r_h$.
This example suggests that the nuclei of galaxies
like M32 or the Milky Way might have been $\sim$ a few
times denser in the past than they are now,
with correspondingly higher rates of stellar
tidal disruption and stellar collisions.

Expansion due to a central black hole has been observed in a 
handful of studies based on fluid \cite{amaro-04},
Monte-Carlo \cite{SM-78,marchant-80,freitag-06},
Fokker-Planck \cite{MCD-91},
and $N$-body \cite{baumgardt-04a,baumgardt-04b} algorithms.
All of these studies allowed stars to be lost into or destroyed
by the black hole; however 
most adopted parameters more
suited to globular clusters than to nuclei,
e.g. a constant-density core.
\citeasnoun{MCD-91} applied the isotropic, multi-mass
Fokker-Planck equation to the evolution of nuclei
containing SBHs, including an approximate loss term
in the form of equation (\ref{eq:fpmod}) to model the
scattering of low-angular-momentum stars into the hole.
Most of their models had what would now be considered
unphysically high densities and the evolution was dominated
by physical collisions between stars.
However in two models with lower densities, they reported
observing significant expansion over $10^{10}$ yr; 
these models had initial central relaxation times of 
$T_r\lap 10^9$ yr when scaled to real galaxies,
similar to the relaxation times near the centers of
M32 and the Milky Way.
The $\rho\sim r^{-7/4}$ form of the density profile
near the SBH was observed to be approximately conserved
during the expansion.
\citeasnoun{freitag-06} carried out Monte-Carlo
evolutionary calculations of a suite of models
containing a mass spectrum, some of which were
designed to mimic the Galactic center star cluster.
After the stellar-mass black holes in their models
had segregated to the center,
they observed a strong, roughly self-similar expansion.
\citeasnoun{baumgardt-04a} followed core collapse
in $N$-body models with and without a 
massive central particle;
``tidal destruction'' was modelled by simply
removing stars that came within a certain distance
of the massive particle.
When the ``black hole'' was present, the cluster 
expanded almost from the start
and in an approximately self-similar way.

These important studies notwithstanding, there
is a crucial need for more work on this problem.
Establishing the self-similarity of the expansion
in the Fokker-Planck or fluid descriptions would be a good start;
such studies \citeaffixed{LBI-83,heggie-85,HS-88}{e.g.}
were an important complement to numerical simulations
in understanding the post-core-collapse
evolution of star clusters.

\subsection{Constraining the Consumption Rate}

Calculations of SBH feeding rates in real galaxies
are subject to many uncertainties, particularly in
collisionless nuclei, due to the wide range 
of possible geometries (\S6.4) and initial conditions (\S6.3).
Additional uncertainties include the form of the nuclear
density profile at $r\ll r_h$ and, of course, the mass of the SBH.
Plots like Figure~\ref{fig:wang8}, which was based on
spherical, steady-state loss cone theory,
should probably be seen as little more than order-of-magnitude 
estimates of the true tidal flaring rate.
In such an uncertain situation, it makes sense to look for observational
evidence of tidal disruptions as a constraint on the theory.
The {\it ROSAT} All-Sky Survey detected soft X-ray outbursts
from a number of galaxies with no previous history of
of nuclear activity.
Roughly half a dozen of these events had the properties of a 
tidal disruption flare (\citename{komossa-02} \citeyear*{komossa-02}
and references therein),
and follow-up optical spectroscopy of the candidate
galaxies confirmed that at least two were subsequently inactive
\cite{gezari-03}.
The mean event rate inferred from these outbursts is roughly
consistent with theoretical predictions \cite{donley-02}.
Some fraction of the X-ray luminosity function of active
galaxies \cite{hasinger-05}
must also be due to stellar tidal disruptions.
Convolving Equation~\ref{eq:SIS} for the disruption rate
with the SBH mass function, and assuming that individual tidal
disruption events have a $L_X\propto t^{-5/3}$ time dependence
as predicted in ``fallback'' models \cite{li-02}, 
one concludes \cite{MMH-06} 
that tidal disruptions can account for
the majority of X-ray selected AGN with soft 
X-ray luminosities below $\sim 10^{43}-10^{44}$ erg s$^{-1}$.
Nearer to home, it might be possible to search for ``afterglows''
of the most recent tidal disruption event at the Galactic
center, which could plausibly have occurred as little
as $\sim 10^3$ yr ago.
Possible examples of such signatures include
X-ray flourescence of giant molecular clouds
\cite{sunyaev-98} and
changes in the surface properties of irradiated stars 
\cite{jimenez-06}.

\section{Dynamics of Binary Black Holes}

Galaxies are believed to grow through the agglomeration
of smaller galaxies and proto-galactic fragments.
If more than one of the fragments contained a SBH,
the two SBHs will form a bound system in the merger product
\cite{BBR-80}.
This scenario has received considerable attention since
the ultimate coalescence of a binary SBH would generate
an observable outburst of gravitational waves \cite{thorne-76}
and possibly electromagnetic radiation as well
\cite{mp-05}.
Binary SBHs are also increasingly invoked to explain the
properties of active galaxies,
including AGN variability \cite{valtaoja-00,xie-03},
the  bending and precession of radio jets \cite{roos-93,romero-00},
$X$- and $Z$-shaped radio lobes \cite{ekers-02,gopal-03},
and the correlation of radio loudness with galaxy
morphology \cite{wc-95,capetti-06a}.

\citeasnoun{komossa-03b} reviews the
observational evidence for binary SBHs.
A handful of galaxies exhibit two active nuclei 
with separations as small as $\sim 1$ kpc 
\cite{komossa-03,ballo-04,hudson-06}, 
much greater however than the parsec-scale separations
that characterize true binaries.
Very recently \cite{rodriguez-06},  
VLBA observations of an elliptical galaxy at $z=0.055$
with two compact central radio sources 
were used to infer the presence of the first,
true binary SBH; the projected separation is only $\sim 7$ pc
and the inferred total mass is $\sim 1.5\times 10^8M_\odot$.

A binary SBH with $a\approx a_h$ (Equation~\ref{eq:ah})
contains a fraction 
$\sim (M_1+M_2)/M_{gal}\approx 10^{-3}$
of the total gravitational energy of its host galaxy,
and such a large binding energy implies a significant change
in the distribution of stars, gas or dark matter at the
center of the galaxy when the binary forms.
If stars are the dominant component, formation of the
binary results in a low-density core, with a displaced
mass of order the mass of the binary.
As discussed in \S3 and \S7.3, luminous elliptical galaxies
always contain such cores, with masses and sizes that
are roughly consistent with predictions of the binary SBH model
\cite{merritt-06}.
Other dynamical effects associated with binary SBHs 
include high-velocity ejection of stars via the gravitational
slingshot \cite{hills-88},
chaos induced in stellar orbits
by the time-dependent potential of the binary
\cite{kandrup-03b},
and the ``gravitational rocket'' effect,
the kick imparted to a coalescing binary due to 
anisotropic emission of gravitational waves
\cite{favata-04,baker-06}, which can displace
a coalesced SBH from its central location
and potentially eject it into intergalactic space.
These mechanisms all go in the direction of ``heating''
the nucleus, and collectively, they set the initial conditions
for evolution of the nucleus after the two SBHs
coalesce.

Binary SBHs may fail to coalesce in some galaxies
because their evolution stalls
at a separation much greater than required for the
efficient emission of gravitational waves \cite{valtonen-96}.
In a galaxy where this occurs, the binary may be present
when a third SBH (or a second binary) falls in, 
resulting in a complicated interaction between the 
multiple SBHs \cite{MV-90}, 
and possibly ejection of one or more from the nucleus
or even from the galaxy.
Such events can not be too frequent or the
tight correlations observed between SBH mass and galaxy
properties would be violated.
Furthermore the total mass density in SBHs in the
local universe is consistent with that inferred
from high-redshift AGN \cite{MF-01b,yt-02},
implying that only a small fraction of SBHs could  have
been ejected from galaxies in the intervening period.

In what follows, the focus will be on the  interaction
of binary and multiple SBHs with (point-mass) stars 
in galactic nuclei and on the implications for nuclear dynamics.
\citeasnoun{living} present a more general review of
the astrophysics of massive black hole binaries.

\subsection{Early evolution of binary SBHs and the generation of mass deficits}

In galaxy mergers in the local universe, typical  mass ratios
are believed to be large, of order $10:1$ \citeaffixed{sesana-04}{e.g.}.
To a good approximation, the initial approach of the two
SBHs can therefore be modelled by assuming that the galaxy hosting
the smaller SBH spirals inward under the influence 
of dynamical friction from the fixed distribution of
stars in the larger galaxy.
Modelling both galaxies as singular isothermal
spheres ($\rho\sim r^{-2}$) and assuming that the smaller
galaxy spirals in on a circular orbit, its tidally-truncated mass
is $\sim \sigma_g^3r/2G\sigma$,
with $\sigma$ and $\sigma_g$ the velocity dispersions
of large and small galaxies respectively \cite{merritt-84}.
Chandrasekhar's (1943) formula then gives for the orbital
decay rate and infall time 
\beq
{dr\over dt} = -0.30 {Gm_2\over\sigma_1r}\ln\Lambda,\ \ \ \ 
t_{infall} \approx 3.3 {r_0\sigma^2\over \sigma_g^3}
\label{eq:df}
\eeq
where $\ln\Lambda$ has been set to 2.
Using Equation~(\ref{eq:ms})
to relate $\sigma$ and $\sigma_g$ to the respective
SBH masses $M_1$ and $M_2$, this becomes
\beq
t_{infall}\approx 3.3{r_0\over\sigma} 
\left({M_1\over M_2}\right)^{0.62},
\eeq
i.e. $t_{infall}$ exceeds the crossing time of the larger
galaxy by a factor $\sim q^{-0.62}$.
Thus for mass ratios $q\gap 10^{-3}$, 
infall requires less than $\sim 10^2 T_{cr}\approx 10^{10}$ yr
 \cite{merritt-00}.
This mass ratio is roughly the ratio between the masses of the
largest ($\sim 10^{9.5}\msun$) and smallest 
($\sim 10^{6.5}\msun$) known SBHs and so it is reasonable
to assume that galaxy mergers will almost always lead to formation of
a binary SBH, i.e. a bound pair, in a time less than $10$ Gyr.

\begin{figure}
\centering
\includegraphics[scale=0.45,angle=0.]{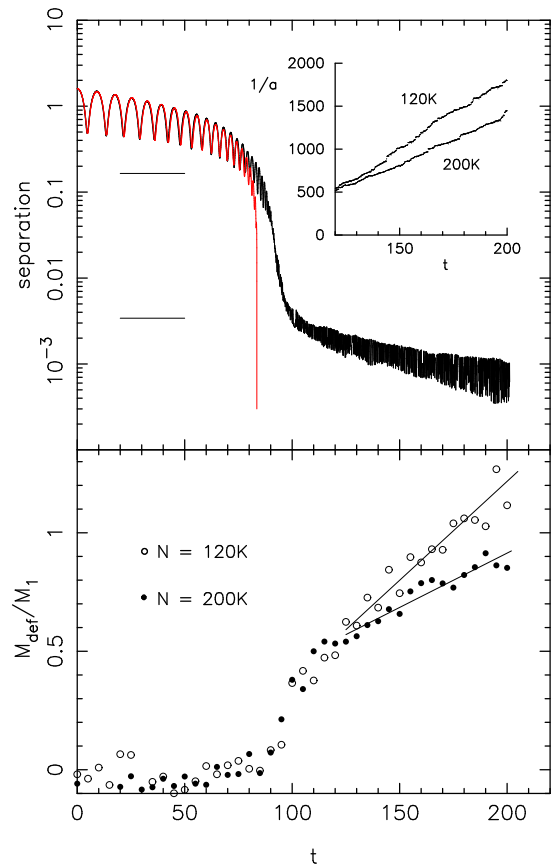}
\caption{{\it Upper panel:} Thick (black) line shows the
$N$-body evolution of a massive binary with 
$M_2/M_1=0.1$; the initial galaxy model had 
a $\rho\sim r^{-1}$ density cusp.
Thin (red) line is the evolution predicted by the
Chandrasekhar's dynamical friction formula assuming a fixed galaxy.
Horizontal lines indicate $r_h$ and $a_h$.
The inset shows the evolution of the inverse semi-major
axis of the binary in  this integration,
and in a second integration with roughly one-half the number
of particles; the latter curve
lies above the former, i.e. the decay occurs more rapidly
for smaller $N$ when $a \lap a_h$, due to the higher
rate of star-star encounters.
In the large-$N$ limit of real galaxies,
the binary hardening rate would drop to zero
at $a\approx a_h$.
{\it Lower panel:} Evolution of the mass deficit 
in the same two $N$-body integrations.
Lines show least-squares fits to 
$t \ge 120$.
(Adapted from \citename{merritt-06} \citeyear*{merritt-06}.)
\label{fig:binary}
}
\end{figure}

Equation (\ref{eq:df}) begins to break down when the
two SBHs approach more closely than $R_{12}\approx r_h$,
the influence radius of the larger hole,
since the orbital energy of $M_2$ is absorbed by the stars, 
lowering their density and reducing the frictional force.
Figure~\ref{fig:binary} illustrates this via 
$N$-body integrations of a $q=0.1$ binary. 
In spite of this slowdown,
the separation between the two SBHs continues to
quickly drop until $a\approx a_h\approx (q/4)r_h$
(Eq.~\ref{eq:ah}) at which separation
the binary's binding energy per unit mass
is $\sigma^2$ and the binary is ``hard'' -- it
ejects stars that pass within a distance $\sim a$
with velocities large enough to remove them from the
nucleus \cite{MV-92,quinlan-96}.

What happens next depends on the density and geometry 
of the nucleus.
In a spherical or axisymmetric galaxy, 
the mass in stars on orbits that intersect the binary
is small, $\lap M_{12}$, and the 
binary rapidly interacts with and ejects these stars.
Once this has occurred, 
no stars remain to interact with the binary and its evolution
stalls.
In non-axisymmetric (e.g. triaxial) nuclei, on the other hand,
the mass in stars on
centrophilic orbits can be much larger, allowing the binary
to continue shrinking past $a_h$.
And in collisional nuclei of any geometry, gravitational scattering
of stars can repopulate depleted orbits.
These different cases are discussed independently below.

The stalling that is predicted to occur in
spherical galaxies
can be reproduced via $N$-body simulations if $N$ is large
enough to suppress two-body relaxation.
One finds 
\beq
{a_{stall}\over r_h} \approx 0.2 {q\over (1+q)^2},
\label{eq:astall}
\eeq
\cite{merritt-06}
with $r_h$ the influence radius of the larger SBH,
defined as the radius containing a stellar mass equal to twice $M_{12}$
{\it after} infall of the smaller SBH has lowered the nuclear
density.
This relation was established using galaxy models with
initial, power-law density profiles, $\rho\sim r^{-\gamma}$,
$0.5\le\gamma\le 1.5$; the coefficient in Equation~(\ref{eq:astall})
depends only weakly on $\gamma$.
Table~\ref{tab:virgo} gives predicted stalling radii for binary
SBHs at the centers of the brightest Virgo cluster galaxies,
based on the galaxy structural data of \citeasnoun{acs6}; 
all of these galaxies exhibit large, low-density cores
which might have been formed by binary SBHs.

\begin{table}
\begin{center}
\caption{Virgo ``Core'' Galaxies \label{tab:virgo}}
\begin{tabular}{ccccccc}
       &       &             &           &       & $a_{\rm stall}$ & $a_{stall}$\\
Galaxy & $B_T$ & $M_\bullet$ & $M_{def}$ & $r_h$ & $q=0.5$         & $q=0.1$   \\
(1) & (2) & (3) & (4) & (5) & (6) & (7) \\
NGC 4472 & -21.8 & $5.94$ & $17.7$ & $130.\ (1.6) $ & $5.6\ (0.070)$ & $2.1\ (0.026)$ \\
NGC 4486 & -21.5 & $35.7$ & $87.3$ & $460.\ (5.7) $ & $20.\ (0.25)$  & $7.6\ (0.095)$  \\
NGC 4649 & -21.3 & $20.0$ & $21.5$ & $230.\ (2.9) $ & $10.\ (0.13)$  & $3.8\ (0.047)$ \\
NGC 4406 & -21.0 & $4.54$ & $3.27$ & $90.\  (1.1) $ & $4.0\ (0.050)$ & $1.5\ (0.019)$  \\
NGC 4374 & -20.8 & $17.0$ & $22.6$ & $170.\ (2.1) $ & $7.6\ (0.094)$ & $2.8\ (0.035)$  \\
NGC 4365 & -20.6 & $4.72$ & $6.00$ & $115.\ (1.4) $ & $5.0\ (0.063)$ & $1.9\ (0.023)$  \\
NGC 4552 & -20.3 & $6.05$ & $6.45$ & $73.\  (0.91)$ & $3.2\ (0.040)$ & $1.2\ (0.015)$ \\
\end{tabular}
\end{center}
\footnotesize
Properties of the brightest Virgo cluster
galaxies.
Col. (1): New General Catalog (NGC) number.
Col. (2): Absolute $B$-band galaxy magnitude.
Col. (3): Black hole mass in $10^8M_\odot$, computed
from the $M_\bullet-\sigma$ relation.
Col. (4): Observed mass deficit in $10^8M_\odot$
from \citeasnoun{acs6}.
Col. (5): Black hole influence radius, defined as the radius containing a mass 
in stars equal to $2M_\bullet$, in pc (arcsec).
Col. (6): Binary stalling radius (Eq.~\ref{eq:astall}) for $q=0.5$, in pc (arcsec), based on Equation~(\ref{eq:astall}).
Col. (7): Binary stalling radius for $q=0.1$.
\end{table}

\begin{figure}
\begin{center}
\includegraphics[width=0.5\textwidth,angle=-90.]{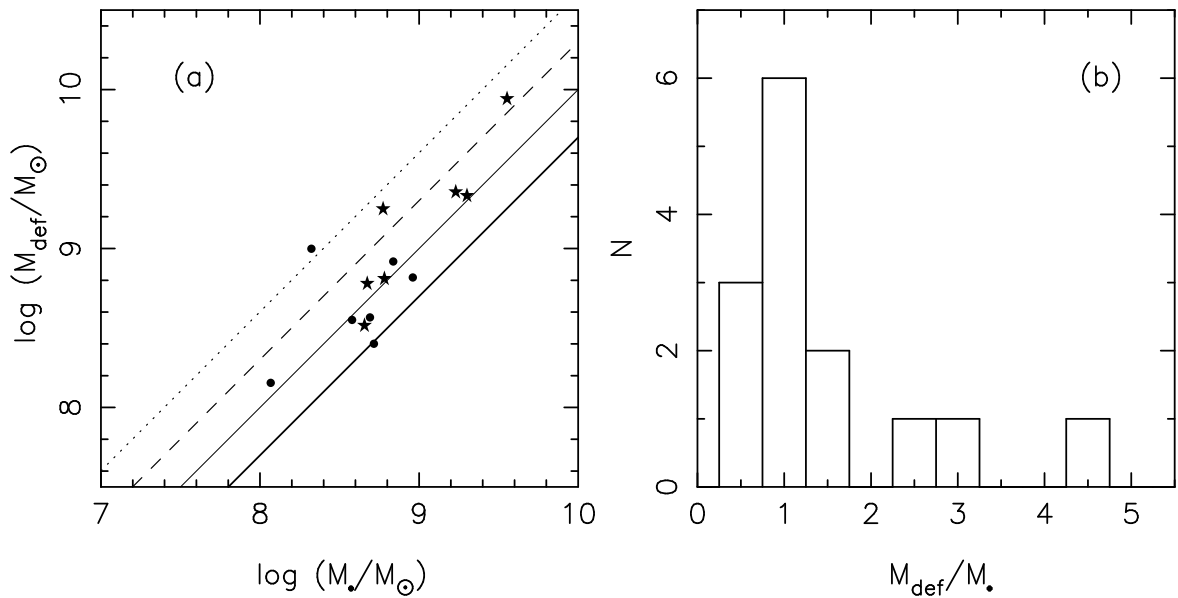}
\caption
{(a) Observed mass deficits, from \citeasnoun{graham-04}
(filled circles) and \citeasnoun{acs6} (stars).
Thick, thin, dashed and dotted lines show 
$M_{def}/M_\bullet = 0.5, 1, 2$ and $4$ respectively.
(b) Histogram of $M_{def}/M_\bullet$ values in (a).
\label{fig:mdefobs}
}
\end{center}
\end{figure}

\begin{figure}
\begin{center}
\includegraphics[width=0.75\textwidth,angle=-90.]{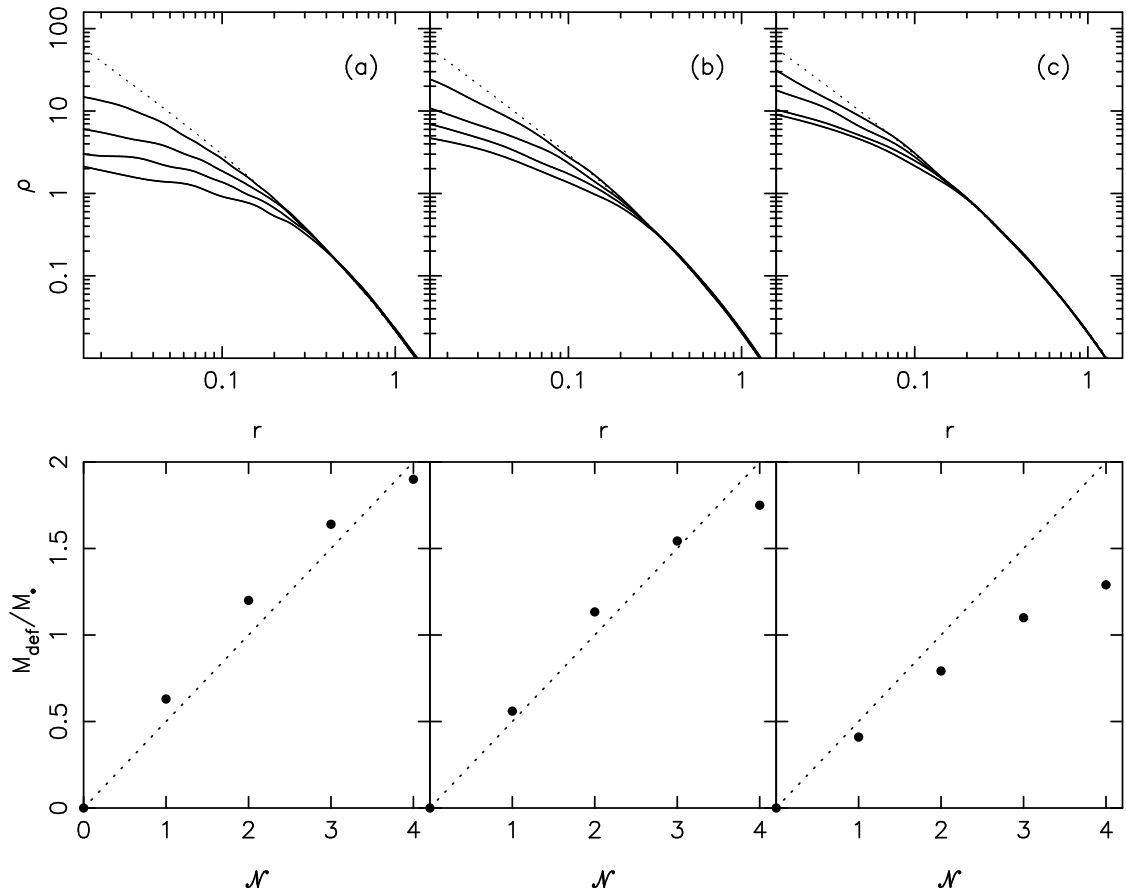}
\caption
{Density profiles (upper) and mass deficits (lower)
in $N$-body simulations of multi-stage mergers.
The initial galaxy (dotted lines, top panels)
contained a central point of mass $0.01M_{gal}$;
subsequent curves show density profiles
after repeated infall (followed by coalescence)
of a second ``black hole'' of mass $0.005$ (a),
$0.0025$ (b) and $0.001$ (c).
In the lower panels, 
points show $M_{def}/M_\bullet$ at $t_{stall}$ and
dotted lines show  $M_{def}/M_\bullet=0.5 {\cal N}$,
where $M_\bullet$ is the accumulated central mass.
(From \citename{merritt-06} \citeyear*{merritt-06}.)
\label{fig:hier}
}
\end{center}
\end{figure}

If the binary does stall at $a\approx a_{stall}$,
it will have given up an energy
\begin{eqnarray}
\Delta E &\approx& -{GM_1M_2\over 2r_h} + {GM_1M_2\over 2a_{stall}} \\
&\approx& -{1\over 2}M_2\sigma^2 + 2M_{12}\sigma^2 \\
&\approx& 2 M_{12}\sigma^2
\end{eqnarray}
to the stars in the nucleus, i.e.,
the energy transferred from the binary to 
the stars is roughly proportional to the 
{\it combined} mass of the two SBHs.
The reason for this counter-intuitive 
result is the $a_{stall}\sim M_2$
dependence of the stalling radius
(Eq.~\ref{eq:astall}): smaller infalling holes
form tighter binaries.
Detailed $N$-body simulations \cite{merritt-06} verify this
prediction:
the mass deficit (Eq.~\ref{eq:defmdef}) is found to be
\beq
{M_{def}\over M_{12}} \approx 0.70 q^{0.2}
\label{eq:mdef}
\eeq
for nuclei with initial density slopes
$1\lap\gamma\lap 1.5$.\footnote{The first $N$-body simulation to follow
the  destruction of a power-law density cusp by a massive binary 
was that of \citeasnoun{mm-01}, however the number
of particles used was so small that the evolution was 
dominated by spurious loss-cone refilling and the estimates of
$M_{def}$ accordingly uncertain.
Subsequent studies \cite{hemsendorf-02,chatterjee-03,makino-04} achieved
nearly  ``empty loss cones'' around the binary by using
mean-field algorithms and/or larger $N$, but these investigations
were all based on galaxy models with pre-existing cores,
again making it difficult to draw useful conclusions about 
the values of $M_{def}$ to be expected in real galaxies.}
Thus, in galaxies with pre-existing density profiles similar to 
those currently observed at the centers of the
Milky Way and M32, Equation~(\ref{eq:mdef}) implies that
mass deficits generated by stalled binaries 
should lie in the relatively narrow range
\beq
0.4\lap {M_{def}\over M_{12}} \lap 0.6,\ \ \ \ 
0.05\lap q \lap 0.5.
\eeq
Observed mass deficits are somewhat larger than this
\cite{mm-02,ravin-02,graham-04,acs6}.
Typical values are $M_{def}\approx M_\bullet$ and 
some galaxies have $M_{def}/M_\bullet$ as large
as $\sim 4$ (Table~\ref{tab:virgo}, Figure~\ref{fig:mdefobs}).
These numbers should be interpreted with caution
since both $M_{def}$ and $M_\bullet$ are subject
to systematic errors, the former from uncertain
$M/L$ corrections, the latter from various
difficulties associated with SBH mass estimation
(\S 4).
On the other hand, bright elliptical galaxies like M49 and M87 
(the first two galaxies in Table~\ref{tab:virgo}) 
have probably undergone numerous mergers
and the mass deficit should increase after each merger, 
even if expressed as a multiple of the final
(accumulated) SBH mass.
If the stellar mass displaced in a single
merger is $\sim 0.5 M_{12}$,
then -- assuming that the two SBHs always coalesce 
before the next SBH falls in --
the mass deficit following ${\cal N}$ mergers 
with $M_2\ll M_1$ is $\sim 0.5{\cal N}M_\bullet$.
$N$-body simulations 
verify this prediction 
(\citename{merritt-06} \citeyear*{merritt-06}, Figure~\ref{fig:hier}).
Mass deficits in the range
$0.5\lap M_{def}/M_\bullet\lap 1.5$
therefore imply $1\lap {\cal N}\lap 3$,
consistent with the number of gas-free
mergers expected for bright galaxies 
\cite{haehnelt-02}.
This correspondence constitutes strong evidence
that the cores of bright elliptical galaxies
are due to heating by binary SBHs.

Mass deficits as large as those in M49 and M87
($3\lap M_{def}/M_\bullet\lap 4$; Table~\ref{tab:virgo})
are harder to explain in this way, although a number
of additional mechanisms associated with binary
SBHs can ``heat'' a nucleus and increase $M_{def}$.
(a) If two SBHs fail to efficiently coalesce
a binary will be present when a third SBH falls in.
This scenario is conducive to smaller values
of $M_\bullet$, since one or more of the SBHs
could eventually be ejected by the gravitational
slingshot \cite{MV-90}; and to larger values  of $M_{def}$,
since multiple SBHs are more efficient than a binary
at displacing stars \cite{merritt-04b}.
Multiple-SBH interactions are discussed in more
detail in \S7.3.
(b) The gravitational-wave rocket effect
is believed capable of delivering kicks
to a coalescing binary as large as $\sim 150$ km s$^{-1}$
\cite{favata-04,blanchet-05,herrmann-06,baker-06}
and possibly much higher \cite{rr-89}.
The stellar density drops impulsively when the SBH 
is kicked out, and again when its orbit
decays via dynamical friction.
Mass deficits produced in this way can be as large as 
$\sim M_\bullet$ \cite{mmfhh-04,copycats-04}.
(c) Binaries might continue to harden beyond
$R_{12}\approx a_h$, as discussed below,
although it is not completely clear what the net
effect on $M_{def}$ would be.

\subsection{Late evolution of binary SBHs and the ``final-parsec problem''}

\begin{figure}
\centering
\includegraphics[scale=0.75]{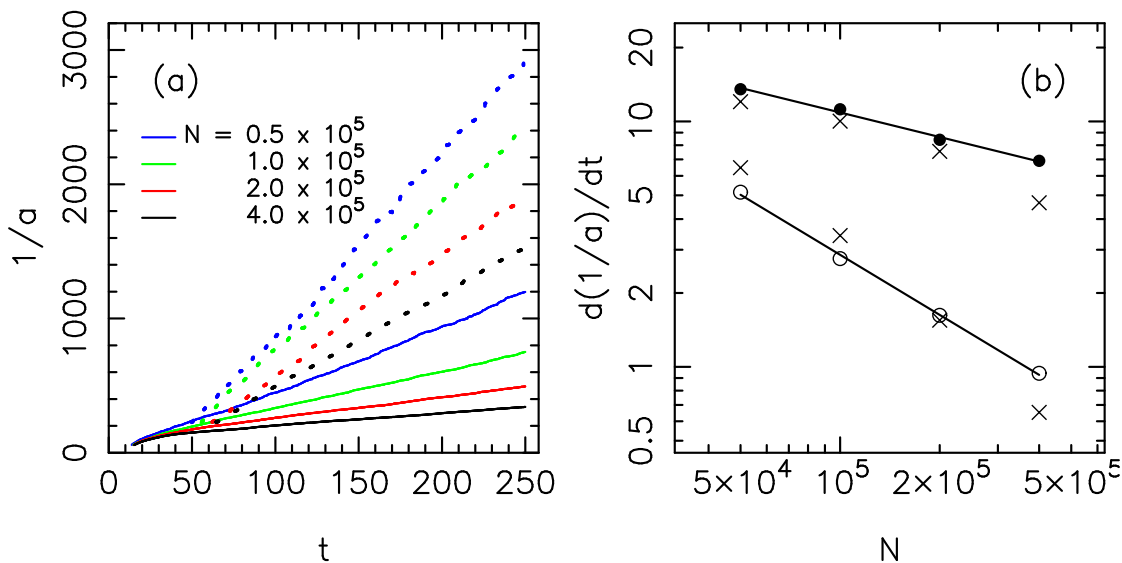}
\caption{
Long-term evolution of the binary semi-major axis (a) and
    hardening rate (b) in a set of high accuracy $N$-body simulations; 
    the initial galaxy model was a low-central-density Plummer sphere.
    Units are $G=M_{gal}=1$, $E=-1/4$, with $E$ the total energy.
    (a) Dashed lines are simulations with binary mass $M_1=M_2=0.005$
    and solid lines are for $M_1=M_2=0.02$, in units where the
    total galaxy mass is one.
    (b) Filled(open) circles are for $M_1=M_2=0.005(0.02)$.
    Crosses indicate the hardening rate predicted
    by a simple model in which the supply of stars to the binary
    is limited by the rate at which they can be scattered
    into the binary's influence sphere by gravitational encounters.
    The simulations with largest ($M_1,M_2$) exhibit the
    nearly $N^{-1}$ dependence expected in the ``empty loss cone''
    regime that is characteristic of real galaxies.
(Adapted from \citename{berczik-05} \citeyear*{berczik-05}.)
\label{fig:plummer}
}
\end{figure}

The rapid phase of binary evolution discussed above
is tractable using $N$-body codes since the
mechanisms that extract angular momentum from
the binary (dynamical friction from the stars, 
gravitational-slingshot ejection of stars by the binary) 
depend essentially on the mass density of 
the objects (stars) interacting with the binary
and not on their individual masses 
(at least for $m_\star\ll M_\bullet$).
Evolution of the binary beyond $a\approx a_h$
can be qualitatively different, and in fact
$N$-body simulations often find that the hardening 
rate $(d/dt)(1/a)$ is a decreasing function of $N$,
the number of ``star'' particles in the simulation,
once $a$ drops below $\sim a_h$
\cite{makino-04,berczik-05,mms-06}.
Figure~\ref{fig:plummer} illustrates the $N$-dependence for
equal-mass binaries in spherical galaxy models.
A straightforward extrapolation of results like these
to real galaxies implies that binary SBHs
would stall at separations of order $10^0$ pc 
(Table~\ref{tab:virgo}) --
the ``final-parsec problem''.

Stalling is expected in the large-$N$ limit since,
in a fixed, smooth potential,
the number of stars on orbits intersecting the binary
is limited, and in fact decreases with time
as the binary shrinks.
Stars can interact with the binary only if their pericenters
lie within $\sim {\cal R}\times a$, where ${\cal R}$ is of order
unity.
Let $L_{lc}={\cal R}a\sqrt{2\left[E-\Phi({\cal R}a)\right]}\approx 
\sqrt{2GM_{12}{\cal R}a}$,
the angular momentum of a star with pericenter ${\cal R}a$.
The binary's ``loss cone'' is the region in phase space defined by
$L\le L_{lc}$.
In a spherical galaxy, the mass of stars in the loss cone is
\begin{eqnarray}
M_{lc}(a)&=&m_*\int dE \int_0^{L_{lc}} dL\ N(E,L^2) \nonumber \\
&=& m_*\int dE\int_0^{L_{lc}^2} dL^2 4\pi^2 f(E,L^2) P(E,L^2) \nonumber \\
&\approx& 8\pi^2 GM_{12}m_* {\cal R}a\int dE f(E) P(E).
\label{eq:def_mlc}
\end{eqnarray}
Here $P$ is the orbital period, $f$ is the number density
of stars in phase space, and $N(E,L^2)dEdL$ is the number
of stars in the integral-space volume defined by $dE$ and $dL$.
In the final line,
$f$ is assumed isotropic and $P$ has been
approximated by the period of a radial orbit of energy $E$.
An upper limit to the mass that is available to interact with 
the binary 
is $\sim M_{lc}(a_h)$, the mass within the loss cone when the
binary first becomes hard; 
this is an upper limit since some stars that are initially within
the loss cone will ``fall out'' as the binary shrinks.
Assuming a singular isothermal sphere for the stellar distribution,
$\rho\propto r^{-2}$, and taking the lower limit of the energy
integral to be $\Phi(a_h)$, 
equation (\ref{eq:def_mlc}) implies
\begin{equation}
M_{lc}(a_h) \approx 3{\cal R}\mu.
\label{eq:mlc}
\end{equation}
The change in $a$ that would result if
the binary interacted with this entire mass can be
estimated using
the mean energy change of a star interacting 
with a hard binary, $\sim 3G\mu/2a$ \cite{quinlan-96}.
Equating the energy carried away by stars with
the change in the binary's binding energy gives
\begin{equation}
{3\over 2}{G\mu\over a} dM \approx {GM_1M_2\over 2}d\left({1\over a}\right)
\label{eq:dm1}
\end{equation}
or
\begin{equation}
\ln\left({a_h\over a}\right) \approx 3{\Delta M\over M_{12}} \approx
{9{\cal R}\mu\over M_{12}}\approx 9{\cal R} {q\over (1+q)^2}
\label{eq:dm2}
\end{equation}
if $\Delta M$ is equated with $M_{lc}$.
Equation~(\ref{eq:dm2}) suggests that
the mass available
to the binary would allow it to shrink by only a 
modest factor below $a_h$. 
In reality, the time scale for the  binary to
shrink is comparable with stellar orbital periods,
and some of the stars with pericenter distances
of order $a_h$ will only reach the binary after
it has shrunk to smaller separations.
And in most galaxies the density profile is
shallower than $\rho\sim r^{-2}$ implying
smaller $M_{lc}$ and less of a change in $a$.

The final parsec problem is a ``problem''
because it implies a low rate of SBH coalescence,
which is disappointing to physicists
hoping to detect the gravitational radiation
emitted during the final plunge \cite{LISA}.
It is also a ``problem'' in the sense that many 
circumstantial lines of evidence suggest that SBH binaries 
{\it do} efficiently coalesce.
(a) Only one, reasonably compelling case for a true
binary SBH exists \cite{rodriguez-06}, 
even though binaries with the same
projected separation ($\sim 7$ pc, which is the
expected stalling radius for a $M_{12}\approx 10^9M_\odot$ 
binary; see Table~\ref{tab:virgo})
could be easily resolved in many other galaxies
by radio interferometry.
(b) Jets in the great majority of radio galaxies
do not show the wiggles expected if the SBH hosting
the accretion disk were orbiting or precessing.
(c) Jets from Seyfert galaxies are randomly oriented
with respect to the disks in their host galaxies
\cite{ulvestad-84}.
This is naturally understood if SBH spins
were randomized at an earlier epoch
by binary coalescences during the
merger events that formed the bulges
\cite{merritt-02,kendall-03,saitoh-04}.
(d) If binary SBHs are common,
mergers will sometimes bring a third SBH into
a nucleus containing an uncoalesced binary,
resulting in three-body ejection of one or more 
of  the holes.
But the total mass density of SBHs in the 
local universe is consistent with that inferred
from high-redshift AGN \cite{MF-01b,yt-02}
implying that ejections are rare.
(e) Tight, empirical correlations between SBH
mass and galaxy properties 
\cite{fm-00,graham-01,marconi-03}
would also be weakened if SBH ejections were common.
(f) The frequency of SBH binary coalescences
estimated from $X$-shaped radio source statistics 
\cite{ekers-02} is roughly consistent with the galaxy merger rate,
implying that the time for coalescence is short compared
to the time between galaxy mergers.

A number of mechanisms have been proposed for efficiently
extracting angular momentum from binary SBHS and avoiding
the ``final parsec problem.''

\subsubsection {Non-axisymmetric geometries}

As discussed above (\S4), steady-state, non-axisymmetric
(e.g. triaxial) configurations are possible for nuclei,
even in the presence of chaos induced by the SBH,
and in fact departures from axisymmetry are often invoked 
to enhance fueling of AGN by gas \cite{shlosman-90}.
Many orbits in a triaxial nucleus are centrophilic, 
passing arbitrarily close 
to the center after a sufficiently long time
\cite{norman-83,gerhard-85}.
This implies feeding rates for a central binary
that can be higher than in the spherical geometry and,
more importantly, independent of $N$ in the large-$N$ 
limit \cite{mp-04}.

\begin{figure}
\includegraphics[scale=0.75]{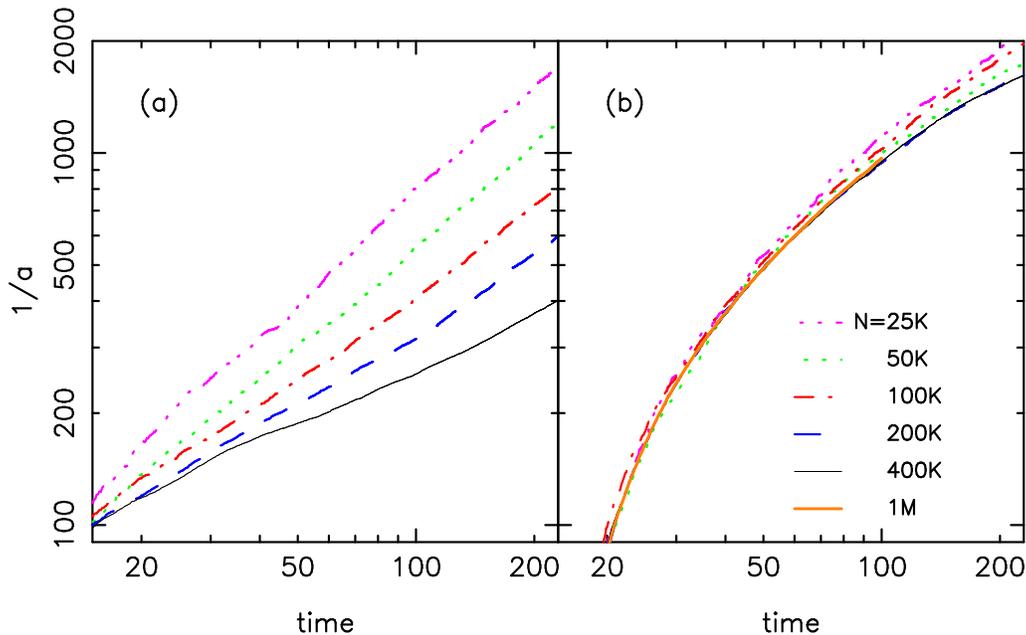}
\caption{
\footnotesize
Efficient merger of binary SBHs in barred galaxies.
Lines show the evolution of the inverse semi-major axis, $1/a$, 
of an equal-mass binary in $N$-body simulations with various $N$.
(a) Spherical, nonrotating galaxy model.  The binary hardening
rate declines with increasing $N$, i.e. the binary would
stall in the large-$N$ limit.
(b) Flattened, rotating version of the same model.
At $t\approx 10$, the rotating model forms a triaxial bar.
Hardening rates in this model are essentially independent
of $N$, indicating that the supply of stars to  the binary 
is not limited by collisional loss-cone refilling.
(From \citename{BMSB-06} \citeyear*{BMSB-06}).
}
\label{fig:berczik1}
\end{figure}

The total rate at which stars pass within a distance ${\cal R}a$ of
a central binary is 
\begin{equation}
\dot M \approx a\int A(E)M_c(E)dE
\end{equation}
where $M_c(E)dE$ is the mass on centrophilic orbits in the
energy range $E$ to $E+dE$, and $A(E)$ was defined above
(Eq.~\ref{eq:aofe}): $A(E)\times d$ is the rate at
which a single star on a centrophilic orbit of energy
$E$ experiences near-center passages with pericenter distances 
$\le d$.
The implied hardening rate is 
\beq
s\equiv {d\over dt} \left({1\over a}\right) 
\approx {2\langle C\rangle\over aM_\bullet}
\int \dot M(E) dE .
\eeq
Here, $\langle C\rangle\approx 1.25$ is the average value
of the dimensionless energy change during a single star-binary
encounter, $C\equiv [M_\bullet/2m_\star](\Delta E/E)$.
This expression can be evaluated for a $\rho\sim r^{-2}$
galaxy using Equation~(\ref{eq:aofesis}) for $A(E)$.
The result is
\beq
s \approx {4\sqrt{6}\over 9} 
{\langle C\rangle K\overline{f_c}\over \sigma r_h^2} 
\int e^{-(E-E_h)/2\sigma^2}dE \approx 2.5 \overline{f_c}{\sigma\over r_h^2}.
\label{eq:sanal}
\eeq
Here $\overline{f_c}$ is an energy-weighted, mean
fraction of centrophilic orbits, and the lower integration
limit was set to $E_h\equiv\Phi(r_h)$.
For the triaxial galaxy modelled in Figure~\ref{fig:berczik1},
the implied value of $s$ is $\sim 40\overline{f_c}$,
consistent with the measured peak value of 
$s\approx 20$.

This simple calculation, like the
$N$-body models on which Figure~\ref{fig:berczik1} was based,
are highly idealized, and it would be premature
to draw strong conclusions 
about the behavior of binary SBHs in more realistic
models of merging galaxies.
However these results do convincingly demonstrate that
binary hardening can be much more efficient
in triaxial geometries than in spherical geometries
due to the qualitatively different character
of  the stellar orbits.

\subsubsection{Collisional loss cone repopulation}

The theory of loss cones around single black holes (\S6)
can be applied, with only minor changes, to binary SBHs.
The orbital separation $a$ of the SBHs in a binary 
is much larger than the tidal disruption radius around 
a single SBH:
\beq
\frac{a}{r_{\rm t}}\approx
10^5 \times \eta^{-2/3} 
\left(\frac{\mh}{10^8M_\odot}\right)^{-1/3} 
\left(\frac{m_*}{M_\odot}\right)^{1/3} 
\left(\frac{r_*}{R_\odot}\right)^{-1} \left(\frac{a}{1\textrm{
    pc}}\right) .
\eeq
Since the physical scale of the loss cone is so
large for a binary,
the angular deflection of a star over
one orbital period will almost always be small
compared with the angular size subtended by the loss cone,
implying that most stars will wander 
``diffusively'' into the binary \cite{mm-03b}.
In the diffusive regime, the 
flux of stars into a sphere of radius $r_t$
scales only logarithmically with $r_t$ (Eq.~\ref{eq:fluxone}), 
and so the rate of supply of stars to a massive binary 
will be of the same order as the loss rate into 
a single SBH of the same mass, or (very roughly)
$\sim M_{12}/T_r(r_h)$.

In the bright elliptical galaxies that
exhibit clear evidence for the ``scouring'' effect of
binary SBHs, nuclear relaxation times are always
extremely long, $\gap 10^{14}$ yr (Fig.~\ref{fig:tr}).
In these galaxies, the mass in stars scattered into a 
central binary in  $10^{10}$ yr would be completely 
negligible compared with $M_{12}$.
Encounter-driven loss cone repopulation is
only likely to be significant in galaxies
with central relaxation times shorter
than $\sim 10^{10}$ yr, since in these galaxies the mass
scattered into the binary over the liftime
of the universe can be comparable with the binary's
mass.
Figures~\ref{fig:tr} and~\ref{fig:wang5} suggest that 
the only galaxies in this regime are those
-- like the Milky Way and M32 -- which
exhibit steep, power-law density profiles at
$r\lap r_h$.
If binary SBHs were ever present in these galaxies,
the low-density cores which they produced have
since presumably ``filled in'' via encounters (\S5.3) or
via new star formation;
hence the rate of scattering of stars into the binary's sphere
of influence would have been much lower in the past
than inferred from the current density profiles.

These arguments were largely confirmed by 
\citeasnoun{yu-02} in her study of
the influence of collisional loss-cone refilling
on binary SBH evolution in a sample of nearby galaxies.
\citeasnoun{yu-02} made a number of simplifying assumptions.
(1) The galaxy's density profile was assumed to be fixed in 
time; the ``scouring'' effect of the binary was ignored.
(2) The distribution of stars around the binary's
loss cone was assumed to be in a collisional
steady state, even in galaxies with nuclear
relaxation times $\gg 10^{10}$ yr.
(3) The influence of stars on the binary was computed
using the results of scattering experiments on 
isolated binaries.
Thus the hardening rate of the binary was computed
from
\begin{equation}
{d\over dt}\left({1\over a}\right) = H{G\rho\over\sigma}
\label{eq:def_H}
\end{equation}
where $\rho$ and $\sigma$ are the stellar density and 
velocity dispersion
and $H=H(a,e,q)$ a dimensionless rate coefficient
\cite{hills-83,hills-92,MV-92,quinlan-96,merritt-01}.
(4) Only changes in orbital angular momenta of the
stars were considered, even though the
stellar energy distribution would also change
significantly in the time $\sim T_r$ required
for significant feeding of the binary.
Under these assumptions, \citeasnoun{yu-02} found that 
binary hardening time scales were longer than a
Hubble time in almost all galaxies, with little
dependence on binary mass ratio.
Yu's more detailed conclusions about the 
properties of ``stalled'' binaries are probably
too rough to be very useful because of her neglect of the
influence of the binary on the density profile.

$N$-body techniques would seem to be better
suited to the collisional-loss-cone problem
since they can easily deal
with the strong, early effects of the binary
on the stellar distribution, and, when coupled 
with regularization schemes, can accurately treat 
binary-star interactions without the need for
rate coefficients derived from scattering experiments.
An example of a mechanism that is not reproduced
in scattering experiments is the ``secondary
slingshot,'' the repeated interaction of a star
with a binary \cite{mm-03b}.
But unless $N$ is very large,
relaxation in $N$-body simulations is so rapid
that the binary's loss cone
remains essentially full, and the diffusive loss
cone repopulation expected in real galaxies
will not be reproduced \cite{mm-03b}.

\begin{figure}
\centering
\includegraphics[scale=0.7,angle=0.]{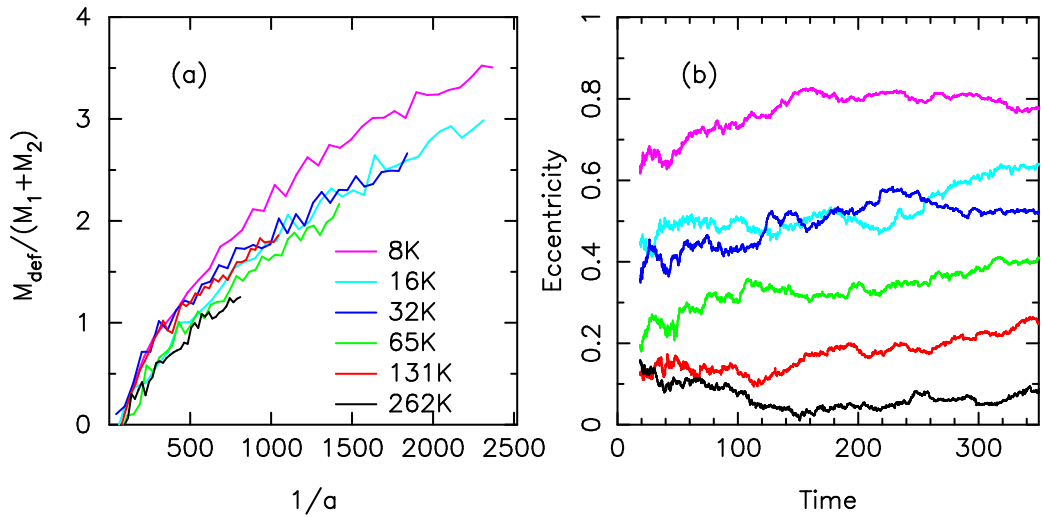}
  \caption{\it Results from a set of $N$-body integrations
    of the long-term evolution a massive binary in a galaxy 
    with a $\rho\sim r^{-0.5}$ density cusp.
    Each curve is the average of a set of integrations
    starting from different random realizations of the same
    initial conditions.
    (a) Evolution of the mass deficit (Eq.~\ref{eq:mdef}).
    For a given value of binary separation $a$,
    the mass deficit is nearly independent of particle
    number $N$
    (b) Evolution of binary eccentricity.
    The eccentricity evolution is strongly $N$-dependent
    and tends to decrease with increasing $N$, suggesting
    that the eccentricity evolution in real binaries would
    be modest.
(Adapted from \citename{mms-06} \citeyear*{mms-06}.)
  \label{fig:szell}
}
\end{figure}

Figure~\ref{fig:szell} shows a promising early step in 
this direction, using $N$-body models with $N$
up to $0.25\times 10^6$; the Mikkola-Aarseth chain 
regularization algorithm was used for close SBH-SBH and SBH-star
interactions \cite{mikkola-90,mikkola-93,aarseth-03b}.
The initial galaxy models had $\rho\sim r^{-0.5}$ near the center,
somewhat shallower than expected
in real galaxies, although steeper than the constant-density
cores assumed in most modelling studies
\citeaffixed{hemsendorf-02,chatterjee-03,makino-04}{e.g.}.
Adopting a steeper initial profile would
have produced models with essentially full loss cones;
in the models of Figure~\ref{fig:szell},
one can show that the loss cone defined by the binary
was only partially filled for the largest $N$ considered
\cite{mms-06}.
The $N$-dependence of the hardening rate was found to be
$(d/dt)(1/a)\sim N^{-0.4}$,
shallower than the $\sim N^{-1}$ dependence expected in real galaxies.
Simulations like these can therefore still not be scaled to 
real galaxies, but they are useful in exploring the
$N$-dependence of the changes induced by the binary on
the galaxy.
For instance,
Figure~\ref{fig:szell} (a) shows that mass deficits
are not strongly $N$-dependent when expressed as a function
of the binary separation $a$.
On the other hand, Figure~\ref{fig:szell} (b) suggests
that the evolution of the binary's eccentricity {\it is}
strongly $N$-dependent.
This may explain the rather disparate results on eccentricity
evolution in various $N$-body studies
\cite{mm-01,hemsendorf-02,aarseth-03}.

\subsubsection{Summary of Late Evolution Regimes}

\begin{table}
\caption{Regimes for Long-Term Evolution of Binary SBHs}
\begin{tabular}{lll} \hline \hline
\emph{Geometry} & \emph{Loss-Cone Regime} & \emph{Decay} \\ 
\hline
Spherical/ & Collisionless & $a^{-1} \propto \textrm{const}$ (stalls)\\
Axisymmetric & & \\
Spherical/ & Collisional (diffusive) & $a^{-1} \propto t/N$ \\
Axisymmetric & & \\
Spherical/ & Collisional (full loss cone) & $a^{-1} \propto t + \textrm{const}$ \\
Axisymmetric & & \\
Triaxial & Collisionless & $a^{-1} \propto t + \textrm{const}$ \\
\hline
\end{tabular}
\label{tab:regimes}
\end{table}

Table~\ref{tab:regimes}
 summarizes the different regimes of binary evolution
in stellar nuclei.
``Collisionless'' refers here (as elsewhere in this article)
to the large-$N$ limit in which
star-star gravitational encounters are inactive and
stars move along fixed orbits, until interacting
with the binary.
Almost all galaxies are expected to be in this regime.
The evolution of a real binary SBH may reflect a combination
of these and other mechanisms, such as interaction with gas
\cite{living}.
There is a close parallel between the final parsec problem
and the problem of quasar fueling: both requre that of order 
$10^8M_\odot$ be supplied to the inner parsec of a galaxy
in a time shorter than the age of the universe.
Nature clearly accomplishes this in the case of quasars,
probably through gas flows driven by torques from stellar bars.
The same inflow of gas could contribute to the decay of a 
binary SBH in a number of ways: by leading to the renewed formation
of stars which subsequently interact with the binary;
by inducing torques which extract angular momentum from the binary;
through accretion, increasing the masses of one or both
of the SBHs and reducing their separation; etc.

\subsection{Multiple Black Hole Systems}

If binary decay stalls, 
an uncoalesced binary may be present 
in a nucleus when a third SBH, or a second binary,
is deposited there following a subsequent merger.
The multiple SBH system that forms will engage in its 
own gravitational slingshot interactions, 
eventually ejecting one or more of the
SBHs from the nucleus and possibly from the galaxy
and transferring energy to the stellar fluid.

If the infalling SBH is less massive than either
of the components of the pre-existing binary,
$M_3<(M_1,M_2)$,
the ultimate outcome is likely to be ejection of the 
smaller SBH and recoil of the binary, with the binary
eventually returning to the galaxy center.
The lighter SBH is ejected with a velocity roughly $1/3$ the relative
orbital velocity of the binary \cite{saslaw-74,hut-92},
and the binary recoils with a speed that is lower
by $\sim M_3/(M_1+M_2)$.
Each close interaction of the smaller SBH with the binary
increases the latter's binding energy by 
$\langle \Delta E/E\rangle \approx 0.4 M_3/(M_1+M_2)$
\cite{hills-80}.
If $M_3>M_1$ or $M_3>M_2$, there will most often
be an exchange interaction, with the lightest SBH
ejected and the two most massive SBHs forming a
binary; further interactions then proceed as in 
the case $M_3<(M_1,M_2)$.

During the three-body interactions, both the semi-major
axis and eccentricity of the dominant binary change
stochastically.
Since the rate of gravitational wave emission is a strong function
of both parameters ($\dot E\propto a^{-4}(1-e^2)^{-7/2}$),
the timescale for coalescence can be enormously shortened.
This may be the most promising way to
coalesce SBH binaries in the low-density nuclei of massive galaxies,
where stalling of the dominant binary is likely.

This process has been extensively modelled 
using the PN2.5 approximation to represent gravitational wave
losses \cite{peters-63} and assuming a fixed
potential for the galaxy 
\cite{valtaoja-89,MV-90,valtonen-94}.
In these studies, 
there was no attempt to follow the pre-merger evolution
of the galaxies or the interaction of the binary SBHs with
stars.
In two short non-technical contributions (submissions for
the IEEE Gordon Bell prizes in 2001 and 2002), 
J. Makino and collaborators
mention two $N$-body simulations of triple
SBH systems at the centers of galaxies using 
the GRAPE-6, and (apparently) a modified version of NBODY1.
Relativistic energy losses were neglected and the SBH particles
all had the same mass.
Plots of the time evolution of the orbital parameters
of the dominant binary show strong and chaotic
eccentricity evolution,
with values as high as $0.997$ reached for short periods.
Such a binary would lose energy by gravity wave emission 
very rapidly, by a factor $\sim 10^8$ at the time of
peak $e$ compared with
a circular-orbit binary with the same semi-major axis.

In a wide, hierarchical triple, $M_3\ll (M_1,M_2)$,
the eccentricity of the dominant binary
oscillates through a maximum value of $\sim \sqrt{1-5\cos^2i/3}$,
$|\cos i|<\sqrt{3/5}$,
with $i$ the mutual inclination angle \cite{kozai-62}.
\citeasnoun{blaes-02}
estimated that the coalescence time of the dominant binary
in hierarchical triples can be reduced by factors of
$\sim 10$ via the Kozai mechanism;
\citeasnoun{iwasawa-05} recently observed Kozai oscillations in 
$N$-body simulations of galaxy models containing
equal-mass triples.

If the binary SBH is hard when the third SBH falls in,
the ejected SBH can gain enough velocity to escape the galaxy.
If the three masses are comparable, even the binary can 
be kicked up to escape velocity.\footnote{This idea
was developed very thoroughly as a model for
two-sided radio galaxies; the radio lobes were associated
with matter entrained with the ejected SBHs
(\citename{valtonen-00} \citeyear*{valtonen-00} and references
therein).
While the ``slingshot'' model for radio galaxies has not
gained wide acceptance, these papers are still an extremely
valuable resource for information about the statistics
of multiple-SBH interactions.}
One study \cite{volonteri-03} estimates (based on a very
simplified model of the interactions) that the recoil
velocity of the smallest SBH is larger than galactic
escape velocities in
99\% of encounters and that the binary escapes in 8\% of
encounters.
Thus a signficant fraction of nuclei could be left with
no SBH, with an offset SBH, or with a SBH whose mass is
lower than expected based on the $\mh-\sigma$ or $\mh-L_{bulge}$
relations.

There is a need for comprehensive $N$-body simulations of multiple-SBH systems
that include gravitational loss terms, accurate (regularized)
interactions between the SBH particles, and self-consistent treatment
of the stars.

\section{Acknowledgements}
\label{section:acknowledgements}
My understanding of the topics discussed here has benefited 
enormously over the years from discussions 
with many of my colleagues, including (but not limited to)
D. Axon,
P. Berczik,
P. Cot\'e,
L. Ferrarese,
D. Heggie,
P. Hut,
S. Komossa,
S. McMillan,
M. Milosavljevic,
S. Portegies Zwart,
G. Quinlan,
F. Rasio,
A. Robinson,
and
R. Spurzem.
This work was supported by grants 
AST-0206031, AST-0420920 and AST-0437519 from the NSF, 
grant NNG04GJ48G from NASA,
and grant HST-AR-09519.01-A from
STScI.  

\clearpage

\bibliography{ms}

\end{document}